\documentclass[12 pt]{article}
\usepackage{amsmath,mathrsfs}
\usepackage{amssymb}
\allowdisplaybreaks
\usepackage{amscd}
\usepackage{graphicx}
\usepackage{tabularx,booktabs}
\usepackage{array}
\newcolumntype{Z}{>{\centering\let\newline\\\arraybackslash\hspace{0pt}}X}
\usepackage[footnotesize,bf,textfont={it},margin=1 cm]{caption}

\def\be{\begin{equation}}
\def\ee{\end{equation}}
\def\frc#1#2{\relax\ifmmode{\textstyle\frac{#1}{#2}} 
                    \else$\frac{#1}{#2}$\fi}         

\def\rI{{{}_{\rm I}}}
\def\rJ{{{}_{\rm J}}}
\def\rK{{{}_{\rm K}}}

\def\hj{{\hat\jmath}}
\def\hk{{\hat k}}
\def\hi{{\hat\imath}}

\newcommand{\LP}{\left(}
\newcommand{\RP}{\right)}

\newcommand{\GF}{{\gamma}^5}
\newcommand{\GZ}{{\gamma}^0}

\newcommand{\GFZ}{\GF \GZ}

\def\fracm#1#2{\hbox{\large{${\frac{{#1}}{{#2}}}$}}}

\def\vCent#1{\vcenter{\hbox{\hss#1\hss}}}

\def\pp{{\mathchoice
          {
              \kern 1pt%
              \raise 1pt
              \vbox{\hrule width5pt height0.4pt depth0pt
                    \kern -2pt
                    \hbox{\kern 2.3pt
                          \vrule width0.4pt height6pt depth0pt
                          }
                    \kern -2pt
                    \hrule width5pt height0.4pt depth0pt}%
                    \kern 1pt
           }
            {
              \kern 1pt%
              \raise 1pt
              \vbox{\hrule width4.3pt height0.4pt depth0pt
                    \kern -1.8pt
                    \hbox{\kern 1.95pt
                          \vrule width0.4pt height5.4pt depth0pt
                          }
                    \kern -1.8pt
                    \hrule width4.3pt height0.4pt depth0pt}%
                    \kern 1pt
            }
            {
              \kern 0.5pt%
              \raise 1pt
              \vbox{\hrule width4.0pt height0.3pt depth0pt
                    \kern -1.9pt  
                    \hbox{\kern 1.85pt
                          \vrule width0.3pt height5.7pt depth0pt
                          }
                    \kern -1.9pt
                    \hrule width4.0pt height0.3pt depth0pt}%
                    \kern 0.5pt
            }
            {
              \kern 0.5pt%
              \raise 1pt
              \vbox{\hrule width3.6pt height0.3pt depth0pt
                    \kern -1.5pt
                    \hbox{\kern 1.65pt
                          \vrule width0.3pt height4.5pt depth0pt
                          }
                    \kern -1.5pt
                    \hrule width3.6pt height0.3pt depth0pt}%
                    \kern 0.5pt
            }
        }}

\def\mm{{\mathchoice
                       {
                             \kern 1pt
               \raise 1pt    \vbox{\hrule width5pt height0.4pt depth0pt
                                  \kern 2pt
                                  \hrule width5pt height0.4pt depth0pt}
                             \kern 1pt}
                       {
                            \kern 1pt
               \raise 1pt \vbox{\hrule width4.3pt height0.4pt depth0pt
                                  \kern 1.8pt
                                  \hrule width4.3pt height0.4pt depth0pt}
                             \kern 1pt}
                       {
                            \kern 0.5pt
               \raise 1pt
                            \vbox{\hrule width4.0pt height0.3pt depth0pt
                                  \kern 1.9pt
                                  \hrule width4.0pt height0.3pt depth0pt}
                            \kern 1pt}
                       {
                           \kern 0.5pt
             \raise 1pt  \vbox{\hrule width3.6pt height0.3pt depth0pt
                                  \kern 1.5pt
                                  \hrule width3.6pt height0.3pt depth0pt}
                           \kern 0.5pt}
                       }}

\def\ad{{\kern0.5pt
                   \alpha \kern-5.05pt \raise5.8pt\hbox{$\textstyle.$}\kern
0.5pt}}

\def\bd{{\kern0.5pt
                   \beta \kern-5.05pt \raise5.8pt\hbox{$\textstyle.$}\kern
0.5pt}}

\def\qd{{\kern0.5pt
                   q \kern-5.05pt \raise5.8pt\hbox{$\textstyle.$}\kern
0.5pt}}
\def\Dot#1{{\kern0.5pt
     {#1} \kern-5.05pt \raise5.8pt\hbox{$\textstyle.$}\kern
0.5pt}}

\catcode`@=11
\def\un#1{\relax\ifmmode\@@underline#1\else
        $\@@underline{\hbox{#1}}$\relax\fi}
\catcode`@=12

\def\a{\alpha}
\def\b{\beta}

\def\d{\delta}
\def\e{\epsilon}

\def\g{\gamma}

\def\l{\lambda}
\def\m{\mu}

\def\r{\rho}
\def\s{\sigma}
\def\t{\tau}

\def\dslash{\not{\hbox{\kern-2pt $\partial$}}}
\def\Dslash{\not{\hbox{\kern-4pt $D$}}}
\def\pslash{\not{\hbox{\kern-2.3pt $p$}}}
 \newtoks\slashfraction
 \slashfraction={.13}
 \def\slash#1{\setbox0\hbox{$ #1 $}
 \setbox0\hbox to \the\slashfraction\wd0{\hss \box0}/\box0 }
 
\font\ro=cmsy10                          
\def\kcr{{\hbox{\ro \char'170}}}                
\def\ktl{{\hbox{\ro \char'170}}}        
\def\ktr{{\hbox{\ro \char'170}}}        
\def\kbl{{\hbox{\ro \char'170}}}        
\def\kbr{{\hbox{\ro \char'170}}}        

\def\plpl{\raise-2pt\hbox{$\raise3pt\hbox{$_+$}\hskip-6.67pt\raise0.0pt
\hbox{$^+$}\hskip 0.01pt$}}
\def\mimi{\raise-2pt\hbox{$\raise3pt\hbox{$_-$}\hskip-6.67pt\raise0.0pt
\hbox{$^-$}\hskip 0.01pt$}} 

\def\bo{{\raise.15ex\hbox{\large$\Box$}}}               
\def\pa{\partial}                                       
\def\TH{{\raise.2ex\hbox{$\displaystyle \bigodot$}\mskip-4.7mu \llap H \;}}
\def\face{{\raise.2ex\hbox{$\displaystyle \bigodot$}\mskip-2.2mu \llap {$\ddot
        \smile$}}}                                      

\def\Tilde#1{\widetilde{#1}}                    
\def\Hat#1{\widehat{#1}}                        
\def\Bar#1{\overline{#1}}                       
\def\leftrightarrowfill{$\mathsurround=0pt \mathord\leftarrow \mkern-6mu
        \cleaders\hbox{$\mkern-2mu \mathord- \mkern-2mu$}\hfill
        \mkern-6mu \mathord\rightarrow$}
\def\dvec#1{\vbox{\ialign{##\crcr
        \leftrightarrowfill\crcr\noalign{\kern-1pt\nointerlineskip}
        $\hfil\displaystyle{#1}\hfil$\crcr}}}           
\def\dt#1{{\buildrel {\hbox{\LARGE .}} \over {#1}}}     

\def\fracm#1#2{\hbox{\large{${\frac{{#1}}{{#2}}}$}}}
\def\frac#1#2{{\textstyle{#1\over\vphantom2\smash{\raise.20ex
        \hbox{$\scriptstyle{#2}$}}}}}                   
\def\sfrac#1#2{{\vphantom1\smash{\lower.5ex\hbox{\small$#1$}}\over
        \vphantom1\smash{\raise.4ex\hbox{\small$#2$}}}} 
\def\bfrac#1#2{{\vphantom1\smash{\lower.5ex\hbox{$#1$}}\over
        \vphantom1\smash{\raise.3ex\hbox{$#2$}}}}       
\def\afrac#1#2{{\vphantom1\smash{\lower.5ex\hbox{$#1$}}\over#2}}    
\def\on#1#2{\mathop{\null#2}\limits^{#1}}               

\def\pa{\partial}      
\newcommand{\bm}[1]{\mbox{\boldmath$#1$}}

\def\ad{{\dot{\alpha}}}
\def\bd{{\dot{\beta}}}

\font\ro=cmsy10                          
\def\kcr{{\hbox{\ro \char'170}}}                
\def\ktl{{\hbox{\ro \char'170}}}        
\def\ktr{{\hbox{\ro \char'170}}}        
\def\kbl{{\hbox{\ro \char'170}}}        
\def\kbr{{\hbox{\ro \char'170}}}        

\parskip=\medskipamount                 
\thicklines                         

\def\border{                                            
        \setlength{\unitlength}{1mm}
        \newcount\xco
        \newcount\yco
        \xco=-21
        \yco=12
        \begin{picture}(140,0)
        \put(\xco,\yco){$\ktl$}
        \advance\yco by-1
        {\loop
        \put(\xco,\yco){$\kcr$}
        \advance\yco by-2
        \ifnum\yco>-240
        \repeat
        \put(\xco,\yco){$\kbl$}}
        \xco=158
        \yco=12
        \put(\xco,\yco){$\ktr$}
        \advance\yco by-1
        {\loop
        \put(\xco,\yco){$\kcr$}
        \advance\yco by-2
        \ifnum\yco>-240
        \repeat
        \put(\xco,\yco){$\kbr$}}
        \put(-20,13){\tiny **University of Maryland * Center for String and
         Particle  Theory* Physics Department***University of Maryland *Center
        for String and Particle  Theory** }
        \put(-20,-241.5){\tiny **University of Maryland * Center for String and
         Particle  Theory* Physics Department***University of Maryland *Center
        for String and Particle  Theory** }
        \end{picture}
        \par\vskip-8mm}

\def\headpic{                                           
        \indent
        \setlength{\unitlength}{.4mm}
        \thinlines
        \par
        \begin{picture}(29,16)
        \put(165,16){\line(1,0){4}}
        \put(170,16){\line(1,0){4}}
        \put(180,16){\line(1,0){4}}
        \put(175,0){\line(1,0){4}}
        \put(180,0){\line(1,0){4}}
        \put(185,0){\line(1,0){4}}
        \put(169,0){\line(0,1){16}}
        \put(170,0){\line(0,1){16}}
        \put(179,0){\line(0,1){16}}
        \put(180,0){\line(0,1){16}}
        \put(184,0){\line(0,1){16}}
        \put(185,0){\line(0,1){16}}
        \put(169,16){\oval(8,32)[bl]}
        \put(170,16){\oval(8,32)[br]}
        \put(179,0){\oval(8,32)[tl]}
        \put(185,0){\oval(8,32)[tr]}
        \end{picture}
        \par\vskip-6.5mm
        \thicklines}
\def\endtitle{\end{quotation}\newpage}                  

\newskip\humongous \humongous=0pt plus 1000pt minus 1000pt
\def\caja{\mathsurround=0pt}
\def\eqalign#1{\,\vcenter{\openup2\jot \caja
        \ialign{\strut \hfil$\displaystyle{##}$&$
        \displaystyle{{}##}$\hfil\crcr#1\crcr}}\,}
\newif\ifdtup

\begin{document}

\def\dt#1{\on{\hbox{\bf .}}{#1}}                
\def\Dot#1{\dt{#1}}

\def\gfrac#1#2{\frac {\scriptstyle{#1}}
        {\mbox{\raisebox{-.6ex}{$\scriptstyle{#2}$}}}}
\def\gg{{\hbox{\sc g}}}
\border\headpic {\hbox to\hsize{
\hfill
{UMDEPP-014-024}}}
\par \noindent
{ \hfill
}
\par

\setlength{\oddsidemargin}{0.3in}
\setlength{\evensidemargin}{-0.3in}
\begin{center}
{\large\bf Adinkras, 0-branes, Holoraumy and
the
SUSY QFT/QM Correspondence }\\[.3in]
Mathew Calkins\footnote{mathewpcalkins@gmail.com}${}^{\dagger}$, D.\ E.\ A.\ 
Gates\footnote{deagates@terpmail.umd.edu}${}^{\dagger}$, S.\, 
James Gates, Jr.\footnote{gatess@wam.umd.edu}${}^{\dagger}$,
and Kory Stiffler\footnote{kmstiffl@iun.edu}${}^*$
\\[0.1in]
${}^\dag${\it Center for String and Particle Theory\\
Department of Physics, University of Maryland\\
College Park, MD 20742-4111 USA}
\\[0.1in] 
and
\\[0.15in] 
${}^*${\it Department of Chemistry, Physics, and Astronomy\\
Indiana University Northwest\\
Gary, Indiana 46408 USA}
\\[0.2in]
{\bf ABSTRACT}\\[.01in]
\end{center}
\begin{quotation}
{We propose the recently defined ``Holoraumy Tensor''  to play a critical role in
defining a metric to establish a correspondence between 4D, $\cal N$-extended 
0-brane-based valise supermultiplet representations and, correspondingly via 
``SUSY Holography,'' on the space of 1D, $N$-extended network-based adinkras.
Using an analogy with the su(3) algebra, it is argued the 0-brane holoraumy
tensor plays the role of the ``d-coefficients'' and provides a newly established tool 
for investigating supersymmetric representation theory.}
\\[1.2in]
\vfill
\endtitle

\setcounter{equation}{0}
\section{Introduction}
$~~~$ 
There are two logical ways to 
construct one dimensional theories that possess the property of supersymmetry 
(SUSY) by starting from:

(a.) a higher dimensional supersymmetrical field theory and
applying \newline $~~~~~~~~~~~$ torus compactification (in the simplest case), or 

(b.) a one dimensional framework and building supersymmetry up without 
 \newline $~~~~~~~~~~~$ any assumptions.

\noindent
These two modes of construction will obviously lead to some area of overlap
and in this realm some interesting questions and possibilities occur.  The former
approach obviously will contain remnants of information about the symmetries of 
the higher dimensional spacetime.  How are these data ``downloaded'' onto the 
theories constructed from a purely adinkra network-based  \cite{adnk1,
DFGHILM, DFGHILM2, DFGHILM3, DFGHILM4,Top&, Top&2, Top&3, Top&4, 
Top&5} approach?  

On the other hand, we have argued the off-shell or auxiliary field problem for 
one dimensional linear representation SUSY theories has a solution \cite{ENUF, 
ENUF2, ENUF3, ENUF4, ENUF5, ENUF6, ENUF7} based on an algebraic structure 
we have denoted by the symbol $\cal {GR}($d, $N)$ or ``Garden Algebras.''  How 
can this information be ``uploaded,'' via the area of overlap between the two 
approaches, in order to more completely analyze the off-shell or auxiliary field 
problem for higher dimensional SUSY theories?  

This area of mutual overlap also suggests the possibility that kinematical 
information about the higher dimensional SUSY field theories can be encoded into 
one dimensional adinkra network-based  models.  We have long called this idea ``SUSY 
Holography.''   In a recent paper, \cite{adnkholor} there was introduced a new 
tensor on the space of adinkras that can be defined for all linear representations.  
We have argued a ``holoraumy tensor'' (in a consciously mixed provenance 
of Greek and German language contributions) is the ``Rosetta Stone'' for the 
translation of information between one dimensional adinkra network-based 
models and higher dimensional field theory ones.

Adinkras are graphical networks for 1D, $N$-extended SUSY representations,
and as such at first there would appear no way for them to carry information about
spin of higher dimensional theories.  In fact, the mechanism known as ``spin
from isospin'' \cite{JR} has long been known in the physics literature and a
variant on this seem workable for overcoming this objection.  Such a mechanism 
seems to be at work within SUSY holography and this variation of the idea is critical.

Among the adinkra graphs, there is a special class defined by its members 
possessing `height' assignments of nodes solely to two distinct levels.  
Members of this \newpage \noindent
restricted class of adinkras have been given the name of 
``valise adinkras.''  It had long been the suspicion of one of the authors (SJG) 
that valise adinkras are in some sense more elementary than all others.  
In recent works, the sense in which the valise adinkras are more elementary 
has come sharply into focus.  When the spinorial derivatives of one dimensional 
SUSY theories are evaluated on the fields of a valise adinkra in a specified 
way this leads to the appearance a tensor (that shares simultaneous attributes 
of the usual Riemann curvature tensor and the usual torsion tensor tensor).  This 
new tensor is the ``holoraumy'' tensor. 

Since the concept of the ``holoraumy'' tensor in supersymmetric representation
theory is a new one, to make it more familiar it is useful to review an analogy 
from group theory.  The Lie algebra su(3) provides a way to do this.  We use 
the Gell-Mann matrix representation of the su(3) generators and denote these 
by ${\bm t}{}_{A}$ leading to the familiar structure constants $f_{A \, B}{}^{C}$
\be
[\,   {\bm t}{}_{A}  ~,~  {\bm t}{}_{B} \, ] ~=~ i \, f_{A \, B}{}^{C} \, {\bm t}{}_{C}  
~~~, 
\label{I1}
\ee
of the algebra.  Next we define eight new matrices via the definitions ${\Tilde {\bm t}}
{}_{A}  \,=\,  -{\bm  t}_{A}^*$ and  find these new matrices satisfy
\be
[\,   {\Tilde {\bm t}}{}_{A}  ~,~  {\Tilde {\bm t}}{}_{B} \, ] ~=~ i \, f_{A \, 
B}{}^{C} \, {\Tilde {\bm t}}{}_{C}  ~~~, 
\label{I3}
\ee
and as this is identical to (\ref{I1}) these eight new matrices also form a
representation of the generators of su(3).  We can introduce a more compact 
notation by further defining $ {{\bm t}}{}_{A}^{(\# 1)}$ $\equiv$  ${{\bm t}}{}_{A}$ 
and $ {{\bm t}}{}_{A}^{(\# 2)}$ $\equiv$  ${\Tilde {\bm t}}{}_{A}$, so 
quadratic Casimir operators for each representation are define by
\be
{C}{}_{(2)}({\cal R}) ~=~ \sum_{A = 1}^8  \, {\bm t}{}_{A}^{({\cal R})}   \,  
{\bm t}{}_{A}^{({\cal R})}   
~~~,
\label{I4}
\ee
which on both takes the value of 16/3 times the 3 $\times$ 3 identity matrix.  Yet,
these two representations of the generators of the su(3) algebra are not
equivalent to one another.  That is there does not exist a unitary transformation
that relates one to the other.  One way to tease out this difference is by 
starting with the introduction ``d-coefficients'' for each representation
\be
\{ \,  {\bm t}{}_{A}^{({\cal R})}  ~,~  {\bm t}{}_{B}^{({\cal R})} \, \} ~=~ \fracm 13 
\delta{}_{A \, B} {\bm {\rm I}}{}_{3 \times 3} ~+~  d{}^{({\cal R})} _{A \, B}
{}^{C} \, {\bm t}{}_{C}^{({\cal R})}  ~~,
\label{I5}
\ee
with $d{}{}^{({\#1})} _{A \, B}{}^{C}$ $\equiv$  $d _{A \, B}{}^{C}$ and
$d{}{}^{({\#2})} _{A \, B}{}^{C}$ $\equiv$  ${\Tilde d}{} _{A \, B}{}^{C}$.
The index $\cal R$ plays the role of a ``representation label.''  We are
free to define a metric (not the Killing metric) ${\tilde g} ({\cal R},\, {\cal 
R}^{\prime})$ on this representation space via the equation
\be
{\tilde g} ({\cal R},\, {\cal R}^{\prime}) ~=~ {\cal N}_0 \sum_{A, \, B, \, C}
 d{}^{({\cal R})} _{A \, B} {}^{C} \,  d{}^{({\cal R}^{\prime})} _{A \, B} {}^{C}
\label{I6}
\ee
where ${\cal N}_0$ is a normalization constant.
The form of this metric leads to a 2 $\times$ 2 matrix
\be
{\tilde g} ({\cal R},\, {\cal R}^{\prime}) ~=~
\left[\begin{array}{cc}
1 & - 1 \\
- 1 & ~ 1 \\
\end{array}\right] ~~~,
\label{I7}
\ee
after using (\ref{I1}) - (\ref{I6}) and an appropriate choice of ${\cal N}_0$.
Thus the metric defined by (\ref{I6}) is capable of distinguishing between
the two representations of su(3).  To construct this new metric required
the d-coefficients and we assert holoraumy tensors are the analogs for 
supersymmetric representations.

Following the first lines of the introduction above about the two ways to 
construct 1D, $N$-extended supermultiplets, there exist two logically 
distinct constructions of the holoraumy tensor.  One follows from the line 
of argument utilizing higher dimensional theory as a starting point and 
the other from the solely one dimensional vantage.  On equating these 
two different definitions (a condition we have called the Adinkra/$\gamma
$-matrix Equation previously \cite{KIAS, KIAS2}) one defines a projection 
operator $\mathscr{P}$ capable of mapping higher dimensional 0-brane-based 
holoraumy results into one dimensional network-based holoraumy results.  
In the following we will explore additional properties of these holoramy 
tensors in some examples.

\setcounter{equation}{0}
\section{Network-Based Valise Adinkra  Representations}
$~~~$ 
We have established a well-defined methodology \cite{DFGHILM2, DFGHILM3, 
DFGHILM4} for creating a set of graphs (called `adinkras') capable of providing 
a basis for constructing linear representations of 1D, $N$-extended off-shell 
supermultiplets.  This methodology is {\em {totally}} divorced from any higher 
dimensional concepts, and accordingly the method contains no a priori relations 
to nor assumptions about symmetries in higher dimensions.  Conceptually
this approached can be based on networks and their associated ``adinkra'' 
adjacency matrices.  Three examples of these are shown below.
$$
\vCent
{\setlength{\unitlength}{1mm}
\begin{picture}(-20,-140)
\put(-89,-70){\includegraphics[width=3in]{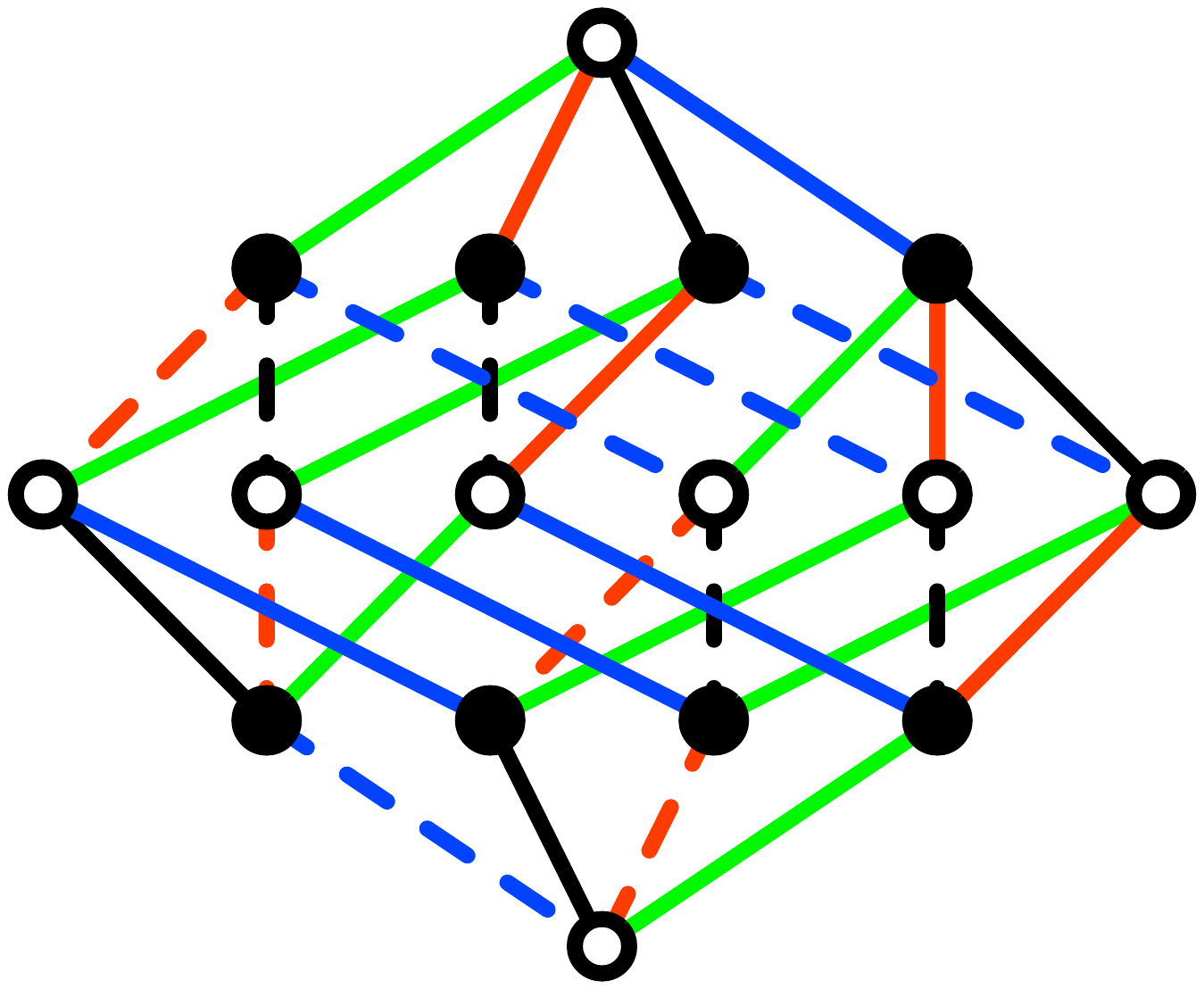}}
\put(-19,-39){\includegraphics[width=1.6in]{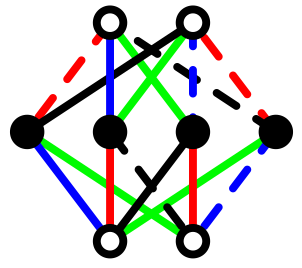}}
\put(32,-30){\includegraphics[width=1.6in]{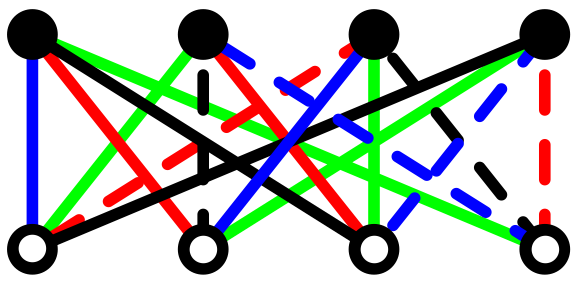}}
\put(-30,-49){\bf {Figure \# 1: Three adinkra graphs }}
\end{picture}}
$$
\newpage 


Of these three adinkras, only the final one has the valise property - all open
nodes appear at the same height in the diagram and all closed nodes
appear at the same height also but which is distinct from the height of the open
nodes.  This third adinkra is a member of the subclass of `valise adinkras.'  
Though the first two diagrams are not valise adinkras, through a series of 
operations designed for the lowering of nodes successively, these two 
adinkras can also be brought to the form of valise adinkras.  Thus, all 
three adinkras shown in the Fig.\ \# 2 are valise adinkras.
$$
\vCent
{\setlength{\unitlength}{1mm}
\begin{picture}(-20,-140)
\put(-89,-58){\includegraphics[width=3in]{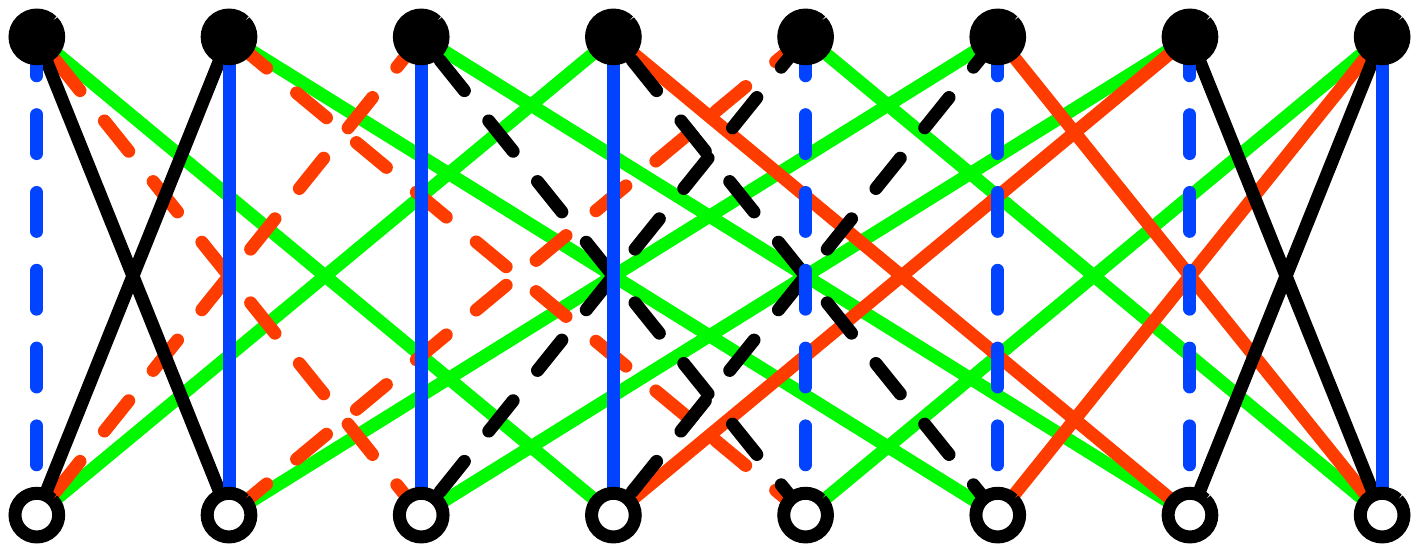}}
\put(-12,-20){\includegraphics[width=1.34in]{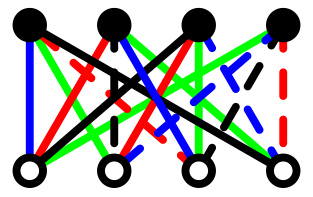}}
\put(32,-20){\includegraphics[width=1.7in]{offTM}}
\put(-50,-30){\bf {Figure \# 2: Three valise adinkra graphs }}
\end{picture}}
$$
\vskip1.1in
\noindent
The parameter $N$ describes the number of equivalence classes of the
links.  In the adinkras, it is shown by the $N$ distinct colors used to draw the
diagram.  Any two links in the same equivalence class possess the same
color.  

There is a subtlety we should mention before moving on.  The definition
of the adinkra graphs given above is a coordinate-independent one.  In
order to make any calculation with an adinkra, one must choose a set of 
coordinate basis elements.  This has been done for the three adinkras 
{$\cal R\,=\, {\rm {\#1}}$}, {${\rm {\#2}}$},  and {${\rm {\#3}}$} shown in Fig.\ 
\# 3
$$
\vCent
{\setlength{\unitlength}{1mm}
\begin{picture}(-20,-140)
\put(-54,0){\bf {{$\cal R\,=$} \# 1}}
\put(25,0){\bf {{$\cal R\,=$} \# 2}}
\put(-74,-36){\includegraphics[width=2.6in]{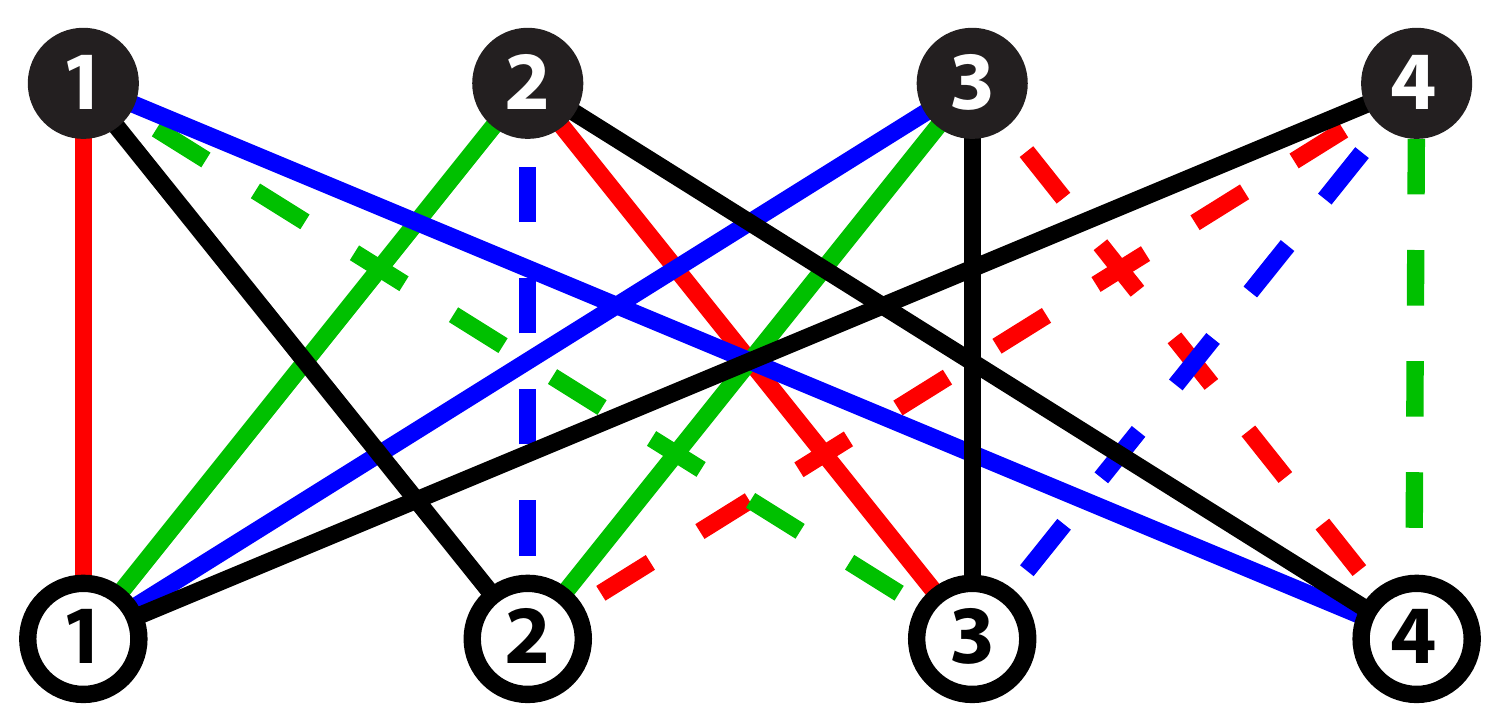}}
\put(6,-36.2){\includegraphics[width=2.6in]{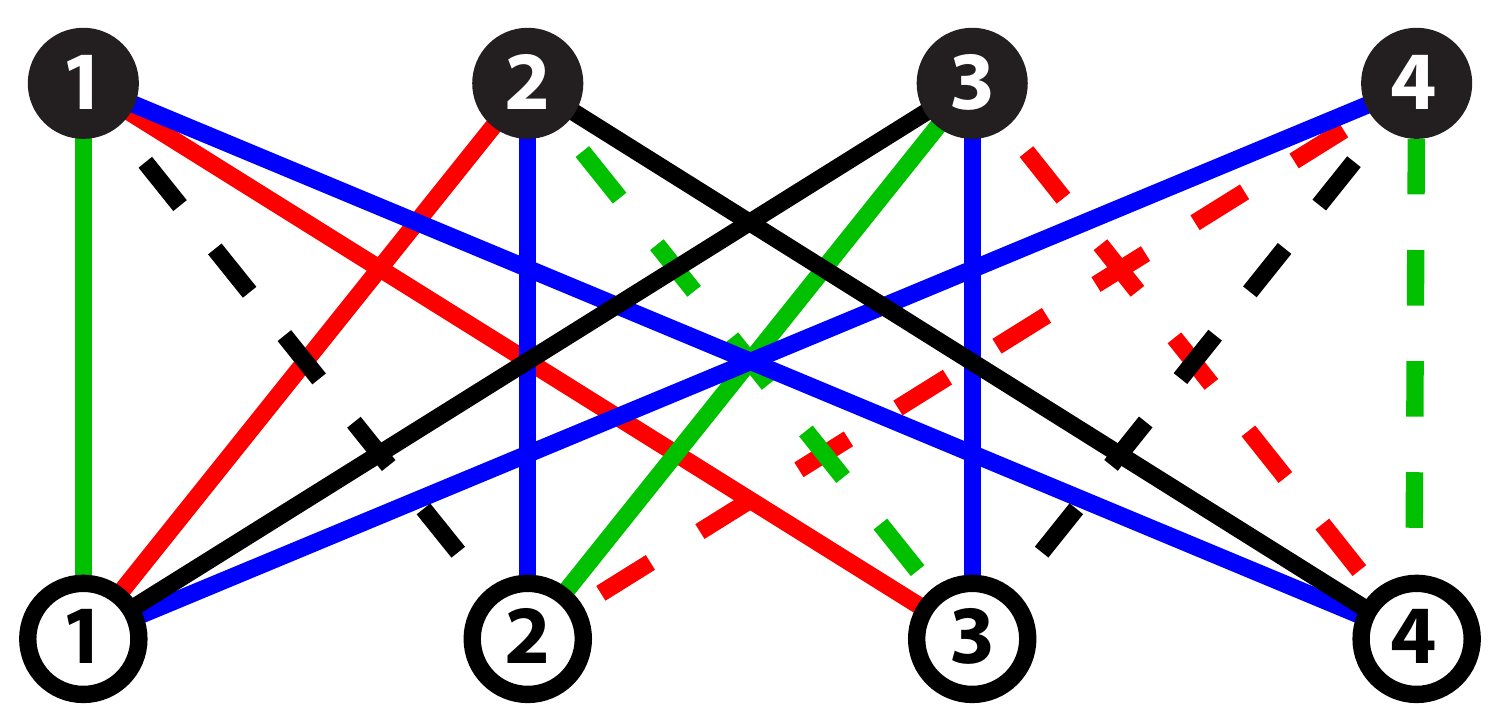}}
\put(-15,-44){\bf {{$\cal R\,=$} \# 3}}
\put(-34,-79){\includegraphics[width=2.6in]{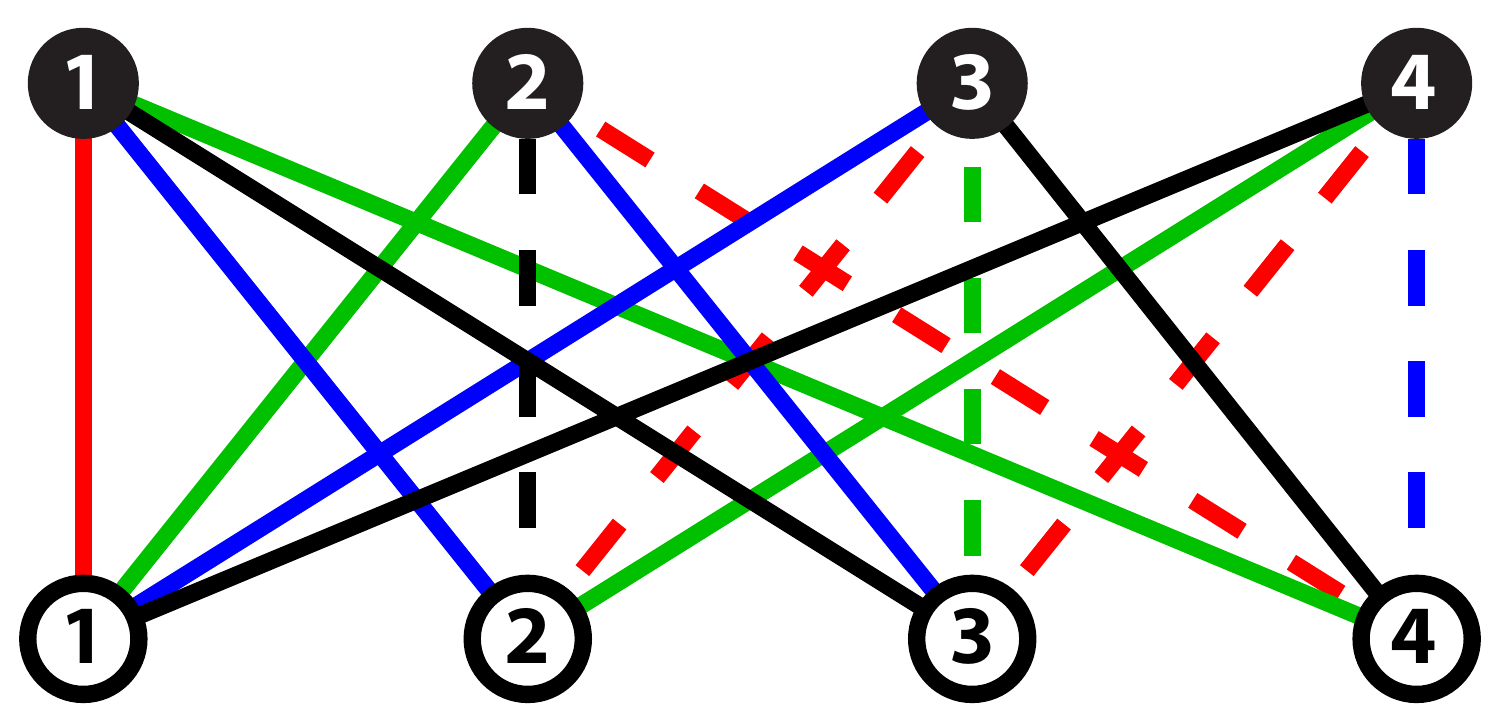}}
\put(-64,-86){\bf {Figure \# 3: Three valise adinkra graphs with node assignment}}
\end{picture}}
$$
\vskip3.2in \noindent
where the values of $\cal R$ can be used to describe which adinkra is
under consideration.  One of these choices of bases is the reference 
basis to which the open nodes are defined.  A second choice of these 
bases is the reference basis to which the closed nodes are defined.  
The final choice of basis is the one to which the links are referred.  The 
first basis choice is a choice of partitioning of the numbers 1 $\dots$ 
d among the bosonic nodes.   The second basis choice is a choice of 
partitioning of the numbers 1 $\dots$ d among the fermionic nodes.  
The final choice one of the numbers 1 $\dots$ $N$ among the classes 
of the links.

Each open node represents a bosonic function (or any of its derivatives) 
of $\tau$ (a time-like parameter) and collectively we denote these by 
$ \Phi_{i} $ with $i$ = 1, $\dots$, d and similarly, each closed node 
represents a fermionic function (or its derivatives) of $\tau$ and 
collectively we denote these by $ \Psi_{\hat k} $ with $\hat k$ = 1, $
\dots$, d.  As the construction of network-based adinkra graphs begins 
with the partition of an appropriate hypercube, the equality in the number 
of bosons and fermions follows as a consequence.  As in our example 
above, any adinkra which is not in the form of a valise can be brought to 
that form by judicious lowering of nodes. 

Valise adinkras are special in the form of the supercovariant derivative takes
when acting on the bosons and fermions in the supermultiplet can be expressed 
as
\be \eqalign{
{\rm D}{}_{{}_{\rm I}} \Phi_{i} ~=~ i \, \left( {\rm L}{}_{{}_{\rm I}}\right) 
{}_{i \, {\hat k}}  \, \, \Psi_{\hat k}  ~~~,~~~
{\rm D}{}_{{}_{\rm I}} \Psi_{\hat k} ~=~ \left( {\rm R}{}_{{}_{\rm I}}\right)
{}_{{\hat k} \, i}  \, \pa_{\tau} \, \Phi_{i}  ~~~,
}  \label{VH1}
\ee
in terms of constants $\left( {\rm L}{}_{{}_{\rm I}}\right) {}_{i \, {\hat k}}$
and $\left( {\rm R}{}_{{}_{\rm I}}\right) {}_{{\hat k} \, i}$ where I = 1, 
$\dots$, $N$.  As we have emphasized many times before (see for
example \cite{ENUF, ENUF2, ENUF3, ENUF4, ENUF5, ENUF6, ENUF7}), when an 
appropriate set of algebraic conditions are imposed on $\left( {\rm L
}{}_{{}_{\rm I}}\right) {}_{i \, {\hat k}}$ and $\left( {\rm R}{}_{{}_{\rm I}}
\right) {}_{{\hat k} \, i}$, the equations in (\ref{VH1}) imply 
\be  \eqalign{
\{\, {\rm D}{}_{{}_{\rm I}}  ~,~ {\rm D}{}_{{}_{\rm J}} \, \} \, \Phi_{i} ~&=~  
i \, 2 \,  \d{}_{{}_{\rm I} \, {}_{\rm J}} \,  \pa_{\tau} \, \Phi_{i}    ~~~, ~~~ 
\{ \, {\rm D}{}_{{}_{\rm I}}  ~,~ {\rm  D}{}_{{}_{\rm J}} \, \} \, \Psi_{\hat k} 
~=~ i\, 2 \,  \d{}_{{}_{\rm I} \, {}_{\rm J}} \, \pa_{\tau} \, \Psi_{\hat k}   
~~~,
}  \label{VH1a}
\ee
i.e.\ the collections of functions ($\Phi_{i}$, $\Psi_{\hat k} $) form a 
supermultiplet.

Using the basis described by the adinkras in Fig.\ \# 3, we find in
the respective cases of the three adinkras
$
\vCent
{\setlength{\unitlength}{1mm}
\begin{picture}(-20,-40)
\put(-32,-42){\bf {Table \# 1: Adinkra Link Color \& Cycles 
in L-matrices }}
\end{picture}}
$
\begin{table}[h] 
\vspace{0.2cm}
\begin{center}
\footnotesize
\begin{tabular}{|c|c|c|c|}  \hline 
$ {~} $ & $~~${$\cal R\,=$} \# 1$~~$ & $~~${$\cal R\,=$} \# 2$~~$ 
& $~~${$\cal R\,=$} \# 3$~~$ \\ \hline
$ RED $ & $ (10)_b \,(243) $ & $ (10)_b \,(1243) $ & $ (14)_b \,(234) $  \\  \hline
$ BLACK $ & $ (0)_b \,(142) $ & $ (6)_b \, (1342) $ & $ (2)_b \,(143) $  \\  \hline
$ GREEN $ & $ (12)_b \,(123) $ & $ (12)_b \,(23) $ & $ (4)_b \,(124) $  \\  \hline
$ BLUE $ & $ (6)_b \,(134) $ & $ (0)_b \,(14) $ & $ (8)_b \, (132) $ \\  \hline
\end{tabular}
\end{center}
\end{table} 
\noindent \vskip.05in  \noindent
where we have used a shorthand notation introduced in \cite{permutadnk,redux} to
denote 4 $\times$ 4 matrices.  The final step needed to determine
the relations of the L-matrices to the graph is the assignment of 
a `rainbow' (as described in \cite{adnkgeo}) where each color in assigned to
a number.  In the following we make the choices $red \,=\, 1$, 
$black \,=\, 2$, $green \,=\, 3$, and $blue \,=\, 4$. 

\section{Network-Based Fermionic Holoraumy Tenors} 
$~~~$ 
When an adinkra is a valise, there is another set of equations that are satisfied.  
These take the forms
\be  \eqalign{
[\, {\rm D}{}_{{}_{\rm I}}  ~,~ {\rm D}{}_{{}_{\rm J}} \, ] \, \Phi_{i}^{({\cal R})} ~&=~ - i
\,2\, \left[ \, {\cal B}{}^{({\cal R})} {}_{{}_{\rm I}}{}_{{}_{\rm J}} \, \right]  {}_{i \, { k}}  \, 
 \pa_{\tau} \, \Phi_{k}^{({\cal R})}     ~\,~~, \cr 
[\, {\rm D}{}_{{}_{\rm I}}  ~,~ {\rm D}{}_{{}_{\rm J}} \, ] \, \Psi_{\hat k}^{({\cal R})}  ~&=~ 
- i\, 2\, \left[ \, {\cal F}{}^{({\cal R})} {}_{{}_{\rm I}}{}_{{}_{\rm J}} \, \right]  {}_{{\hat k} \, 
{\hat \ell}}  \,  \pa_{\tau} \, \Psi_{\hat \ell}^{({\cal R})}     ~~~.
}  \label{VH2}
\ee
In the work of \cite{adnkholor}, we have given the quantities $\left[ \, {\cal B}{}^{({\cal R})} 
{}_{{}_{\rm I}}{}_{{}_{\rm J}} \, \right]  {}_{i \, { k}} $ and $\left[ \, {\cal F}{}^{({\cal R})} 
{}_{{}_{\rm I}}{}_{{}_{\rm J}} \, \right]  {}_{{\hat k} \, {\hat \ell}}$ respectively 
the names of the ``bosonic holoraumy tensor'' and the ``fermionic holoraumy 
tensor.''  We include the ``representation label'' $\cal R$ within these equations
to recognize that each of the three adinkras in Fig.\ \# 3 has distinct
L-matrices (and R-matrices) depending on the adinkra under consideration.

From (\ref{VH1}) and (\ref{VH2}), it follows
\be  \eqalign{ {~~}
\left[ \, {\rm D}{}_{\rm I} ~,~  {\rm D}{}_{\rm J} \, \right] \, \Phi_{i}^{({\cal R})}   
~&=~ - i\,2\, \left[ \, {\cal B}{}^{({\cal R})} {}_{{}_{\rm I}}{}_{{}_{\rm J}} \, \right]  
{}_{i \, { k}}  \,  \pa_{\tau} \, \Phi_{k}^{({\cal R})} \, ~=~  - \, i\, 2 \, [ \, V{}^{({\cal 
R})}_{\rI\rJ}\, ] {}_i{\,}^k \, \left( \pa_{ \tau} \Phi_{k}^{({\cal R})}   \right) ~~,  \cr
\left[ \, {\rm D}{}_{\rm I} ~,~  {\rm D}{}_{\rm J} \, \right] \, \Psi_{\hat{k}}^{({\cal 
R})}  ~&=~ - i\, 2\, \left[ \, {\cal F}{}^{({\cal R})} {}_{{}_{\rm I}}{}_{{}_{\rm J}} 
\, \right]  {}_{{\hat k} \, {\hat \ell}} \,  \pa_{\tau} \, \Psi_{\hat \ell}^{({\cal R})}  
~=~ - \, i\, 2 \, [ \, {\Tilde V}{}^{({\cal 
R})}_{\rI\rJ}\, ] {}_{\hat k}{\,}^{\hat \ell} \, \left( \pa_{ 
\tau} \Psi_{\hat \ell}^{({\cal R})}   \right) ~~~,
}  \label{VH3}
\ee
where
\be \eqalign{
\Big[ \, V{}_{\rI\rJ}^{({\cal R})} \, \Big] {}_i{\,}^k ~&=~  \frac12 \, \Big[ \, (\,{\rm 
L}_\rI\,{}^{({\cal R})} )_i{}^\hj\>(\,{\rm R}_\rJ\, {}^{({\cal R})} )_\hj{}^k ~-~ 
(\,{\rm L}_\rJ\, {}^{({\cal R})} )_i{}^\hj\>(\,{\rm R}_\rI {}^{({\cal R})} \,)_\hj{}^k 
\, \Big] 
~~~\,, ~~~\cr
\left[ \, {\Tilde V}{}_{\rI\rJ}^{({\cal R})}  \,\right] {}_\hi{\,}^\hk ~&=~   \frac12 \, 
\left[ \, (\,{\rm R}_\rI {}^{({\cal R})} \,)_\hi{}^j\>(\, {\rm L}_\rJ {}^{({\cal R})} \,
)_j{}^\hk ~-~ (\,{\rm R}_\rJ  {}^{({\cal R})} \, )_\hi{}^j\>(\,{\rm L}_\rI
{}^{({\cal R})} 
\,)_j{}^\hk  \, \right]
~~.
}  \label{VH4}
\ee
On the space of the three different adinkra representation seen in Fig.\ \# 3, we
can define a `metric'  via the definition
\be \eqalign{
{\cal G}\left[ \, ({\cal R})\,  , \, ({\cal R}^{\prime})\, \right] ~=~ \fracm 1{\,48\,} \, 
\sum_{\rI  , \, \rJ} {\rm {Tr}} \left[ \, {\Tilde V}{}_{\rI\rJ}^{({\cal R})} \, {\Tilde V
}{}_{\rI\rJ}^{({\cal R}^{\prime})}   \, \right]  ~~~~, }    \label{M6}
\ee
so a few computations reveal a 3 $\times$ 3 matrix
\be \eqalign{
{\cal G}\left[ \, ({\cal R})\,  , \, ({\cal R}^{\prime})\, \right] ~=~
\left[\begin{array}{ccc}
~1 & ~0 &  ~0  \\
~0 & ~~1 &  - \, \fracm13  \\
~0 & - \, \fracm13 &  ~~ 1 \\
\end{array}\right]   ~~~.
}  \label{M2}
\ee
In the work of \cite{permutadnk} we showed there exists 1,536 ways to construct 
adinkra networks of four colors, four open nodes, and four closed nodes. 
The ones shown in Fig.\ \# 3 are only three of this multitude.  So in principle,
for each of these 1,536 networks one could assign a value of $\cal R$
so \# 1 $\le$ $\cal R$ $\le$ \# 1,536 and for each of these values there
exists a calculable $\Tilde V$-matrix.  Thus, the metric defined by (\ref{M6})
can be extended over the entirety of a 1,536 dimensional space.

Let us close by noting in the discussion above, there is no reference
at all to higher dimensional supersymmetrical theories nor their symmetries.

\section{0-Brane-Based Valise Supermultiplets} 
$~~~$
By now in a number of our past works (e.\ g.\ \cite{G-1}), we have discussed the 
process by which we pass from a standard Lorentz-covariant four dimensional 
formulation of a supermultiplet to obtain a corresponding QM supermultiplet in 
valise form, where equations with the following form are valid 
\be \eqalign{
{\rm D}{}_a \, \Phi{}_{\Lambda}  ~=~ i\,  \left( {\rm L}{}_{\Lambda}\right){}_{a}
{}^{\Hat {\Lambda}} \, \, \Psi{}_{\Hat {\Lambda}}
~~~,~~~
{\rm D}{}_{a} \Psi{}_{\Hat {\Lambda}} ~=~  \left( {\rm R}{}^{\Lambda} \right)
{}_{\Hat {\Lambda}}{\,}_{a} \, \pa_{\tau} \, \Phi{}_{\Lambda}  ~~~,
} \label{covVAL}
\ee 
and where the explicit forms of $\Phi{}_{\Lambda} $ and $\Psi{}_{\Hat {\Lambda}}$ 
vary for each multiplet.  In particular for the CS we have
\be \eqalign{ {~~~}
\Phi{}_{\Lambda} ~&=~    \left(\, A, \, B, \, F, \, G  \,\right)   ~~,~~~
\Psi{}_{\Hat {\Lambda}} ~=~ \left( \, \psi{}_a  \,\right)   ~~~~~, \cr
} \ee
for the VS we have
\be \eqalign{ {~~~~}
\Phi{}_{\Lambda} ~&=~     \left(\, A{}_m, \, {\rm d}  \,\right)   ~~~~~\,~~~,~~~~
\Psi{}_{\Hat {\Lambda}} ~=~ \left( \, \lambda{}_a  \,\right)   ~~~~~, \cr
} \ee
for the TS we have
\be \eqalign{ {~~~~~}
\Phi{}_{\Lambda} ~&=~     \left(\, \varphi , \, B{}_{m \,n}   \,\right)   ~~~~~\,~,~~~~
\Psi{}_{\Hat {\Lambda}} ~=~ \left( \, \chi{}_a  \,\right)   ~~~~~, \cr
} \ee
for the AVS we have
\be \eqalign{ {~~~~~}
\Phi{}_{\Lambda} ~&=~     \big(\, U{}_m, \, {\Tilde {\rm d}}  \,\big)   ~~~~~~~,~~~~
\Psi{}_{\Hat {\Lambda}} ~=~ \big( \,  {\Tilde {\lambda}}{}_a  \,\big)   ~~~~~\,~, \cr
} \ee
and for the ATS we have
\be \eqalign{ {~~~~~}
\Phi{}_{\Lambda} ~&=~     \left(\, {\Tilde \varphi} , \, {\Tilde B}{}_{m \,n}   \,\right)   
~~~~~\,~,~~~~ \Psi{}_{\Hat {\Lambda}} ~=~ \left( \, {\Tilde \chi}{}_a  \,\right)   
~~~~~. \cr
} \ee
These equations demonstrate the indices $\Lambda$ range over distinct
bosonic representations of the SO(1, 3).  While in all of these examples the
${\Hat {\Lambda}}$ index corresponds to solely spinor indices of SO(1, 3), 
for more general supermultiplets (e.\ g.\ supergravity, etc.) this index can
range over distinct fermionic representations of SO(1, 3).  Below the explicit 
forms of the ${\rm L}{}_{\Lambda}$ and ${\rm R}{}^{\Lambda}$ coefficients 
can be read out for the cases of \vskip.1in
\noindent
(a.) Chiral Supermultiplet (CS);
\be
\eqalign{
{\rm D}_a A  ~&=~ \psi_a  ~~,~~
{\rm D}_a B   ~=~ i \, ( \gamma^5 )_a{}^b \psi_b  ~~,~~  
{\rm D}_a F  ~=~ ( \gamma^0)_a{}^b \, \psi_b  ~~, ~~
{\rm D}_a G   ~=~ i\, ( \gamma^5 \gamma^0 )_a{}^b \, \psi_b  ~~, 
\cr
{\rm D}_a \psi_b  ~&=~ i\, ( \gamma^0 )_{ab}  \left( \,\partial_{
\tau} A  \, \right) - ( \gamma^5 \gamma^0)_{ab}  \left( \, 
\partial_{\tau} B  \, \right) ~-~ i C_{ab}  \left( \,  \partial_{\tau} F  \, \right) + 
( \gamma^5 )_{ab}  \left( \, \partial_{\tau} G   \, \right) 
~~~,
} \label{BV2ZZ} 
\ee
\noindent
(b.) Vector Supermultiplet (VS);
\be
\eqalign{
{\rm D}{}_a A_m & ~=~ (\gamma_m)_a{}^b \lambda_b  ~~~, ~~~ {\rm D}_a 
{\rm d}  \, =\,  i (\gamma^5 \gamma^0)_a{}^b \, \lambda_b 
 ~~~~~~\,~~~~, \cr
{\rm D}{}_a \lambda_b & ~=~ - i \, (  \gamma^0 
\gamma^m )_{ab} \,  \left( \,  \pa_{\tau} A_m  \, \right)~+~ (\gamma^5
)_{ab} \,  \left( \, \pa_{\tau} {\rm d} \, \right) 
~~~,  
}  \label{BV4ZZ} 
\ee
\noindent
(c.) Tensor Supermultiplet (TS);
\be
\eqalign{
{\rm D}{}_a \varphi ~&=~ \chi_a   ~~~, ~~~
{\rm D}{}_a B_{m \, n}  ~=~ - \tfrac{1}{4} ([\gamma_m,\, 
\gamma_n])_a{}^b \chi_b  ~\,~~, \cr
{\rm D}{}_a \chi_b ~&=~ i (\gamma^0)_{ab} \,  \partial_{\tau}
\varphi - i  \tfrac{1}{2} (\gamma^0 \, [\gamma^m,\, \gamma^n]
)_{ab}  \, \partial_\tau B_{m \, n}  
 ~~~\,~,
}  \label{BV6ZZ} 
\ee
\noindent
(d.) Axial vector Supermultiplet (AVS); and
\be  \eqalign{
{\rm D}{}_a U_{m} & ~=~ i \, (\gamma^5 \gamma_{m})_a{}^b {\Tilde 
{\lambda}}{}_b   ~~,~~
 {\rm D}_a {\Tilde {\rm d}}  ~=~  -\,  ( \gamma^{0})_a{}^b \, \pa_{\tau} 
{\Tilde {\lambda}}{}_b  ~~,  
 \cr
{\rm D}{}_a {\Tilde {\lambda}}{}_b & ~=~   (\gamma^5  \gamma^{0}
\gamma^{m} )_{ab} \,   \left( \,  \pa_{\tau} U_m  \, \right)  \, ~+~ i C{}_{ab} \,   
{\Tilde {\rm d}} ~~,~~}  \label{BVa4ZZ}
\ee
\noindent
(e.) Axial tensor Supermultiplet (ATS)
\be
\eqalign{
{\rm D}{}_a {\Tilde \varphi} ~&=~   i \, ( \gamma^5 )_a{}^b  {\Tilde \chi}{}_b   
~~~, ~~~ {\rm D}{}_a  {\Tilde B}{}_{m \, n}  ~=~ -i\,  \tfrac{1}{4} (  \gamma^5 
[\gamma_m,\, \gamma_n])_a{}^b  {\Tilde \chi}{}_b  ~\,~~, \cr
{\rm D}{}_a  {\Tilde \chi}{}_b ~&=~ - (\gamma^0 \gamma^5)_{ab} \,  \partial_{
\tau}{\Tilde \varphi} +   \tfrac{1}{2} (\gamma^0 \gamma^5 \, [\gamma^m,\, 
\gamma^n])_{ab}  \, \partial_\tau  {\Tilde B}{}_{m \, n}   
 ~~~~~~~~~~~~.
}  \label{BV6ZZz} 
\ee

From (\ref{covVAL}) it follows
\be \eqalign{  {~~~~}
\left[  \, {\rm D}{}_a ~,~ {\rm D}{}_b \,  \right]  \, \Phi{}_{\Lambda}  ~&=~ -\,  i\, \left[ \, 
\left( {\rm L}{}_{\Lambda}\right){}_{a}{}^{\Hat {\Lambda}} \, \left( {\rm R}{}^{\Delta} 
\right){}_{\Hat {\Lambda}}{\,}_{b} ~-~   \left( {\rm L}{}_{\Lambda}\right){}_{b}{}^{\Hat 
{\Lambda}} \, \left( {\rm R}{}^{\Delta} \right){}_{\Hat {\Lambda}}{\,}_{a} \, \right] \, 
\pa_{\tau} \, \Phi{}_{\Delta}   ~~~~, \cr
\left[  \, {\rm D}{}_a ~,~ {\rm D}{}_b \,  \right]  \, \Psi{}_{\Hat {\Lambda}}  ~&=~ -\,  i 
\, \left[ \,  \left( {\rm R}{}^{\Delta} \right){}_{\Hat {\Lambda}}{\,}_{a} 
 \, \left( {\rm L}{}_{\Delta}\right){}_{b}{}^{\Hat {\Delta}} ~-~
  \left( {\rm R}{}^{\Delta} \right){}_{\Hat {\Lambda}}{\,}_{b} 
 \, \left( {\rm L}{}_{\Delta}\right){}_{a}{}^{\Hat {\Delta}} 
\, \right] \,  \pa_{\tau} \, \Psi{}_{\Hat {\Delta}   ~~~. }
}  \label{covVAL1}
\ee 
The first one of these is a calculation that requires taking the antisymmetric 
part of some products of strings of $\gamma$-matrices.  The second one of 
these is of the form of Fierz identities of some products of strings of $\gamma
$-matrices.  So the character of the coefficients appearing is very different 
from the similar looking coefficients in (\ref{VH4}).  At a minimum as the coefficients
in (\ref{covVAL1}) are constructed from $\gamma$-matrices, information
about the symmetries of spacetime are inherited in the 1D supermultiplets
discussed in this chapter.   The question is, ``How is such information
to be recovered in the supermultiplets constructed from network-based adinkras?''

\section{0-Brane Fermionic Holoraumy Tenors} 
$~~~$ 
Direct calculations show the operator equation
\be
 \{ \, {\rm D}{}_a ~,~ {\rm D}{}_b \, \} ~=~ i \, 2 \, (\gamma^0){}_{a \, b} \, \pa_{\tau}
 ~~~,
\label{BH1}
\ee
is satisfied on all fields in  (\ref{BV2ZZ}) - (\ref{BV6ZZz}) as expected by SUSY.  
However, as each of the equations in  (\ref{BV2ZZ}) - (\ref{BV6ZZz}) describes 
a valise supermultiplet, there is something else we can do.  The anti-commutator 
in (\ref{BH1}) can be replaced by a commutator.

Our conventions are such that when using the commutator (not the 
anti-commutator as in (\ref{BH1})) of two spinor covariant derivatives evaluated 
on the fermions of a covariant valise supermultiplet we write
\be  \eqalign{ {~~~~}
[ \, {\rm D}{}_{a}  ~,~ {\rm D}{}_{b} \, ] ~&=~ - \, i\, 2\, (\, \mathscr{F}{}_{{a } \, 
{b}} \,) \, \pa_{\t} ~~~,
} \label{BH2}
\ee
where $ (\, \mathscr{F}{}_{{a } \, {b }} \,) $ denotes a set of representation dependent 
constants (these will be discussed in more detail via examples below) and when 
evaluated on the bosons of a valise supermultiplet we write
\be  \eqalign{ {~~~~}
[ \, {\rm D}{}_{a }  ~,~ {\rm D}{}_{b} \, ] ~&=~ - \, i\, 2\, (\, \mathscr{B}{}_{{a} \, {b}} \,) 
\, \pa_{\t} ~~~,
} \label{BH3}
\ee
where $ (\, \mathscr{B}{}_{{a} \, {b }} \,) $ denotes a set of representation dependent 
constants once more.  In the remainder of this chapter, we are going to focus on
the fermionic holoraumy tensor in (\ref{BH2}) and defer discussion of the bosonic
holoraumy tensor in (\ref{BH2}) until an appendix.

For the fermionic fields in (\ref{BV2ZZ}) - (\ref{BV6ZZz}) we find
\be \eqalign{ 
( \, \mathscr{F}{}_{a}{}^{b} {}^{(CS)} \, ){}_c {}^d ~&=~   -   (\gamma^5 \gamma^m
)_{a}{}^{b} \, (\gamma^{5} \gamma^{0} \gamma_{m})_c^{~d} ~~~,
\cr   
( \, \mathscr{F}{}_{a}{}^{b}{}^{(VS)}\, ){}_c{}^{d} ~&=~ +\delta_{a}{}^{b} (
\gamma^0)_c^{~d} + (\gamma^5)_{a}{}^{b} (\gamma^5\gamma^0)_{c
}^{~d} + (\gamma^5\gamma^0)_{a}{}^{b}(\gamma^5)_c^{~d}   ~~~, \cr   
( \, \mathscr{F}{}_{a}{}^{b}{}^{(TS)}\, ){}_c{}^{d} ~&=~ - \delta_{a}{}^{b} (
\gamma^0)_{c}^{~d} +  (\gamma^5)_{a}{}^{b} (\gamma^5 \gamma^0)_{
c}^{~d} - (\gamma^5\gamma^0)_{a}{}^{b} (\gamma^5)_{c}^{~d} ~~~, \cr
( \, \mathscr{F}{}_{a}{}^{b}{}^{(AVS)}\, ){}_c{}^{d} ~&=~ -\delta_{a}{}^{b} (
\gamma^0)_c^{~d} - (\gamma^5)_{a}{}^{b} (\gamma^5\gamma^0)_{c
}^{~d} + (\gamma^5\gamma^0)_{a}{}^{b}(\gamma^5)_c^{~d} ~~~,    \cr
( \, \mathscr{F}{}_{a}{}^{b}{}^{(ATS)}\, ){}_c{}^{d} ~&=~ + \delta_{a}{}^{b} 
(\gamma^0)_{c}^{~d} -  (\gamma^5)_{a}{}^{b} (\gamma^5 \gamma^0)_{
c}^{~d} - (\gamma^5\gamma^0)_{a}{}^{b} (\gamma^5)_{c}^{~d} ~~~.
}  \label{BH4}
\ee
A further set of identities given by
\be \eqalign{ 
( \, \mathscr{F}{}_{a}{}^{b}{}^{(CS)}\, ){}_c{}^{d} &=~ (\gamma^5)_{c}{}^{e} \,
 ( \, \mathscr{F}{}_{a}{}^{b}{}^{(CS)}\, ){}_e{}^{f} \, (\gamma^5)_{f}{}^{d}  ~~~, \cr
( \, \mathscr{F}{}_{a}{}^{b}{}^{(AVS)}\, ){}_c{}^{d} &=~ (\gamma^5)_{c}{}^{e} \,
 ( \, \mathscr{F}{}_{a}{}^{b}{}^{(VS)}\, ){}_e{}^{f} \, (\gamma^5)_{f}{}^{d} ~~~, \cr
( \, \mathscr{F}{}_{a}{}^{b}{}^{(ATS)}\, ){}_c{}^{d} &=~ (\gamma^5)_{c}{}^{e} \,
 ( \, \mathscr{F}{}_{a}{}^{b}{}^{(TS)}\, ){}_e{}^{f} \, (\gamma^5)_{f}{}^{d} 
~~~,
}  \label{BH4p}
\ee
are also valid.  These equations have clear interpretations. Under the
conjugation operation defined by a similarity transformation using $\gamma^5$,
the holoraumy of the valise chiral supermultiplet is self-conjugate, while
the holoraumy of the vector (tensor) valise supermultiplet is conjugate
to the holoraumy of the axial-vector (axial-tensor) valise supermultiplet. 

A second type of conjugation of the covariant holoraumy tensors can
be defined by replacing the $\gamma^5$ in (\ref{BH4p}) by $\gamma^0$.
From the equations above (\ref{BH4}), it also follows
\be \eqalign{  {\,}
 (\gamma^0)_c^{~e}  \, ( \, \mathscr{F}{}_{a}{}^{b} {}^{(CS)} \, ){}_e {}^f \,  
 (\gamma^0)_f^{~d} ~&=~  +   (\gamma^5 \gamma^m)_{a}{}^{b} 
\, (\gamma^{5} \gamma^{0} \gamma_{m})_c^{~d} ~~~,
\cr   
(\gamma^0)_c^{~e}  \, ( \, \mathscr{F}{}_{a}{}^{b} {}^{(VS)} \, ){}_e {}^f \,  
(\gamma^0)_f^{~d} ~&=~ - \delta_{a}{}^{b} (\gamma^0)_c^{~d} 
+ (\gamma^5)_{a}{}^{b} (\gamma^5\gamma^0)_{c}^{~d} + (\gamma^5
\gamma^0)_{a}{}^{b}(\gamma^5)_c^{~d}   
~~~, \cr   
(\gamma^0)_c^{~e}  \, ( \, \mathscr{F}{}_{a}{}^{b} {}^{(TS)} \, ){}_e {}^f \,  
(\gamma^0)_f^{~d} ~&=~ + \delta_{a}{}^{b} (\gamma^0)_c^{~d} 
+ (\gamma^5)_{a}{}^{b} (\gamma^5\gamma^0)_{c}^{~d} - (\gamma^5
\gamma^0)_{a}{}^{b}(\gamma^5)_c^{~d}   ~~~, \cr 
(\gamma^0)_{c}{}^{e} \,
( \, \mathscr{F}{}_{a}{}^{b}{}^{(AVS)}\, ){}_e{}^{f} \, (\gamma^0)_{f}{}^{d} 
~&=~ + \delta_{a}{}^{b} (\gamma^0)_c^{~d} - (\gamma^5)_{a}{}^{b} (
\gamma^5\gamma^0)_{c}^{~d} + (\gamma^5\gamma^0)_{a}{}^{b
}(\gamma^5)_c^{~d}  ~~~, \cr   
(\gamma^0)_c^{~e}  \, ( \, \mathscr{F}{}_{a}{}^{b} {}^{(ATS)} \, ){}_e {
}^f \,  (\gamma^0)_f^{~d} ~&=~ - \delta_{a}{}^{b} (\gamma^0)_c^{~d} 
- (\gamma^5)_{a}{}^{b} (\gamma^5\gamma^0)_{c}^{~d} - (\gamma^5
\gamma^0)_{a}{}^{b}(\gamma^5)_c^{~d}  
~~~.
}  \label{BH5}
\ee

Given the results in (\ref{BH4}) and (\ref{BH5}), we can define an inner product on 
the covariant 0-brane space of representations.  With the covariant 0-brane reduced 
supermultiplets $(CS)$, $(VS)$, $(TS)$, $(AVS)$, and $(ATS)$, we can introduce 
a ``representation index'' denoted by $({\cal R})$ that simply takes on each of the 
values $(CS)$, $(VS)$, $(TS)$, $(AVS)$, and $(ATS)$.  We can regard these as 
five distinct vectors in a representation space.  We can introduce a inner product 
on this vector space.  Let $({\cal R})$ and $({\cal R}^{\prime})$ denote any two of 
these representation, then we can introduce a inner product, denoted by ${\Tilde 
{\cal G}}\left[( {\cal R}), ({\cal R}^{\prime}) \right]$  on this space through the definition:
\be \eqalign{
{\Tilde {\cal G}} \left[ \, ({\cal R})\,  , \, ({\cal R}^{\prime})\, \right] ~=~ 
{\Tilde {\cal G}}\left[ \, ({\cal R}^{\prime})\,  , \, ({\cal R})\, \right] 
~=~   \fracm 1{48} \, 
 (\gamma^0)_c^{~e}  \, ( \, \mathscr{F}{}_{a}{}_{b} {}^{({\cal R})} \, ){}_e {}^f \,  
 (\gamma^0)_f^{~d} \, ( \, \mathscr{F}{}^{a}{}^{b} {}^{({\cal R}^{\prime})} \, )
 {}_d {}^c   ~~~~.
}  \label{M1}
\ee
Several direct calculations lead to a result cast into the form of
a 5 $\times$ 5 matrix
\be \eqalign{
{\Tilde {\cal G}}\left[ \, ({\cal R})\,  , \, ({\cal R}^{\prime})\, \right] ~=~
\left[\begin{array}{ccccc}
~1 & ~0 &  ~0 & ~0 &  ~0\\
~0 & ~~1 &  - \, \fracm13 &  - \, \fracm13 &    - \, \fracm13 \\
~0 & - \, \fracm13 &  ~~ 1 &  - \, \fracm13 &   - \, \fracm13 \\
~0 & ~- \, \fracm13 &   - \, \fracm13 & ~1 &   - \, \fracm13 \\
~0 &  - \, \fracm13 &  - \, \fracm13 &  - \, \fracm13 &  ~1 \\
\end{array}\right]   ~~~,
}  \label{M2z}
\ee
where the rows and columns are respectively labeled by $(CS)$, $(VS)$, $(TS)$, 
$(AVS)$, and $(ATS)$.  Given the definition in (\ref{M1}), we see the $(CS)$, 
$(VS)$, $(TS)$, $(AVS)$, and $(ATS)$ representations are equivalent to unit 
vectors since,
\be \eqalign{  {~~~~}
{\Tilde {\cal G}}\left[ \, (CS)\,  , \, (CS)\, \right] ~&=~ 
{\Tilde {\cal G}} \left[ \, (VS)\,  , \, (VS)\, \right] ~~~~\,~=~
{\Tilde {\cal G}}\left[ \, (TS)\,  , \, (TS)\, \right] ~~~~\,~=~  \cr
~&=~ {\Tilde {\cal G}}\left[ \, (AVS)\,  , \, (AVS)\, \right] ~=~
{\Tilde {\cal G}}\left[ \, (ATS)\,  , \, (ATS)\, \right] ~=~ 1 ~~~.
}  \label{M3}
\ee
It is striking that the only entries in (\ref{M2z}) are (-1/3, 0 ,  1) and
these are exactly the values found in the metric introduced
in (\ref{M2}) for the three network-based adinkras.

Furthermore, we can define an angle between any two of the representations 
$({\cal R})$ and $({\cal R}^{\prime})$ via the definition
\be {
cos \left\{ \theta [({\cal R})\,  , \, ({\cal R}^{\prime})] \right\} ~=~{ { {\Tilde {\cal 
G}}\left[ \, ({\cal R})\,  , \, ({\cal R}^{\prime})\, \right] } \over {~ {\sqrt{
{\Tilde {\cal G}} \left[ \, ({\cal R})\,  , \, ({\cal R})\, 
\right]}} \, {\sqrt{ {\Tilde {\cal G}} \left[ \, ({\cal R}^{\prime})\,  , \, ({\cal R}^{\prime
})\, \right]}}~~ } } ~~~.
}   \label{M4}
\ee
We have thus obtained a geometrical viewpoint of the five minimal off-shell 4D, 
$\cal N$ = 1 supersymmetry representations.  The unit vector representing the 
CS representation is orthogonal to the unit vectors representing the VS, TS, AVS, 
and ATS representations.  The angles between the remaining representations 
can be read from the matrix given in (\ref{M2z}) to have a common value of $
\theta_{TV}$ where
\be
cos (\theta_{TV}) ~=~ -  \, \fracm 13  ~~~~.
\label{VTthetan}
\ee
On the subspace of the CS, VS, and TS representations, this is illustrated as
given previously in the work of \cite{KIAS, KIAS2}.
$$
\vCent
{\setlength{\unitlength}{1mm}
\begin{picture}(-20,-140)
\put(-40,-68){\includegraphics[width=3.4in]{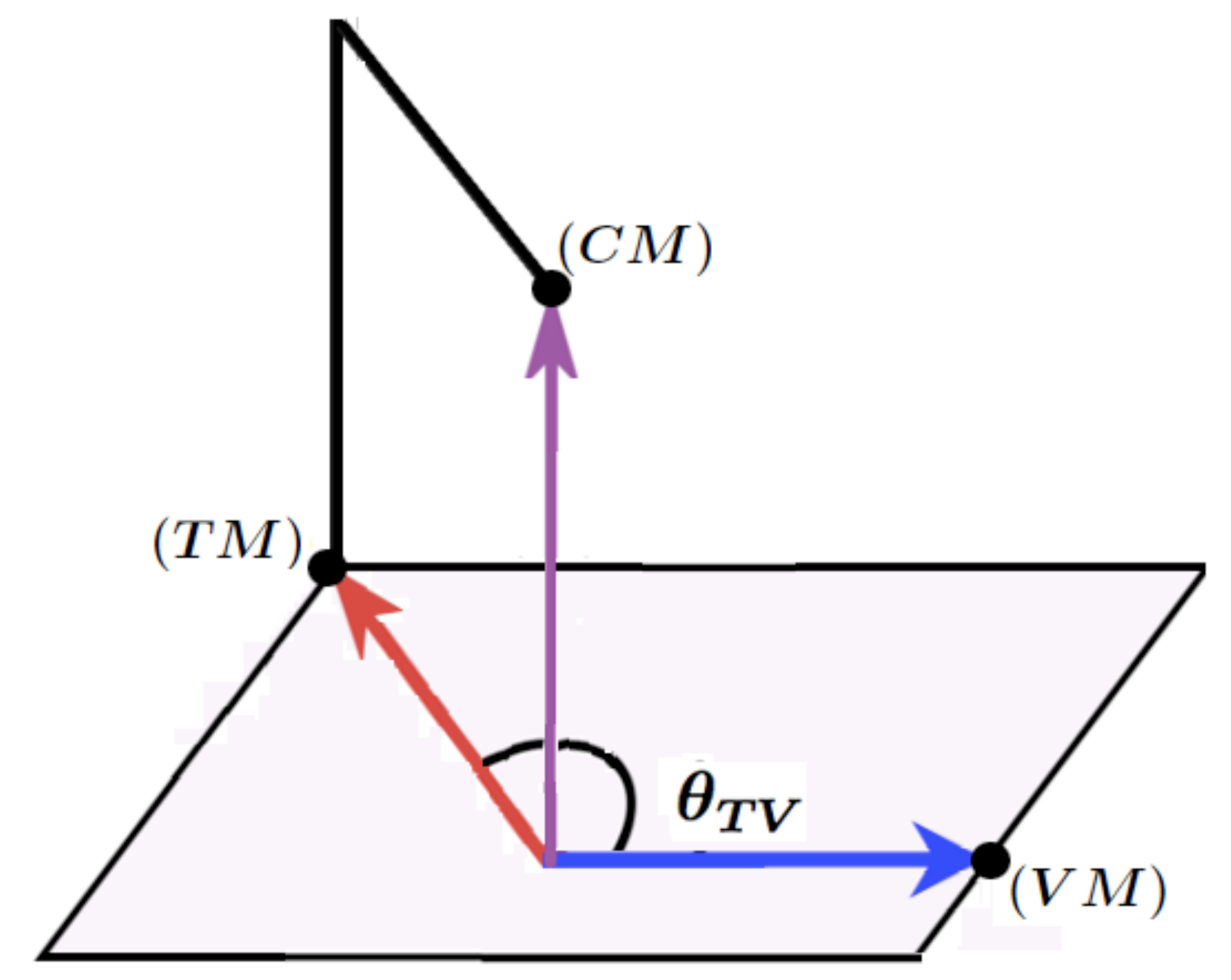}}
\put(-74,-76){\bf {Figure \# 4: Illustration of the CS-VS-TS subspace using the 
$\Tilde {\cal G}$ metric}}
\end{picture}}
$$
\vskip3.0in \noindent
Though the diagram above in Fig.\ \# 4 is the same as the one in \cite{KIAS, 
KIAS2}, the interpretation here is very different.  In the previous work, this 
diagram was based on the calculations using network-based adinkras (see 
Fig.\ \# 3) and using the ${\cal G}$-metric defined by equation (\ref{M6}).  The 
$\Tilde {\cal G}$-metric was not known then and the image  (Fig.\ \# 4) is 
calculated based on (\ref{M1}) and 0-brane-based valise supermultiplets.  
The fact that the two images are identical is due to SUSY holography.

\newpage
It is possible on more general 4D, $\cal N$ = 1 supermultiplets the fermionic 
holoraumy tensor could be of the form
\be \eqalign{ 
( \, \mathscr{F}{}_{a}{}^{b} {}^{({\cal R})} \, ){}_c {}^d ~&=~      (\gamma^5 
\gamma^m)_{a}{}^{b} \, {\Tilde \ell}_{m \, n}  \,  (\gamma^{5} \gamma^{0} 
\gamma^{n})_c^{~d}  \cr   
~&~~~~ 
+\delta_{a}{}^{b} \, \ell_{1 \, 1} \, (\gamma^0)_c^{~d} + \delta_{a}{}^{b}
\, \ell_{1 \, 2} (\gamma^5\gamma^0)_{c}^{~d} +   \delta_{a}{}^{b} \, \ell_{1 
\, 3} \, (\gamma^5)_c^{~d}   \cr   
 ~&~~~~ +(\gamma^5)_{a}{}^{b} \, \ell_{2 \, 1} \, (\gamma^0)_c^{~d} 
 + (\gamma^5)_{a}{}^{b} \, \ell_{2 \, 2} (\gamma^5\gamma^0)_{c}^{~d} + 
  (\gamma^5)_{a}{}^{b} \, \ell_{2 \, 3} \, (\gamma^5)_c^{~d}   \cr   
 ~&~~~~ +(\gamma^5 \gamma^0)_{a}{}^{b} \, \ell_{3 \, 1} \, (\gamma^0)_c^{
 ~d} + (\gamma^5 \gamma^0)_{a}{}^{b} \, \ell_{3 \, 2} (\gamma^5\gamma^0
 )_{c}^{~d} +  (\gamma^5 \gamma^0)_{a}{}^{b} \, \ell_{3 \, 3} \, (\gamma^5
 )_c^{~d}   \cr  
 ~&~~~~ + \left[ \, \delta_{a}{}^{b} \, {\Hat \ell}{}_{1 \, m}  + (\gamma^5)_{a}{
 }^{b} \, {\Hat \ell}{}_{2 \, m} + (\gamma^5 \gamma)_{a}{}^{b} \, {\Hat \ell}{
 }_{3 \, m} \,  \right] \, (\gamma^{5} \gamma^{0} \gamma^{m})_c^{~d}  \cr  
 ~&~~~~ + (\gamma^{5} \gamma^{0} \gamma^{m})_{a}{}^{b} \, \left[ \, 
 {\Bar \ell}{}_{m \, 1} \, \delta_c^{~d}  +  {\Bar \ell}{}_{m \, 2} \,  (\gamma^{
 5} \gamma^{0})_c^{~d} + {\Bar \ell}{}_{ m \, 3} \,  (\gamma^{5} )_c^{~d}   
 \, \right]  ~~~,
}  \label{BH4z}
\ee
containing the 36 real parameters ${\ell}_{m \, n}$, ${\Tilde \ell}_{m \, n}$,
${\Hat \ell}{}_{m \, n} $ and ${\Bar \ell}{}_{m \, n} $.  It will require further studies 
of 0-brane valise supermultiplets to determine if such representations exist.

The results in this chapter provide yet another confirmation of the idea of ``SUSY
Holography.''  It is possible to define other inner products on the space of 0-brane 
SUSY representations.  However, the one in (\ref{M1}) matched with a proposal 
made previously in the work of \cite{KIAS, KIAS2} and is the same (up to a choice 
of normalization) at that given in \cite{KIAS, KIAS2}.   In (\ref{M1}), the representations 
are those obtained by 0-brane reduction from higher dimensional supermultiplets.  
In (\ref{M6}), the representations are those obtained directly from network-based 
adinkra graphs.  The results for the angles  between the different representations 
match perfectly.

\section{An Intermezzo For Matchings} 
$~~~$ 
By starting from the network viewpoint one has a set of bosons
$ \Phi_{i}$ corresponding to half the nodes in a network and
a set of fermions $\Psi_{\hat k}$ corresponding to the other half.
The $i$-indices on the bosons $ \Phi_{i}$ (the ${\hat k}$-indices on
the fermions $\Psi_{\hat k}$) in the equations
\be \eqalign{
{\rm D}{}_{{}_{\rm I}} \Phi_{i} ~=~ i \, \left( {\rm L}{}_{{}_{\rm I}}\right) 
{}_{i \, {\hat k}}  \, \, \Psi_{\hat k}  ~~~,~~~
{\rm D}{}_{{}_{\rm I}} \Psi_{\hat k} ~=~ \left( {\rm R}{}_{{}_{\rm I}}\right)
{}_{{\hat k} \, i}  \, \pa_{\tau} \, \Phi_{i}  ~~~,
}  \label{VH1Z}
\ee
simply counts the bosons (fermions).  As such, they {\it a} {\it priori} {\it contain}
{\it no} {\it information} {\it about} {\it {space-time}} {\it symmetries}.

On the other hand, starting from a field theoretical viewpoint, one 
has a set of bosons $\Phi{}_{\Lambda}$ and fermions $\Psi_{\Hat {\Lambda}}$.
The ${\Lambda}$-indices on the bosons $\Phi{}_{\Lambda}$
(the ${\Hat {\Lambda}}$-indices on the fermions $\Psi{}_{\Hat {\Lambda}}$) in
the equations
\be \eqalign{
{\rm D}{}_a \, \Phi{}_{\Lambda}  ~=~ i\,  \left( {\rm L}{}_{\Lambda}\right){}_{a}
{}^{\Hat {\Lambda}} \, \, \Psi{}_{\Hat {\Lambda}}
~~~,~~~
{\rm D}{}_{a} \Psi{}_{\Hat {\Lambda}} ~=~  \left( {\rm R}{}^{\Lambda} \right)
{}_{\Hat {\Lambda}}{\,}_{a} \, \pa_{\tau} \, \Phi{}_{\Lambda}  ~~~,
} \label{covVALZ}
\ee
{\it are} spacetime indices (i.\ e.\ scalar, vector, etc. for $\Lambda$ and
spinor, etc. for $\Hat{\Lambda}$) and thus provide explicit representations
of space-time symmetries.

So though it is true there are many more constructions possible involving 
the adjacency matrices associated with networks realizations of off-shell 
1D, $N$-extended SUSY, a subset of the models described by (\ref{VH1Z}) 
must in fact be identical to the models described by (\ref{covVALZ}).  For 
this subset, it follows there must be ``hidden'' relations between the 
network-based ${\rm L}{}_{{}_{\rm I}}$ and ${\rm R}{}_{{}_{\rm I}}$ matrices 
on one side of a correspondence and the ${\rm L}{}_{\Lambda}$ and ${
\rm R}{}^{\Lambda}$ coefficients on the other side.  This implies within the 
subset, the system described by (\ref{VH1Z}) must be a hologram of the 
system described by (\ref{covVALZ}).  Hence, SUSY holography must be 
realized.

In the work of \cite{KIAS, KIAS2}, an equation called the ``Adinkra/$\gamma
$-matrix Equation'' was presented without comment.  From our present vantage 
point, this equation can be explained more fully.  As we saw in chapters 
two and three, beginning with network-based valise adinkras it is easy to 
calculate their associated $V$-matrices and $\Tilde V$-matrices.

In chapters four and five, it was shown that application of 0-brane reduction and
node-lowering field redefinitions can be applied to 4D, $\cal N$ = 1 supermultiplets
and result in equations that have the same general form as those that arise from
a valise adinkra.  This allows the calculations of the covariant holoraumy tensors
$\mathscr{B}{}_{{a} \, {b}}$ and $\mathscr{F}{}_{{a} \, {b}}$.   The next task must be 
to unearth the mathematical tools permitting supermultiplets from 
networks to be mapped onto supermultiplets obtained by 0-brane 
reduction.  Three such tools have been identified for this purpose. 

\vskip.1in \indent
(a.) The Adinkra/$\gamma$-matrix Equation \newline \indent $~~~~~~$
In the work of \cite{KIAS, KIAS2}, the assertion appears there
exist a projection 
 \newline \indent $~~~~~~$
operator $\mathscr{P}$ such that
\be {
\mathscr{P}(\mathscr{F}{}_{{a} \, {b}}) ~=~ \left[ \, {\cal F}{}_{{}_{\rm I}}
{}_{{}_{\rm J}} \, \right] 
} \label{JHproj1}
\ee
\indent $~~~~~~$
so the components of the covariant fermionic holoraumy tensor $\mathscr{F
}{}_{{a} \, {b}}$ 
 \newline \indent $~~~~~~$
can be set equal to the entries of the $\Tilde V$-matrices.  This 
provides a sort 
 \newline \indent $~~~~~~$
of ``Rosetta Stone'' that performs the translation between the higher 
 \newline \indent $~~~~~~$
dimensional SUSY field theory on one side of the correspondence and
 \newline \indent $~~~~~~$
the ``Garden Algebra,'' network-based adinkras, and codes on the other 
 \newline \indent $~~~~~~$
side.  

According to the observations in \cite{adnkgeo}, this translation on the 
adinkra side will also be important to understand how supermultiplets 
emerge from Riemann surfaces and Bayli pairs.  
 
This mapping is not one-to-one as was understood from our initial discussions
of adinkras.  In the work of \cite{adnk1}, there appears an image of a pyramid
meant to elicit the notion that the number of distinct SUSY representation 
decreases as the dimension of the superspace increases.  This was not meant 
in a metaphorical sense.

One way to see all of this playing out is in the context of the minimal off-shell
representations of 4D, $\cal N$ = 1 supermultiplets.  Counting variants \cite{
Varep}, there are only eight distinct such supermultiplets.  On the other hand, 
if one does not permit any similarity transformations in counting all possible 
four color, four-open node and four-closed node adinkras there are 1,536 such 
objects.  So on this subclass of objects the operator $\mathscr{P}$ projects the 
eight supermultiplets into the `sea' of 1,536 adinkras.
 
\vskip.1in \indent
(b.) Coxeter Algebra/Hodge Duality Orbit Matching \newline \indent $~~~~~~$
In the work of \cite{permutadnk}, it was shown using the action of the Hodge 
\newline \indent $~~~~~~$
duality map acting on the fields in equations (\ref{covVALZ}) leads 
to shadows
\newline \indent $~~~~~~$
of these maps that can be detected on the fields in equations (\ref{VH1Z})
\newline \indent $~~~~~~$
and the L-matrices in Table \# 1.  With this identification, Coxeter 
\newline \indent $~~~~~~$
algebras 
were found to play a 
hidden role in
organizing the 1,536 
\newline \indent $~~~~~~$
networks  so the action of the Hodge $*$-operator on
the 0-brane- \newline \indent $~~~~~~$
based equations of (\ref{VH1Z}) induces the existence of 
shadow of the Hodge 
\newline \indent $~~~~~~$
$*$-operator on the adinkra network-based equations of (\ref{covVALZ}) 
implying
\newline \indent $~~~~~~$
the equation ${
\mathscr{P}({}^*\mathscr{F}{}_{{a} \, {b}}) ~=~ \left[ \, {}^*{\cal F}{}_{{}_{\rm I}}
{}_{{}_{\rm J}} \, \right] } $.

$$
\vCent
{\setlength{\unitlength}{1mm}
\begin{picture}(-20,-140)
\put(-50,-114){\includegraphics[width=4.2in]{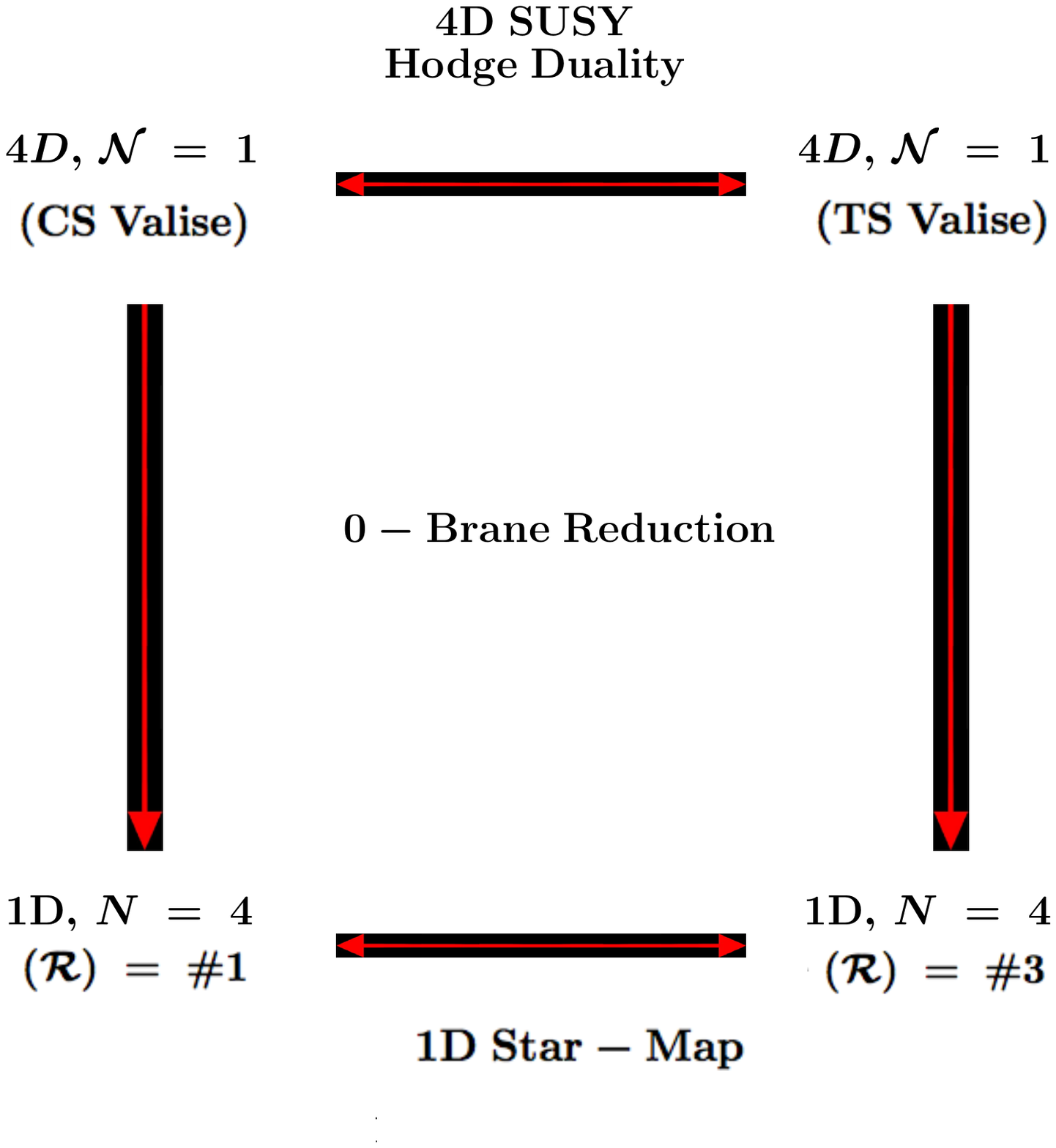}}
\put(-44,-75){\bf {Figure \# 5: Illustration of Hodge $*$-operator}}
\end{picture}}
$$
\vskip2.8in \noindent
The diagram in Fig.\ \# 5 shows the orbit matching for the CS-TS
pair.  The orbit on the network-based side shown at the bottom of
the image has been extended to the entirety of the space of 1,536 
network-based adinkras as shown in Fig.\ \# 6 below taken from the work
in \cite{permutadnk}.
$$
\vCent
{\setlength{\unitlength}{1mm}
\begin{picture}(-20,-140)
\put(-50,-64){\includegraphics[width=3.4in]{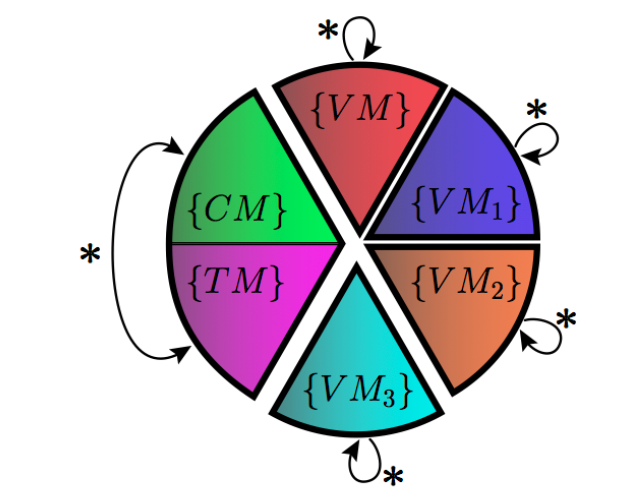}}
\put(-50,-70){\bf {Figure \# 6: Orbits of Network-Based Adinkras}}
\end{picture}}
$$
\vskip2.6in \noindent

This current work has provided a tool that appears to be the most powerful in
identifying network-based adinkras with 0-brane based supermultiplets.
This tool is to impose matching conditions based on the representation space
metrics ${\Tilde {\cal G}}[({\cal R}), \, ({\cal R}^{\prime})] $ and ${\cal G}[({\cal R}), 
\, ({\cal R}^{\prime})] $ which leads to a projection operator.

\indent
(c.) Representation Space Metric Matching \newline \indent $~~~~~~$
The work of \cite{adnkholor} fully enunciated an idea that when the two sets
of
\newline \indent $~~~~~~$
equations in the forms of (\ref{VH1Z}) and (\ref{covVALZ}) are
known, then it is pos- \newline \indent $~~~~~~$
sible to calculate (based on these) two respective
fermionic holo- \newline \indent $~~~~~~$
raumy tensors and require their matching as the condition to when
 \newline \indent $~~~~~~$ 
the
network-based construction is the shadow of the 0-brane con-
 \newline \indent $~~~~~~$ struction.
 \be {
\mathscr{P}({\Tilde {\cal G}}\left[ \, ({\cal R})\,  , \, ({\cal R}^{\prime})\, \right]) ~=~ 
{\cal G}\left[ \, ({\cal R})\,  , \, ({\cal R}^{\prime})\, \right]
} \label{JHproj2}
\ee

Thus, we believe these results constitute the essential elements of a proof 
that the kinematics of higher dimensional 4D, $\cal N$ = 1 supersymmetric 
multiplets are embedded into the structure of network-based adinkras.

\section{A Detailed Presentation On Holoraumy \& Dimensional Reduction} 
$~~~$ 
In this chapter, we will take a detailed look at the interplay of dimensional 
reduction and the holoraumy tensor associated with multiplets in different 
dimensions by looking at the examples of the 4D, $\cal N$ = 1 vector 
supermultiplet and the 2D, $\cal N$ = 2 twisted chiral multiplet to be 
contrasted with the simple reduction of the 4D, $\cal N$ = 1 chiral supermultiplet. 

\subsection{Implementing Torus Compactification} 
$~~~$ 
We will naturally be led to a 2-manifold if we assume all the four dimensional 
fields do NOT depend on $x^2$ and $x^3$.  For the  $\cal N$ = 1 vector supermultiplet 
in 4D we thus have
\be \eqalign{
{\rm D}_a \, A{}_{\mu} ~&=~  (\gamma_\mu){}_a {}^b \,  \l_b  ~~~, \cr
{\rm D}_a \l_b ~&=~   - \,i \, \fracm 14 ( [\, \gamma^{\mu}\, , \,  \gamma^{\nu} 
\,]){}_a{}_b \, (\,  \partial_\mu  \, A{}_{\nu}    ~-~  \partial_\nu \, A{}_{\mu}  \, )
~+~  (\gamma^5){}_{a \,b} \,    {\rm d} ~~,  \cr
{\rm D}_a \, {\rm d} ~&=~  i \, (\gamma^5\gamma^\mu){}_a {}^b \, 
\,  \partial_\mu \l_b  ~~~, \cr
}   \label{Khy3}   \ee
as our starting point.  After a number of steps explained in detail in an appendix,
we find
$$
\eqalign{
{\rm D}_a \, {\widehat A} ~&=~  (\gamma_2){}_a {}^b \, {\widehat \psi}_b  ~~,~~
{\rm D}_a \, {\widehat B} ~=~  (\gamma_3){}_a {}^b \, {\widehat \psi}_b  
~~~~~~~~~~\,~~~~~\,~~, ~~~
 \cr
 {\rm D}_a {\widehat \psi}_b ~&=~  i \, ( \gamma^{2}  \gamma^{0} ){}_a{}_b 
\, (\,  \partial_0  \, {\widehat A}   \, )   + \,i \,  (  \gamma^{2} \gamma^{1} ){}_a
{}_b \, (\,   \partial_1 \, {\widehat A}  \, )
 ~~  \cr
~&~ ~~~~+~   i \, (  \gamma^{3} \gamma^{0} ){}_a{}_b \, 
(\,  \partial_0  \, {\widehat B}    \, )  + \,i \,  ( 
\gamma^{3} \gamma^{1} ){}_a{}_b \, (\,  \partial_1  \, {\widehat B}    
\, )  \cr
~&~ ~~~~   - \,i \,  ( \gamma^{0} \gamma^{1} 
){}_a{}_b \, {\widehat F} ~+~  (\gamma^5){}_{a \,b} \, {\widehat G}
 ~~, } 
 $$
 \be
\eqalign{
{\rm D}_a \, {\widehat F} ~&=~  (\gamma_1){}_a {}^b \, (\, \partial_0 {\widehat \psi}_b
\, ) ~-~  (\gamma_0){}_a {}^b \, ( \, \partial_1 {\widehat \psi}_b \, ) ~~,~~ 
{~~~~~~~~~~~~~\,~~~~~}
 \cr
{\rm D}_a \, {\widehat G} ~&=~  i \, (\gamma^5\gamma^0){}_a {}^b \, 
\, ( \partial_0{\widehat \psi}_b  )  ~+~  i \, (\gamma^5\gamma^1){}_a {}^b \, 
\, ( \partial_1{\widehat \psi}_b  ) 
~~~, \cr
} \label{Khy8b}
\ee
and this is a formulation of the  $\cal N$ = 2 twisted chiral supermultiplet (TCS) in two 
dimensions.  The process above is precisely the one that led in 1984 to the discovery
of the twisted chiral supermultiplet \cite{TCS}.

On the other hand, it is simple to demand the fields of the 4D chiral supermultiplet
\be \eqalign{
{\rm D}_a A  ~=~&~ \psi_a   ~~~~~~~~~~~~~~~~~~~~~~~~~~~~\,~, \cr
{\rm D}_a B  ~=~ &~i \, (\gamma^5){}_a{}^b \, \psi_b   ~~
~~~~~~~~~~~~~~~~~~, \cr
{\rm D}_a \psi_b  ~=~ &~i\, (\gamma^\mu){}_{a \,b}\, \, \left( \pa_{\m} A  
\right) ~-~ ~ (\gamma^5\gamma^\mu){}_{a \,b} \,\, \left( \pa_{\m} B  
\right) ~\cr
&-~ ~i \, C_{a\, b} 
\,F   ~+~ ~ (\gamma^5){}_{ a \, b} G   ~~, \cr
{\rm D}_a F  ~=~&~ (\gamma^\mu){}_a{}^b \,\, \left( \pa_{\m} \, 
\psi_b    \right)
~~~~~~~~~~~~~~, \cr
{\rm D}_a G  ~=~ &~i \,(\gamma^5\gamma^\mu){}_a{}^b \,\, 
\left( \pa_{\m} \,  \psi_b   \right) ~~~~~~~~\,~,
}   \label{Khy9}
\ee
should only depend on $x^0$ and $x^1$ yielding
\be
 \eqalign{
{\rm D}_a A ~&=~ \psi_a  ~~~~~~~~~~~~~~~~~~~~~~~~~
~~~~~~~~~~~~~~~~~~~~~~~~,   \cr
{\rm D}_a B ~&=~ i \, (\gamma^5){}_a{}^b \, \psi_b  ~~~~~~
~~~~~~~~~~~~~~~~~~~~~~~~~~~~~~~~~~, \cr
{\rm D}_a \psi_b ~&=~ i\, (\gamma^0){}_{a \,b}\, \left(  \partial_0 A 
\right) ~+~   i\, (\gamma^1){}_{a \,b}\, \left(  \partial_1 A 
\right) \cr 
& ~~~\,\,-~  (\gamma^5\gamma^0){}_{a \,b} \, \left( \partial_0 B \right) 
 ~-~  (\gamma^5\gamma^1){}_{a \,b} \, \left( \partial_1 B \right) 
\cr 
& ~\,~\,~-\,~ i \, C_{a\, b} \,F  ~+~  (\gamma^5){}_{ a \, b} G  
~~~~~~~~~~~~~~~~~~~~~~~~, \cr
{\rm D}_a F ~&=~  (\gamma^0){}_a{}^b \, \left( \partial_0 \, \psi_b 
\right) ~+~ (\gamma^1){}_a{}^b \, \left( \partial_1 \, \psi_b 
\right)   ~~~~~~~~~~~~, ~~~  \cr
{\rm D}_a G ~&=~ i \,(\gamma^5\gamma^0){}_a{}^b \, \left( \partial_0 \,  
\psi_b \right) ~+~  i \,(\gamma^5\gamma^1){}_a{}^b \, \left( \partial_1 \,  
\psi_b \right)
~~~.
}   \label{Khy10}
\ee
$$~~$$

\subsection{Reducing the 2D CS \& TCS to 1D with Node Lowering} 
$~~~$
in order to find the covariant holoraumy tensors associated with the
two dimensional, $\cal N$ = 2 chiral and twisted chiral supermultiplets,
they each need to be reduced to one dimensional supermultiplets and
node-lowering must be applied to their respective auxiliary fields.  Once 
the respective holoraumy tensors are found, then the metric on the space 
of these two representations can be calculated.  The reduction of the CS 
and TCS to one dimension simply amounts to demanding all fields 
depend solely on $x^0$. 

Under this circumstance, we find for the
CS
\be
 \eqalign{
{\rm D}_a A ~&=~ \psi_a  ~~~, ~~~
{\rm D}_a B ~=~ i \, (\gamma^5){}_a{}^b \, \psi_b  ~~~~~\,~~~,~~~~~~~~~~
~~~ \cr
{\rm D}_a \psi_b ~&=~ i\, (\gamma^0){}_{a \,b}\, \left(  \pa_{\tau} A 
\right) -~  (\gamma^5\gamma^0){}_{a \,b} \, \left( \pa_{\tau} B \right) 
\cr 
& ~\,~\,~-\,~ i \, C_{a\, b} \,F  ~+~  (\gamma^5){}_{ a \, b} G  ~~~~~~~~~~~~~~~, \cr
{\rm D}_a F ~&=~  (\gamma^0){}_a{}^b \, \left( \pa_{\tau} \, \psi_b 
\right)   ~~~~\,~~~,   \cr
{\rm D}_a G ~&=~ i \,(\gamma^5\gamma^0){}_a{}^b \, \left( \pa_{\tau} \,  
\psi_b \right) 
~~~,  \cr
}   \label{Khy11}
\ee
and we find for the TCS
\be
\eqalign{
{\rm D}_a \, {\widehat A} ~&=~  (\gamma_2){}_a {}^b \, {\widehat \psi}_b  ~~,~~
{\rm D}_a \, {\widehat B} ~=~  (\gamma_3){}_a {}^b \, {\widehat \psi}_b  
~~~~~~~~~~~\,~~, ~~~
 \cr
 {\rm D}_a {\widehat \psi}_b ~&=~  i \, ( \gamma^{2}  \gamma^{0} ){}_a{}_b 
\, (\,  \pa_{\tau}  \, {\widehat A}   \, )  
~+~   i \, (  \gamma^{3} \gamma^{0} ){}_a{}_b \, 
(\,  \pa_{\tau}  \, {\widehat B}    \, )  \cr
~&~ ~~~~   - \,i \,  ( \gamma^{0} \gamma^{1} 
){}_a{}_b \, {\widehat F} ~+~  (\gamma^5){}_{a \,b} \, {\widehat G}
 ~~~~~~~~~~~~~~~~~~,   \cr
{\rm D}_a \, {\widehat F} ~&=~  (\gamma_1){}_a {}^b \, (\, \pa_{\tau} {\widehat \psi}_b
\, )  ~~~~~~~,~~
 \cr
{\rm D}_a \, {\widehat G} ~&=~  i \, (\gamma^5\gamma^0){}_a {}^b \, 
\, ( \pa_{\tau}{\widehat \psi}_b  ) 
~~~. \cr
}    \label{Khy12}
\ee

We introduce node lowering by implementing the re-definition of the 
``auxiliary fields'' in each supermultiplet according to
\be
F ~\to ~ \pa_{\tau} F ~~~,~~~ G ~\to ~ \pa_{\tau} G ~~~,~~~ 
{\widehat F} ~\to ~ \pa_{\tau} {\widehat F} ~~~,~~~ {\widehat G} ~\to ~ \pa_{\tau} 
{\widehat G}
 ~~~,
\ee
so (\ref{Khy11}) becomes
\be
  \eqalign{
{\rm D}_a A ~&=~ \psi_a  ~~~\,~~~~\,~~~, ~~~
{\rm D}_a B ~=~ i \, (\gamma^5){}_a{}^b \, \psi_b  ~~~~~~~~~~~\, ,~~~
~~~  \cr
{\rm D}_a F ~&=~  (\gamma^0){}_a{}^b \,  \psi_b    ~~~, ~~~
{\rm D}_a G ~=~ i \,(\gamma^5\gamma^0){}_a{}^b \, \psi_b
~~~~~~~~~,
\cr
{\rm D}_a \psi_b ~&=~ i\, (\gamma^0){}_{a \,b}\, \left(  \pa_{\tau} A 
\right) ~-~  (\gamma^5\gamma^0){}_{a \,b} \, \left( \pa_{\tau} B \right) 
\cr 
& ~\,~\,~-\,~ i \, C_{a\, b} \, \left( \pa_{\tau} F \right)  ~+~  
(\gamma^5){}_{ a \, b} \, \left( \pa_{\tau} G \right) ~~~~~~~~~~~~~,}  \label{Khy13}
\ee
and for (\ref{Khy12}) we find
\be
\eqalign{
{\rm D}_a \, {\widehat A} ~&=~  (\gamma_2){}_a {}^b \, {\widehat \psi}_b  ~~,~~
{\rm D}_a \, {\widehat B} ~=~  (\gamma_3){}_a {}^b \, {\widehat \psi}_b  
~~\,~~~~~~~~\,~~, ~~~   \cr
{\rm D}_a \, {\widehat F} ~&=~  (\gamma_1){}_a {}^b \,  {\widehat \psi}_b
 ~~,~~
{\rm D}_a \, {\widehat G} ~=~  i \, (\gamma^5\gamma^0){}_a {}^b \, {\widehat \psi}_b  
~~~~~~~~\,,
 \cr
 {\rm D}_a {\widehat \psi}_b ~&=~  i \, ( \gamma^{2}  \gamma^{0} ){}_a{}_b 
\, (\,  \pa_{\tau}  \, {\widehat A}   \, )  ~+~   i \, (  \gamma^{3} \gamma^{0} ){}_a{}_b \, 
(\,  \pa_{\tau}  \, {\widehat B}    \, ) \cr
~&~ ~~~~   - \,i \,  ( \gamma^{0} \gamma^{1} 
){}_a{}_b \, (\,  \pa_{\tau}  \, {\widehat F}    \, ) 
 ~+~  (\gamma^5){}_{a \,b} \, (\,  \pa_{\tau}  \, {\widehat G}    \, )  
 ~~~~.   \cr
}  \label{Khy14}
\ee

Given the equations in (\ref{Khy14}), the holoraumy tensor associated with
the 0-brane reduced 2D, $\cal N$ = 2 twisted chiral supermultiplet can be 
calculated and is found to be the same as the 0-brane reduced 4D, 
$\cal N$ = 1 vector supermultiplet as shown below
\be \eqalign{ 
( \, \mathscr{F}{}_{a}{}^{b} {}^{(CS)} \, ){}_c {}^d ~&=~   -   (\gamma^5 \gamma^m
)_{a}{}^{b} \, (\gamma^{5} \gamma^{0} \gamma_{m})_c^{~d} ~~~,
\cr   
( \, \mathscr{F}{}_{a}{}^{b}{}^{(TCS)}\, ){}_c{}^{d} ~&=~ +\delta_{a}{}^{b} (
\gamma^0)_c^{~d} + (\gamma^5)_{a}{}^{b} (\gamma^5\gamma^0)_{c}^{~d} 
+ (\gamma^5\gamma^0)_{a}{}^{b}(\gamma^5)_c^{~d}   ~~~.
}  \label{Khy15}
\ee
Now a few calculations show
\be \eqalign{ 
&( \, \mathscr{F}{}_{a}{}_{b} {}^{(CS)} \, ){}_e {}^f \, ( \, \mathscr{F}{}_{c}{}_{d}{
}^{(TCS)}\, ){}_f{}^{h}  ~-~ 
( \, \mathscr{F}{}_{c}{}_{d} {}^{(TCS)} \, ){}_e {}^f \, ( \, \mathscr{F}{}_{a}{}_{b}{
}^{(CS)}\, ){}_f{}^{h} ~=~ 0 ~~~,   \cr
&{~~~~~~~~~~~~}( \, \mathscr{F}{}_{a}{}_{b} {}^{(CS)} \, ){}_e {}^f \, (\, \mathscr{
F}{}_{c}{}_{d}{}^{(TCS)}\, ){}_f{}^{e} ~=~ 0 ~~~. 
}  \label{Khy16}
\ee
The holoraumy tensors of the 2D, $\cal N$ = 2 chiral supermultiplet are found to 
commute with those of the 2D, $\cal N$ = 2 twisted chiral supermultiplet.   The 
final equation in (\ref{Khy16}) implies starting from these two distinct 2D, $\cal 
N$ = 2 chiral supermultiplets are ``orthogonal,'' i.e.\ 
\be \eqalign{  {~~~~}
{\Tilde {\cal G}}\left[ \, (CS)\,  , \, (TCS)\, \right] ~&=~ 0 ~~~,
}  \label{M3z}
\ee
just as the 4D, $\cal N$ = 1 chiral supermultiplet is ``orthogonal'' to the 
4D, $\cal N$ = 1 vector supermultiplet.

\section{How Ignoring SUSY Holography Can Lead To Violation of 
Lorentz Symmetry In Uplifting} 
$~~~$ 
Let us begin by considering the two adinkra graphs shown below.
$$
\vCent
{\setlength{\unitlength}{1mm}
\begin{picture}(-10,0)
\put(-70,-20){\includegraphics[width=2.6in]{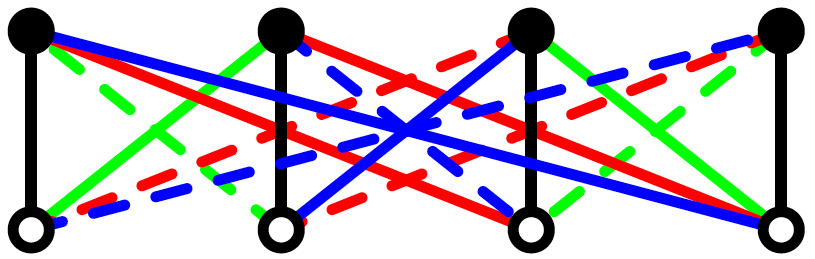}}
\put(-44,-26){Adnk-A}
\put(6,-20){\includegraphics[width=2.6in]{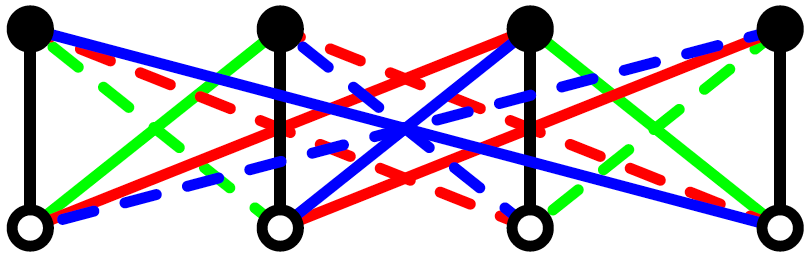}}
\put(32,-26){Adnk-B}
\put(-70,-34){\bf {Figure \# 7: Two valise adinkra networks with opposite red-parity}}
\end{picture}}
\nonumber
$$
\vskip 1.2in
\noindent It is seen they are identical but with one exception.  All the solid 
red links in Adnk-A are replaced by dashed red links in Adnk-B and vice versa.

Now we can use the adinkra in (Adnk-A) to find the associated L-matrices (and 
R-matrices which are simply the transposed versions of the L-matrices).  After
such explicit expressions are found (see Appendix E), the corresponding 
$\Tilde V$ matrices are calculated.  The results of these calculations further imply
\be \eqalign{  {~~~~~~~~}
{\Tilde V}_{1 \, 2}^{({\rm {Adnk-A}})} ~&=~ {\Tilde V}_{3 \, 4}^{({\rm {Adnk-A}})} ~~,~~  
{\Tilde V}_{2 \, 3}^{({\rm {Adnk-A}})} ~=~ -\, {\Tilde V}_{1 \, 4}^{({\rm {Adnk-A}})} ~~,~~ \cr
{\Tilde V}_{1 \, 3}^{({\rm {Adnk-A}})} ~&=~ {\Tilde V}_{2 \, 4}^{({\rm {Adnk-A}})}  ~~~.
}   \label{Lvgvf1}
\ee
Similarly, 
\be \eqalign{  {~~~~~~~~}
{\Tilde V}_{1 \, 2}^{({\rm {Adnk-B}})} ~&=~ - \, {\Tilde V}_{3 \, 4}^{({\rm {Adnk-B}})} ~~,~~
{\Tilde V}_{2 \, 3}^{({\rm {Adnk-B}})} ~=~ {\Tilde V}_{1 \, 4}^{({\rm {Adnk-B}})} ~~,~~ \cr
{\Tilde V}_{1 \, 3}^{({\rm {Adnk-B}})} ~&=~  -\, {\Tilde V}_{2 \, 4}^{({\rm {Adnk-B}})}  ~~~.
}   \label{Lvgvf2}
\ee
Furthermore, these same calculations make it clear
$$\eqalign{  {~~}
{\Tilde V}_{1 \, 2}^{({\rm {Adnk-A}})} ~&=~ - \, {\Tilde V}_{1 \, 2}^{({\rm {Adnk-B}})} ~~,~~
{\Tilde V}_{1 \, 3}^{({\rm {Adnk-A}})} ~=~ - \, {\Tilde V}_{1 \, 3}^{({\rm {Adnk-B}})} ~~,~~ \cr
{\Tilde V}_{1 \, 4}^{({\rm {Adnk-A}})} ~&=~   {\Tilde V}_{1 \, 4}^{({\rm {Adnk-B}})}  ~~~,
}   \label{Lvgvf3}
$$
\be {
{\Tilde V}_{I \, J}^{({\rm {Adnk-A}})} \, {\Tilde V}_{K \, L}^{({\rm {Adnk-B}})} 
~-~  {\Tilde V}_{K \, L}^{({\rm {Adnk-B}})}   \, {\Tilde V}_{I \, J}^{({\rm {Adnk-A}})} ~\ne~ 0  ~~~.
} \label{BVc}
\ee
If one makes the claim either (Adnk-A) or (Adnk-B) is the 0-brane reduction 
of the chiral supermultiplet and the other is the 0-brane reduction of the twisted 
chiral multiplet, then one is forced to one of two conclusions:

(a.) the assignment of these adinkra as `shadows' of the 2D, $\cal N$
= 2 \newline \indent $~~~~~$ supermultiplets is inconsistent, or

(b.) SUSY holography is violated.

\noindent
The reason one is driven to these is we have previously argued the
commutator of the ${\Tilde V}$-matrices is tantamount to the evaluation of
the fermionic holoraumy tensor on the supermultiplet via SUSY holography.
Via the first equation of (\ref{Khy16}) and SUSY holography we conclude 
the commutator in (\ref{BVc}) should be zero.

In Appendix C, it is shown how the original derivation of the 2D, $\cal N$ 
= 2 twisted chiral supermultiplet is obtained from the 4D, $\cal N$ 
= 1 vector supermultiplet in the real basis used in this work.  We now wish
to build on this derivation to show that applying the traditional understanding
of how the 2D, $\cal N$ = 2 twisted chiral supermultiplet (CS) is related to the 
4D, $\cal N$ = 1 vector supermultiplet (VS) together with the assertion that one
of the adinkras shown in (Adnk-A) and (Adnk-B) corresponds to the CS and
the other to the VS means implies a violation of Lorentz invariance.

The closure of the off-shell (and on-shell) algebra is not dependent on the
choice of $\gamma$-matrices used.  It is only their algebraic properties
that are important.  Thus the calculation of the holoraumy on the 0-brane
side has the same form when expressed in terms of $\gamma$-matrices.
The calculation of the holoraumy on the adinkra side will depend on the
$V$ and $\Tilde V$ matrices of the adinkra used.

So, if one claims the chiral supermultiplet has (Adnk-A) as its adinkra
shadow, the projection of the 0-brane-based holoraumy tensor onto the 
adinkra-based holoraumy tensor implies there must exist constants ${\cal 
S}^{i \, j}_1 $, ${\cal S}^{i \, j}_2 $, and ${\cal S}^{i \, j}_3 $
so
\be  \eqalign{
\gamma^{1 \, 2} ~&=~ {\cal S}^{1 \, 2}_1 \, {\Tilde V}{}_{1 2}^{({\rm {Adnk-A}})} ~+~  
{\cal S}^{1 \, 2}_2 \, {\Tilde V}{}_{1 3}^{({\rm {Adnk-A}})}
~+~   {\cal S}^{1 \, 2}_3 \, {\Tilde V}{}_{1 4}^{({\rm {Adnk-A}})}   ~~~,   \cr
\gamma^{2 \, 3} ~&=~ {\cal S}^{2 \, 3}_1 \, {\Tilde V}{}_{1 2}^{({\rm {Adnk-A}})} ~+~ 
{\cal S}^{2 \, 3}_2 \, {\Tilde V}{}_{1 3}^{({\rm {Adnk-A}})}
~+~  {\cal S}^{2 \, 3}_3 \, {\Tilde V}{}_{1 4}^{({\rm {Adnk-A}})}   ~~~,   \cr
\gamma^{3 \, 1} ~&=~ {\cal S}^{3 \, 1}_1 \, {\Tilde V}{}_{1 2}^{({\rm {Adnk-A}})} ~+~ i\, 
{\cal S}^{3 \, 1}_2 \, {\Tilde V}{}_{1 3}^{({\rm {Adnk-A}})}
~+~ {\cal S}^{3\, 1}_3 \, {\Tilde V}{}_{1 4}^{({\rm {Adnk-A}})}   ~~~.
}  \label{Ss1}
\ee
and this equation defines an explicit representation.

If it is asserted the adinkra in (Adnk-B) is the valise shadow of the 0-brane
reduced valise version of the vector supermultiplet shown in (3.), the holoraumy 
tensor of (13) on the adinkra side must be related to the holoraumy tensor in 
(4) that originates from the VS supermultiplet.  In that case there must exist constants 
${\cal S}_1 $, $\dots$,  ${\cal S}_{9} $ such that
\be   \eqalign{
\gamma^{0} ~&=~  {\cal S}_1 \, {\Tilde V}{}_{1 2}^{({\rm {Adnk-B}})} ~+~ i \,
{\cal S}_2 \, {\Tilde V}{}_{1 3}^{({\rm {Adnk-B}})}
~+~ {\cal S}_3 \, {\Tilde V}{}_{1 4}^{({\rm {Adnk-B}})}    ~~~,  \cr
\gamma^{0} \gamma^{5} ~&=~  {\cal S}_4 \, {\Tilde V}{}_{1 2}^{({\rm {Adnk-B}})} ~+~ 
{\cal S}_5 \, {\Tilde V}{}_{1 3}^{({\rm {Adnk-B}})}
~+~ {\cal S}_6 \, {\Tilde V}{}_{1 4}^{({\rm {Adnk-B}})}    ~~~,  \cr
\gamma^{5} ~&=~ {\cal S}_7 \, {\Tilde V}{}_{1 2}^{({\rm {Adnk-B}})} ~+~ 
 {\cal S}_8 \, {\Tilde V}{}_{1 3}^{({\rm {Adnk-B}})}
~+~ {\cal S}_9 \, {\Tilde V}{}_{1 4}^{({\rm {Adnk-B}})}    ~~~. 
}   \label{Ss2} \ee
Furthermore, since no explicit representation of the $\gamma$-matrices
need be chosen, this equation defines an explicit representation.

The freedom in choosing the $\cal S$-coefficients is not as large
as it would first appear.  The commutator algebra of the $\gamma$ matrices
can easily be derived from (5).  On the other hand, the commutation relations
of the $\Tilde V$ matrices are determined from the adinkras and their adjacency
matrices.

Combining the results in (\ref{Lvgvf3}),  (\ref{Ss1}),  and  (\ref{Ss2}) leads
to a major problem. 
Once one determines a choice of the $\cal S$ coefficients, then one can
calculate the traces
\be  \eqalign{
t^{0 \, \, i \, j} ~&=~ {\rm {Tr}} \left[ \,  \gamma^0 \, \gamma^{i \, j} 
\, \right]  ~~~~~~~~,~~~  \cr
t^{5 \, \, i \, j} ~&=~ {\rm {Tr}} \left[ \,  \gamma^5 \, \gamma^{i \, j} \, 
\right]  ~~~~~~~~,~~~  \cr
t^{0 \, 5 \, \, i \, j} ~&=~ {\rm {Tr}} \left[ \,  \gamma^0 \,  \gamma^5 \, 
 \gamma^{i \, j} \, \right]  ~~~\,~.   }
 \label{TRs}
\ee
For any proper choice of $\gamma$-matrices all of these traces vanish.
Also for any proper choice of $\gamma$-matrices we have
\be  {
 \left[ \,  \gamma^0 ~,~  \gamma^{i \, j} \, \right]  ~=~ 0 ~~~,~~~
 \left[ \,  \gamma^5 ~,~ \gamma^{i \, j} \, \right]  ~=~ 0 ~~~,~~~
 \left[ \,  \gamma^0 \gamma^5 ~,~  \gamma^{i \, j} \, \right]  ~=~ 0 ~~~.  }
 \label{LV}
\ee
However, when one

(a.) 
uses the choice of adinkras shown in (Adnk-A) and in (Adnk-B), 

(b.)
asserts these are the valise adinkra shadows
of the chiral \newline $~~~~~~~~~~~$ 
supermultiplet and the vector supermultiplet, 

(c.) 
implies the set $\{ {\Tilde V}_{1 \, 2}^{({\rm {Adnk-A}})} , \,  {\Tilde V}_{1 
\, 3}^{({\rm {Adnk-A}})} , \,  {\Tilde V}_{1 \, 4}^{({\rm {Adnk-A}})}  \, \}$ 
becomes \newline $~~~~~~~~~~~$  linearly dependent on the set
$\{ {\Tilde V}_{1 \, 2}^{({\rm {Adnk-B}})} , \,  {\Tilde V}_{1 \, 3}^{({\rm {
Adnk-B}})} , \,  {\Tilde V}_{1 \, 4}^{(Adnk-B)}  \, \}$  \newline 
$~~~~~~~~~~~$ as shown in (\ref{Lvgvf3}),

(d.)
forces the set $\{ \gamma^{i \, j}  \, \}$ to be linearly dependent on the set 
$\{ \gamma^0 , \,  \gamma^0 \gamma^5 , \,  \gamma^5  \, \}$

(e.)
leads to
the non-vanishing of the traces in (\ref{TRs}), and 

(f.)
ends by
violating the conditions in (\ref{LV}).
\vskip.1in \noindent
Lorentz invariance is violated!  How can all this be avoided?

If one repeats all the calculations based on the network-based adinkras 
in Fig.\ \# 7 (see also Appendix E), all the problems disappear.  Explicitly 
and by using the same rainbow assignment.  We find the results
\be \eqalign{  {~~~~}
{\Tilde V}_{1 \, 2}^{({\rm {Adnk-C}})} ~&=~ - \,{\Tilde V}_{3 \, 4}^{({\rm {
Adnk-C}})} ~~,~~ {\Tilde V}_{2 \, 3}^{({\rm {Adnk-C}})} ~=~ -\, {\Tilde V}_{
1 \, 4}^{({\rm {Adnk-C}})} ~~,~~ \cr
{\Tilde V}_{1 \, 3}^{({\rm {Adnk-C}})} ~&=~ {\Tilde V}_{2 \, 4}^{({\rm {
Adnk-C}})}  ~~~~~\,~~, \cr
{\Tilde V}_{1 \, 2}^{({\rm {Adnk-D}})} ~&=~ {\Tilde V}_{3 \, 4}^{({\rm {
Adnk-D}})} ~~~~~~,~~
{\Tilde V}_{2 \, 3}^{({\rm {Adnk-D}})} ~=~  {\Tilde V}_{1 \, 4}^{({\rm {
Adnk-D}})} ~~~~~~,~~ \cr
{\Tilde V}_{1 \, 3}^{({\rm {Adnk-D}})} ~&=~ - \, {\Tilde V}_{2 \, 4}^{({\rm {
Adnk-D}})}  ~~~.
}   \label{Lvgvf1Z}
\ee

$$
\vCent
{\setlength{\unitlength}{1mm}
\begin{picture}(-10,0)
\put(7,-29.6){\includegraphics[width=2.3in]{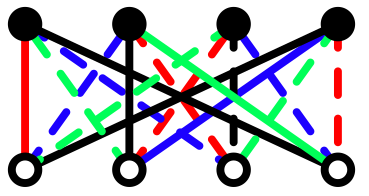}}
\put(28,-37){Adnk-D}
\put(-60,-30){\includegraphics[width=2in]{offCMv}}
\put(-41.8,-37){Adnk-C}
\put(-44,-44){\bf {Figure \# 8: Two valise adinkra networks}}
\end{picture}}
\nonumber
$$ \vskip 1.6in \noindent
The $\Tilde V$ matrices associated with the (Adnk-C)
are linearly independent of the $\Tilde V$ matrices associated with 
the (Adnk-D). Further it is possible to derive
\be {
{\Tilde V}_{I \, J}^{({\rm {Adnk-C}})} \, {\Tilde V}_{K \, L}^{({\rm {Adnk-D}})} 
~-~  {\Tilde V}_{K \, L}^{({\rm {Adnk-D}})}   \, {\Tilde V}_{I \, J}^{({\rm {
Adnk-C}})} ~=~ 0  ~~~.
} \label{BVgd}
\ee
which shows SUSY holography is restored.  It is also the case 
\be \eqalign{
{\rm {Tr}} \left[ \, {\Tilde V}_{I \, J}^{({\rm {Adnk-C}})} \, {\Tilde V}_{K \, L}^{(
{\rm {Adnk-D}})} \, \right]  ~=~ 0 ~~~, 
} \label{BVTrgd}
\ee
which are the necessary and sufficient conditions to avoid the violations
of Lorentz symmetry.  

We close this discussion by noting a necessary condition for two adinkras 
to represent the shadows of two distinct higher dimensional supermultiplets 
is that their respective L-matrices must be linearly independent.  One way 
to do this is to choose the permutation cycles related to the two quartets of 
L-matrices to lie in different sectors of the Venn diagram shown in (Fig.\ 2) 
of \cite{permutadnk}.   If the shadows of two distinct higher dimensional 
supermultiplets possess the same permutation cycles, then a proper choice 
of Boolean factors can also be used to maintain linear independence.

\section{Holoraumy Tensors for the Valise Real Scalar Supermultiplet} 
$~~~$
The transformation laws for the fields of the real scalar supermultiplet in 
valise form are given by the equations
$$
\begin{aligned}
D_a K & = \zeta _a \\
D_a M & = \frac{1}{2} \Lambda _a - \frac{1}{2} \left( \gamma^0 \right)_a^{
\ d} \zeta _d \\
D_a N & = - i \frac{1}{2} \left( \gamma^5 \right) _a ^{\ d} \Lambda _d + i 
\frac{1}{2} \left( \gamma ^5 \gamma ^0 \right) _a ^{\ d} \zeta _d \\
D_a U_0 & = i \frac{1}{2} \left( \gamma^5 \gamma _0 \right) _a ^{\ d} 
\Lambda_d - i \frac{1}{2} \left( \gamma^5 \right) _a ^{ \ d} \zeta _d 
{~~~~~~~~~~~~~~~~~~~} {~~~~~~~~~~~~~~~~~~~} \\
\end{aligned}
$$
\be
\begin{aligned}
D_a U_m & = i \frac{1}{2} \left( \gamma^5 \gamma _m \right) _a ^{\ d} 
\Lambda _d - i \frac{1}{2} \left( \gamma^5 \gamma^0 \gamma_m 
\right)_a ^{ \ d} \zeta _d \\
D_a {\rm d} & = - \left( \gamma ^0 \right) _a ^{\ d} \Lambda _d \\
D_a \zeta _b & = i \left( \gamma ^0 \right) _{ab}  \pa_{\tau}K + \left( 
\gamma^5 \gamma^\mu \right) _{ab} \pa_{\tau}U_\mu  + i C_{ab} \pa_{
\tau}M  + \left( \gamma^5 \right) _{ab} \pa_{\tau}N  \\
D_a \Lambda _b & = i \left( \gamma ^0 \right) _{ab}  \pa_{\tau}M  + \left( 
\gamma^5 \gamma^0 \right) _{ab} \pa_{\tau}N + \left( \gamma^5 \gamma^0 
\gamma^\nu \right) _{ab} \pa_{\tau}U_\nu  + i C_{ab} \pa_{\tau}{\rm d} 
\end{aligned}
\label{R1}
\ee

and to find bosonic holoraumy tensor for the real scalar supermultiplet we note
$$
\begin{aligned}
\left[ \, {\rm D}_a ~,~ {\rm D}_b \, \right] K & =  i 2 C_{ab}  \pa_{\tau}M + 2 
\left( \g^5 \right)_{ab}  \pa_{\tau}N  + 2 \left( \g^5 \g^0 \right)_{ab} \pa_{
\tau}U_0  + 2 \left( \g^5 \g^m \right)_{ab} \pa_{\tau}U_m {~}   \\
\left[ \, {\rm D}_a ~,~ {\rm D}_b \, \right] M & = 2 \left( \gamma^5 \gamma^0
\right){}_{ab}  \pa_{\tau}N - 2  \LP \GF \RP_{ab}  \pa_{\tau}U _0 
+ i C_{ab}  \pa_{\tau} ({\rm d} - K)   \\
\left[ \, {\rm D}_a ~,~ {\rm D}_b \, \right] N & = - 2 \LP \GFZ 
\RP_{ab}  \pa_{\tau}M 
+ i 2 C_{ab}  \pa_{\tau}U _0 + \LP \GF \RP_{ab}  \pa_{\tau}({\rm d} - K)  \\
\end{aligned}
$$
\be
\begin{aligned}
\left[ \, {\rm D}_a ~,~ {\rm D}_b \, \right] U_0 & = 2 \LP \GF \RP_{ab}  \pa_{
\tau}M - i 2 C_{ab}  \pa_{\tau} N + \LP \GFZ \RP_{ab} \pa_{\tau}\LP {\rm 
d} - K \RP \\
\left[ \, {\rm D}_a ~,~ {\rm D}_b \, \right] U_m & = - \LP \GF \g_m \RP_{ab} 
\pa_{\tau} \LP {\rm d} + K \RP -  i \LP \GZ \left[ \g_m \, , \,\g^n \right] \RP_{
ab}  \pa_{\tau}U _n   \\
\left[ \, {\rm D}_a ~,~ {\rm D}_b \, \right] {\rm d} & = - 2 i C_{ab}  \pa_{\tau}M 
- 2 \LP \GF \RP _{ab}  \pa_{\tau}N - 2 \LP \GFZ \RP_{ab}  \pa_{\tau}U_0  + 2 
\LP  \GF \g^m \RP_{ab}  \pa_{\tau}U _m     
\end{aligned}
\label{R2}
\ee
It is clear the linear combinations (d $\pm \, K$) possess some significance 
suggesting these be rewritten in the forms
\be
\begin{aligned}
\left[ \, {\rm D}_a ~,~ {\rm D}_b \, \right] ({\rm d} + K) & =   4 \left( \g^5 
\g^m \right)_{ab} \pa_{\tau}U_m  \\
\left[ \, {\rm D}_a ~,~ {\rm D}_b \, \right] M & = 2 \left( \gamma^5 \gamma^0
\right){}_{ab}  \pa_{\tau}N - 2  \LP \GF \RP_{ab}  \pa_{\tau}U _0 
+ i C_{ab}  \pa_{\tau} ({\rm d} - K)   \\
\left[ \, {\rm D}_a ~,~ {\rm D}_b \, \right] N & = - 2 \LP \GFZ  \RP_{ab}  \pa_{
\tau}M + i 2 C_{ab}  \pa_{\tau}U _0 + \LP \GF \RP_{ab}  \pa_{\tau}({\rm d} 
- K)  \\
\left[ \, {\rm D}_a ~,~ {\rm D}_b \, \right] U_0 & = 2 \LP \GF \RP_{ab}  \pa_{
\tau}M - i 2 C_{ab}  \pa_{\tau} N + \LP \GFZ \RP_{ab} \pa_{\tau}\LP {\rm d} - 
K \RP \\
\left[ \, {\rm D}_a ~,~ {\rm D}_b \, \right] U_m & = - \LP \GF \g_m \RP_{ab} 
\pa_{\tau} \LP {\rm d} + K \RP -  i \LP \GZ \left[ \g_m \, , \,\g^n \right] \RP_{
ab}  \pa_{\tau}U _n   \\
\left[ \, {\rm D}_a ~,~ {\rm D}_b \, \right] ({\rm d} - K) & = - 4 i C_{ab}  \pa_{\tau}
M - 4 \LP \GF \RP _{ab}  \pa_{\tau}N - 4 \LP \GFZ \RP_{ab}  \pa_{\tau}U_0 
\end{aligned} 
\label{R3}
\ee
These equations are interesting as they clearly identify two distinct subsets 
\be
subset_1 ~=~ \{ \, ({\rm d} + K), \, U_m  \, \}  ~~~,~~~ subset_2 ~=~ \{ \, ({\rm 
d} - K), \, M, \, N, \, U_0  \, \} 
\label{R4}
\ee
Calculations for the fermionic holoraumy tensor for the real scalar supermultiplet 
yield
\be
\begin{aligned}
\left[ \, {\rm D}_a ~,~ {\rm D}_b \, \right] \zeta_c & = i C_{ab} \,  (\GZ )_c ^{ 
\ d} \, \pa_{\tau}\left[  ~  \zeta_d -  (\GZ )_d ^{ \ e}  \Lambda  _e  ~ \right] \\
\ & \ \ ~~- i ( \GF )_{ab}\,  ( \GFZ )_c ^{ \ d}   \, \pa_{\tau}\left[ ~ \zeta _d - ( 
\gamma^0 )_d^{ \ e} \Lambda _e ~ \right]  \\
\ & \ \ ~~+i ( \GF \g^0 )_{ab}\,  ( \GF  )_c ^{\ d}   \,  \pa_{\tau}\left[~ \zeta_d - 
  (\g^0 )_d ^{ \ e} \Lambda_e  ~\right]   \\
\ & \ \ ~~+i ( \GF \g^m )_{ab}\, ( \GF \GZ \g_{m} )_c ^{\ d}   \,  \pa_{\tau} \left[~   
 \zeta_d  + (\gamma^0 ){}_d ^{ \ e} \Lambda_d  ~ \right]    \\
\left[ \, {\rm D}_a ~,~ {\rm D}_b \, \right] \Lambda_c & =  i C_{ab}\,  ( \GZ 
)_c ^{ \ d} \,  \pa_{\tau} \left[~  (\GZ )_d ^{ \ e} (  {\zeta}_e -  (\GZ )_e ^{ \ f}  
 \Lambda_f  )  ~ \right] \\
\ & \ \ ~~+ i ( \GF )_{ab} \,   ( \GFZ )_c ^{ \ d} \, \pa_{\tau} \left[~  (\GZ )_d ^{ \ 
e} ( {\zeta}_e -  (\GZ )_e ^{ \ f}   \Lambda_f  )  ~\right]   \\
\ & \ \ ~~- i ( \GF \GZ )_{ab}\,   ( \GF  )_c ^{ \ d} \,  \pa_{\tau} \left[ ~  (\GZ 
)_d^{ \ e} ( {\zeta}_e -  (\GZ )_e ^{ \ f}   \Lambda_f  )  ~ \right]
\\
\ & \ \ ~~- i ( \GF \g^m )_{ab} \,   ( \GF \g^0 \g_{m} )_c ^{ \ d}  \,  \pa_{\tau} 
\left[~   (\GZ )_d ^{ \ e} (  {\zeta}_e +  (\GZ )_e ^{ \ f}   \Lambda_f  )  ~\right] 
\end{aligned}
\label{R5}
\ee
or alternatively
\be
\begin{aligned}
\left[ \, {\rm D}_a ~,~ {\rm D}_b \, \right] \left[ \, \zeta_c  +  (\gamma^0){}_c
{}^d  \Lambda{}_d \, \right]
\ &= +i 2 ( \GF \g^m )_{ab}\, ( \GF \GZ \g_{m} )_c ^{\ d}   \,  \pa_{\tau} \left[~   
 \zeta_d  + (\gamma^0 ){}_d ^{ \ e} \Lambda_d  ~ \right]    \\
\left[ \, {\rm D}_a ~,~ {\rm D}_b \, \right]  \left[ \, \zeta_c  -  (\gamma^0){}_c{}^d  
\Lambda{}_d \, \right] & =  - i 2 C_{ab}\,  ( \GZ )_c ^{ \ d}   \,  
\pa_{\tau} \left[~  (  {\zeta}_d -  (\GZ )_d ^{ \ f}   \Lambda_f  )  ~ \right] \\
\ & \ \ ~~+ i 2 ( \GF )_{ab} \,   ( \GFZ )_c ^{ \ d} \, \pa_{\tau} \left[~ (  
{\zeta}_d -  (\GZ )_d ^{ \ f}   \Lambda_f  )  ~\right]   \\
\ & \ \ ~~+ i 2 ( \GF \GZ )_{ab}\,   ( \GF  )_c ^{ \ d} \,  \pa_{\tau} \left[ ~  (  
{\zeta}_d -  (\GZ )_d ^{ \ f}   \Lambda_f  )  ~ \right]
\end{aligned}
\label{R6}
\ee
We observe
\be
{\rm D}{}_a ({\rm d} \pm K) ~=~ - \left[ \, (\gamma^0){}_a{}^b  \Lambda{}_b ~
~\mp~ \zeta{}_a   \, \right] 
\label{R7}
\ee
along with the expressions for the fermionic holoraumy tensor above, this 
suggests the definitions
\be
\eqalign{ {~~~}
\Phi{}_{\Lambda} ~&\equiv~    \left(\, ({\rm d} - K), \, M, \, N, \, U_0 \, {\large 
|} \, U_m, \, ({\rm d} + K)  \,\right)   ~~~~~,~  \cr
\Psi{}_{\Hat {\Lambda}} ~&\equiv~ \left( \,  \zeta{}_a + (\gamma^0){}_a{
}^b \Lambda{}_b  \,  {\large |} \, \zeta{}_a - (\gamma^0){}_a{}^b \Lambda{
}_b   \,\right)   ~~~~, \cr
} \label{R8}
\ee
and we combine these equations to find,
\be
\begin{aligned}
\left[ \, {\rm D}_a ~,~ {\rm D}_b \, \right] \Phi{}_{ {\Lambda}} & = - \, i\, 2\, 
(\, \mathscr{B}{}_{{a } \, {b}} \,){}_{{\Lambda}} {}^{ {\Delta}}  \, \pa_{\t}  
\Phi{}_{{\Delta}}
\end{aligned}
\label{R9}
\ee
\be
\begin{aligned}
\left[ \, {\rm D}_a ~,~ {\rm D}_b \, \right] \Psi{}_{\Hat {\Lambda}} & = - \, i\, 2\, 
(\, \mathscr{F}{}_{{a } \, {b}} \,){}_{\Hat {\Lambda}} {}^{\Hat {\Delta}} \, \pa_{\t}  
\Psi{}_{\Hat {\Delta}}
\end{aligned}
\label{R10}
\ee
for the explicit representation-dependent forms of $(\, \mathscr{B}{}_{{a } {
b}} \,){}_{ {\Lambda}} {}^{ {\Delta}}$ and $(\, \mathscr{F}{}_{{a } {b}} \,){}_{\Hat 
{\Lambda}} {}^{\Hat {\Delta}}$.  The basis for $\Phi{}_{\Lambda}$ and $\Psi{
}_{\Hat {\Lambda}}$ has the property that both $(\, \mathscr{B}{}_{{a } {b}} 
\,){}_{ {\Lambda}} {}^{ {\Delta}}$ and $(\, \mathscr{F}{}_{{a } {b}} \,){}_{\Hat {
\Lambda}} {}^{\Hat {\Delta}}$ are block diagonal with 4 $\times$ 4 matrices 
on the main diagonal of the 8 $\times$ 8 matrices.   This completes the 
derivation of the holoraumy tensors on the 0-brane side of the correspondence.  

The splitting of the real scalar supermultiplet into the two distinct subsets 
seen in (\ref{R8}) can be illustrated\footnote{This image appeared in 
``Symbols of Power'' in \cite{SoP}.} as below using adinkras.
$$
\vCent
{\setlength{\unitlength}{1mm}
\begin{picture}(-20,-140)
\put(-77,-37){\includegraphics[width=5.9in]{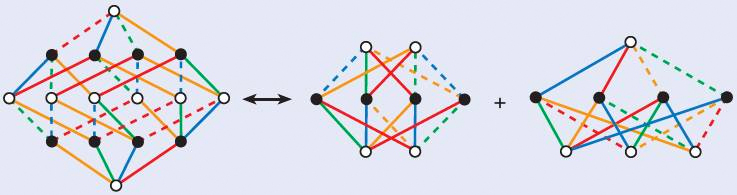}}
\put(-54,-46){\bf {Figure \# 9: Illustration of Diagonalized Holoraumy }}
\end{picture}}
$$
\vskip1.7in

The subset of fields shown to the left of the ``vertical bars'' in (\ref{R8}) constitute 
the fields seen in the middle adinkra of Fig.\ \# 9 while subset of fields shown to 
the right of the ``vertical bars'' constitute the fields seen in the rightmost adinkra.  
The leftmost adinkra corresponds to the real scalar supermultiplet.   It can be 
seen by the equations in (\ref{R6}), the splitting is accomplished by choosing 
between the basis defined by the linear combination of fermions defined by 
$[\,  \zeta_c  \pm  (\gamma^0){}_c{}^d  \Lambda{}_d \, ]$ as these linear 
combinations control how the fermions in the original real scalar superfield 
must be rearranged to obtain the fermions in the sub-multiplets.  On the 
network-based adinkra side, this same splitting is accomplished by the use
of error-correcting codes \cite{DFGHILM2, DFGHILM3, DFGHILM4}.  Thus 
we find the result the use of error-correcting codes on the network-based 
adinkra side corresponds to the splitting of SO(1,3) fermions into covariant 
SO(3) ones on the 0-brane reduced side.

We believe this is a very telling sign of the linkage between 1D, $N$-extended
SUSY together with error-correcting codes and the emergence of Lorentz 
symmetry in the higher dimensional space-time manifolds.

On the adinkra side, we find the following results.  We define the adinkra 
node ``vectors'' via
\be \eqalign{ {~~~}
\Phi{}_{i} ~&=~    \left(\, {\rm d}, \, M + U_2 , \, U_0 + U_1 , \, N + U_3 , \,  
- N + U_3 , \,  U_0 + U_1 , \,  - M + U_2 , \,  K  \,\right)   ~~~~~,~  \cr
\Psi{}_{\Hat {k}} ~&=~ \left( \,  \Lambda_1 , \,  \Lambda_2 , \,  \Lambda_3 , 
\,  \Lambda_4 , \, \zeta_1 , \,  \zeta_2 , \,  \zeta_3 , \,  \zeta_4
 \,\right)   ~~~~, 
} \label{R11}
\ee
and using this definition we find the L-matrices and R-matrices are given
be
\be  \eqalign{  {~~~~~~~~~~\,}
{\rm L}_1 ~=~ (128)_b\{ (12)(38)(57)  \} ~~&,~~ {\rm L}_2 ~=~ (22)_b\{ (34)
(53)(67)  \}  
~~~~~~~~~~~~, \cr
{\rm L}_3 ~=~ (42)_b\{ (14)(28)(31)  \} ~~\,~&,~~ {\rm L}_4 ~=~ (230)_b\{  
(13)(27)(46)(51) \}  
~~~~~,
} \label{R12}
\ee
\be
\eqalign{  {~~~~~~~~~~\,}
{\rm R}_1 ~=~ (64)_b\{ (12)(36)(58)  \} ~~&,~~ {\rm R}_2 ~=~ (11)_b\{ (35)
(43)(68) \}  ~~~~~~~~~~~~, \cr
{\rm R}_3 ~=~ (168)_b\{ (13)(24)(78)  \} ~~\,~&,~~ {\rm R}_4 ~=~ (122)_b\{  
(15)(26)(31)(47) \}  ~~~~~,
} \label{R13}
\ee
For the $V$-matrices and $\Tilde V$-matrices we find,
\be
 \eqalign{ {~~\,}
V{}_{\rI\rJ} ~=~  \frac12 \, 
 (\,{\rm L}_\rI \, {\rm R}_\rJ  ~-~ {\rm L}_\rJ \,{\rm 
R}_\rI\,)  ~~~,~ ~~
{\Tilde V}{}_{\rI\rJ} ~=~   \frac12 \,
 (\,{\rm R}_\rI \, {\rm L}_\rJ  ~-~ {\rm R}_\rJ\ \,{\rm 
 L}_\rI\,)   ~~,
} \label{R14}
\ee
and enumerate the non-vanishing and non-redundant values:
\be
\begin{aligned}
{V}_{12} & = (149)_b\{(12)(34)(56)(78)\} \\
{V}_{13} & = (163)_b \{(14)(23)(58)(67)\} \\
{V}_{14} & = (184)_b \{ (16)(25)(38)(47)\} \\
{V}_{23} & = (57)_b \{ (13)(24)(57)(68)\} \\
{V}_{24} & = (30)_b \{ (15)(26)(37)(48) \} \\
{V}_{34} & = (141)_b \{ (17)(28)(35)(46) \}
\end{aligned}
\label{R15}
\ee

\be
\begin{aligned}
\Tilde{V}_{12} & = (46)_b\{(12)(37)(48)(56)\} \\
\Tilde{V}_{13} & = (226)_b\{(18)(24)(36)(57)\} \\
\Tilde{V}_{14} & = (165)_b\{(17)(23)(46)(58) \} \\
\Tilde{V}_{23} & = (204)_b \{(14)(28)(35)(67) \} \\
\Tilde{V}_{24} & = (139)_b \{(13)(27)(45)(68)\} \\
\Tilde{V}_{34} & = (184)_b \{(15)(26)(34)(78) \}    ~~~.
\end{aligned}
\label{R16}
\ee
For the algebra of the $V$-matrices and $\Tilde V$-matrices we find,
\be
\frac{1}{2}[V_{IJ},V_{KL}] = - \delta _{IK} V_{JL} + 
\delta _{IL} V_{JK} + \delta _{JK} V_{IL} - \delta _{JL} V_{IK}  ~~~,
\label{R17}
\ee
\be
\frac{1}{2}[\Tilde{V}_{IJ},\Tilde{V}_{KL}] = - \delta _{IK} \Tilde{V}_{JL} + 
\delta _{IL} \Tilde{V}_{JK} + \delta _{JK} \Tilde{V}_{IL} - \delta _{JL} \Tilde{
V}_{IK}  ~~~,
\label{R18}
\ee
and the factors of one-half appearing in these informs us these
constitute spinor representations of SO(4).  We can form the 
following linear combinations of the $\Tilde V$-matrices
\be
\Tilde{V}_{I\, J}^{(\pm)}  ~=~ \fracm 12 
\left[ \, \Tilde{V}_{I\, J} ~\pm~ \fracm 12 \epsilon{}_{I \, J \, K \, L} \, \Tilde{V}_{
K \, L} \, \right]
\label{R19}
\ee
and due to (\ref{R17}) we find
\be
\left[ \, \Tilde{V}_{I\, J}^{(\pm)} ~,~ \Tilde{V}_{K \, L}^{(\mp)}  \right] ~=~ 0
\label{R20}
\ee
It is now of interest to note the possibility to define two sets of matrices 
denoted by $[ {\cal S}{}_{\rI \, \rJ\, \rK}^{(\pm)} ]$ and defined by the expressions
\be
{V}_{J\, K}^{(\pm)}  {\rm L}_\rI
~-~ {\rm L}_\rI \Tilde{V}_{J\, K}^{(\pm)} ~=~ \left[ {\cal S}{}_{\rI \, \rJ\, \rK}^{(\pm)} 
\right]    ~~~,
\label{R21}
\ee
since the equivalent expression of the 0-brane side contain information
about spin and R-symmetry spin of the fields in the real scalar superfield.
We plan to return to this topic in a future effort.

\section{Extending Holoraumy to 4D, $\bm {\cal N}$ $>$ 1 SUSY} 
$~~~$ 
All our previous discussion of holoraumy has been within the arena of 
four dimensional (or less) simple supersymmetry.  In this section, we 
outline the general extension of these ideas to theories with more 
supercharges.

We begin by noting a simple counting argument.  In order to describe an 
extended supersymmetry in spacetime, we replace the supercovariant
derivative via the rule
\be
{\rm D}{}_{a}  ~~~\to ~~~ {\rm D}{}_{a \, i}  ~~~,
\label{H1}
\ee
where the $i$ subscript counts the number of four dimensional spacetime 
supercharges. It thus ranges as 1$, \, \dots , \, {\cal N}$, where now $\cal N$ 
can be any integer.  The algebra of this extended SUSY covariant derivative 
is simply
\be
\{ \, {\rm D}{}_{a \, i}  ~,~ {\rm D}{}_{b \, j} \, \} ~=~ i \, 2 \, \delta_{i \, j} \, (\gamma^{\mu})
{}_{a \, b} \, \pa_{\mu} ~~~,
\label{H2}
\ee
for all  {\em {off-shell}} SUSY representations.  Via ``0-brane reduction'' \cite{
redux} and ``node lowering'' all {\em {off-shell}} representations of this algebra 
can expressed in such a way that this anti-commutation relationship becomes
\be
\{ \, {\rm D}{}_{a \, i}  ~,~ {\rm D}{}_{b \, j} \, \} ~=~ i \, 2 \, \delta_{i \, j} \, (\gamma^{0})
{}_{a \, b} \, \pa_{\tau} ~~~.
\label{H3}
\ee

In many places in the SUSY literature, it is customary to deform the anti-commutator
relation in (\ref{H2}) so it reads
\be   \eqalign{
\{ \, {\rm D}{}_{a \, i}  ~,~ {\rm D}{}_{b \, j} \, \} ~&=~ i \, 2 \, \delta_{i \, j} \, (\gamma^{\mu})
{}_{a \, b} \, \pa_{\mu} ~+~ ( \,\Delta{}_{{a i} \, {b j} }{}^{[{\Hat 1}]}\, ) {\cal Z}{}_{[{\Hat 
1}]} ~+~ ( \,\Delta{}_{{a i} \, {b j} }{}^{[{\Hat 2}]}\, ) {\cal Z}{}_{[{\Hat 2}]}        \cr
&~~~~+~ ( \,\Delta{}_{{a i} \, {b j} }{}^{[{\Hat 3}]}\, ) {\cal Z}{}_{[{\Hat 3}]} ~+~ 
( \,\Delta{}_{{a i} \, {b j} }{}^{[{\Hat 4}]}\, ) {\cal Z}{}_{[{\Hat 4}]}   ~+~ \dots ~~
~~~,  {~~~~~~~}
} \label{H4}
\ee
where the operators ${\cal Z}{}_{[{\Hat 1}]}$, ${\cal Z}{}_{[{\Hat 2}]}$,  ${\cal Z}{}_{[{
\Hat 3}]}$, ${\cal Z}{}_{[{\Hat 4}]}$, etc. are called ``off-shell'' central charges if they 
vanish when the equations of motion, derivable from some Lagrangian, are imposed.  
The quantities $( \,\Delta{}_{{a i} \, {b j} }{}^{[{\Hat 1}]}\, )$, etc. are simply some set of
constants.  Although we do not have a general proof about the behavior of ``off-shell'' 
central charges under 0-brane reduction, we do have at least one example which 
suggests something interesting about the result one can expect generally.

In the work of \cite{adnkKYeoh}, the Wess-Fayet hypermultiplet \cite{Wss} (which has
widely been used to represent $\cal N$ = 2 supermatter), was studied under 0-brane
reduction and ``node lowering.''  The result of the study indicated in this limit,
one finds
\be   \eqalign{
{\cal Z}{}_{[{\Hat 1}]} ~=~  {\rm { \sc z}}{}_{[{\Hat 1}]} \, \pa_{\tau} ~~,~~
{\cal Z}{}_{[{\Hat 2}]} ~=~  {\rm {\sc z}}{}_{[{\Hat 2}]} \, \pa_{\tau} ~~,~~ \dots ~~~,
} \label{H5}
\ee
for some constants ${\rm {\sc z}}{}_{[{\Hat 1}]} $, $ {\rm {\sc z}}{}_{[{\Hat 2}]} $, ${\rm  
{\sc z}}{}_{[{\Hat 3}]} $, $\dots$ on this supermultiplet.  In the following, we make 
the {\em {assumption}} this behavior is valid for {\em {all}} theories which possess 
off-shell central charges.  Under this assumption, when (\ref{H4}) is subjected to 
0-brane reduction and node-lowering we expect the equation
\be   \eqalign{
\{ \, {\rm D}{}_{a \, i}  ~,~ {\rm D}{}_{b \, j} \, \} ~&=~ i \, 2 \, \delta_{i \, j} \, (\gamma^{0})
{}_{a \, b} \, \pa_{\tau} ~+~ ( \,\Delta{}_{{a i} \, {b j} }{}^{[{\Hat 1}]}\, ) 
 {\rm {\sc z}}{}_{[{\Hat 1}]} \, \pa_{\tau} ~+~ ( \,\Delta{}_{{a i} \, {b j} }{}^{[{\Hat 2}]}\, ) 
  {\rm {\sc z}}{}_{[{\Hat 2}]} \, \pa_{\tau}      \cr
&~~~~+~ ( \,\Delta{}_{{a i} \, {b j} }{}^{[{\Hat 3}]}\, ) {\rm {\sc z}}{}_{[{\Hat 3}]} \, \pa_{\tau} 
~+~ ( \,\Delta{}_{{a i} \, {b j} }{}^{[{\Hat 4}]}\, )  {\rm {\sc z}}{}_{[{\Hat 4}]} \, \pa_{\tau}  
~+~ \dots ~~~~~,  {~~~~}
} \label{H6}
\ee
to emerge.   In the case of the hypermultiplet, we also found the conditions
\be  \eqalign{
{~~~~~}
&\delta^{i \, j} \, (\gamma^{\mu}){}^{a \, b} \, ( \,\Delta{}_{{a i} \, {b j} }{}^{[{\Hat 1}]}\, ) 
~=~ \delta^{i \, j} \, (\gamma^{\m}){}^{a \, b} \, ( \,\Delta{}_{{a i} \, {b j} }{}^{[{\Hat 2}]}\, ) 
~=~ \delta^{i \, j} \, (\gamma^{\mu}){}^{a \, b} \, ( \,\Delta{}_{{a i} \, {b j} }{}^{[{\Hat 3}]}\, ) 
~=~   \cr
&\delta^{i \, j} \, (\gamma^{\mu}){}^{a \, b} \, ( \,\Delta{}_{{a i} \, {b j} }{}^{[{\Hat 4}]}\, ) 
~=~ \dots ~=~ 0  ~~~.
} \label{H7}
\ee

Once more we assert this is also the case for any general 0-brane reduction and
node-lowering on any SUSY representation.  The result in (\ref{H7}) and the
fact for all the $\Delta$-symbols we have the identity
\be  \eqalign{
( \,\Delta{}_{{a i} \, {b j} }{}^{[{\Hat {\rm I}}]}\, )  ~=~ ( \,\Delta{}_{{b j} \, {a i} }{}^{[{\Hat 
{\rm I}}]}\, ) ~~,
} \label{H8}
\ee
implies we can count their linearly independent parts easily.  Since these 
quantities are bilinear in spinor indices $a$ and $b$ they can be expanded over 
the six elements
\be    \eqalign{
C_{a \, b}  ~~,~~ (\gamma^{5}){}_{a \, b}  ~~,~~  (\gamma^{5} \gamma^{\mu}){}_{
a \, b} ~~, } \label{H9}
\ee
(which are antisymmetric on the exchange of the $a$ and $b$ indices) and
the ten elements
\be    \eqalign{
 (\gamma^{\mu}){}_{a \, b}  ~~,~~  (\gamma^{\mu} \, \wedge \, \gamma^{\nu}){}_{
 a \, b} ~~, } \label{H10}
\ee
which are symmetric on the exchange of the $a$ and $b$.  In order to realize 
the symmetry as in (\ref{H8}) we must pair the six elements in (\ref{H9}) with 
an antisymmetric tensor $\mathscr{A}{}_{i \, j}$ to form the objects
\be    \eqalign{
C_{a \, b} \, \mathscr{A}{}_{i \, j} ~~,~~ (\gamma^{5}){}_{a \, b} \, \mathscr{A}{}_{
i \, j}  ~~,~~  (\gamma^{5} \gamma^{\mu}){}_{a \, b} \, \mathscr{A}{}_{i \, j}~~,
} \label{H11}
\ee
which gives a total of $6$ $\times$ ${\cal N}\, (\, {\cal N} - 1 \, )/2$ tensors.  In a 
similar manner in order to realize the symmetry as in (\ref{H8}) we must pair the 
10 elements in (\ref{H10}) with an symmetric tensor $\mathscr{S}{}_{i \, j}$ to form 
the objects
\be    \eqalign{
(\gamma^{\mu} ){}_{a \, b} \, \mathscr{S}{}_{i \, j}  ~~,~~  (\gamma^{\mu} \wedge 
\gamma^{\nu}){}_{a \, b} \, \mathscr{S}{}_{i \, j}~~,
} \label{H12}
\ee
which gives a total of $10$ $\times$ ${\cal N}\, (\, {\cal N} + 1 \, )/2$ tensors.  We 
also note the objects in (\ref{H12}) can be replaced by 
\be    \eqalign{
 (\gamma^{\mu} ){}_{a \, b} \, \delta{}_{i \, j}  ~~,~~ 
  (\gamma^{\mu} ){}_{a \, b} \, \mathscr{S}{}_{i \, j}^{(T)}  
~~,~~  (\gamma^{\mu} \wedge \gamma^{\nu}){}_{a \, b} \, \mathscr{S}{}_{i \, j}~~,
} \label{H13}
\ee
where $\delta{}^{i \, j} \, \mathscr{S}{}_{i \, j}^{(T)}  $ = 0. Thus, without loss of 
generality, the $\Delta$-tensors appearing in (\ref{H6}) - (\ref{H8}) can expanded 
in the product tensors in (\ref{H11}) and (\ref{H13}). 

Note two identities
 \be    \eqalign{ {~~~~~~}  \left(\begin{array}{c}
d_F \\
2 \\
\end{array}\right) 
\,  \left(\begin{array}{c}
{\cal N} \\
2 \\
\end{array}\right) 
 ~+~  \left(\begin{array}{c}
d_F + 1 \\
2 \\
\end{array}\right) 
\,  \left(\begin{array}{c}
{{\cal N} + 1 } \\
2 \\
\end{array}\right)  
 ~&=~ 
 \left(\begin{array}{c} 
d_F \, {\cal N} ~+~ 1 \\
2 \\
\end{array}\right)  ~~,
}  \label{BY1}
\ee
\be    \eqalign{
 \left(\begin{array}{c}
d_F + 1 \\
2 \\
\end{array}\right) 
\,  \left(\begin{array}{c}
{\cal N} \\
2 \\
\end{array}\right) 
 ~+~  \left(\begin{array}{c}
d_F  \\
2 \\
\end{array}\right) 
\,  \left(\begin{array}{c}
{{\cal N} + 1 } \\
2 \\
\end{array}\right)  
~&=~ 
 \left(\begin{array}{c}
d_F \, {\cal N}  \\
2 \\
\end{array}\right)  ~~,
}  \label{BY2}
\ee
Adding together the number of independent constant tensors in (\ref{H11}) 
and (\ref{H12}) gives 
\be    \eqalign{
 \left(\begin{array}{c}
4 \\
2 \\
\end{array}\right) 
\,  \left(\begin{array}{c}
{\cal N} \\
2 \\
\end{array}\right) 
 ~+~  \left(\begin{array}{c}
5 \\
2 \\
\end{array}\right) 
\,  \left(\begin{array}{c}
{{\cal N} + 1 } \\
2 \\
\end{array}\right) 
~=~  \left(\begin{array}{c}
{4 {\cal N} + 1 } \\
2 \\
\end{array}\right) 
 ~~.
} \label{H14}
\ee
This is a special case of the identity in (\ref{BY1}) for $d_F$ = 4.  On the 
other hand, if we interchange the factors of 6 and 10 in (\ref{H14}), we obtain
\be    \eqalign{
 \left(\begin{array}{c}
5 \\
2 \\
\end{array}\right) 
\,  \left(\begin{array}{c}
{\cal N} \\
2 \\
\end{array}\right) 
 ~+~  \left(\begin{array}{c}
4 \\
2 \\
\end{array}\right) 
\,  \left(\begin{array}{c}
{{\cal N} + 1 } \\
2 \\
\end{array}\right) 
~=~  \left(\begin{array}{c}
{4 {\cal N}  } \\
2 \\
\end{array}\right) 
 ~~,
} \label{H15}
\ee
and this is a special case of the identity in (\ref{BY2}) for $d_F$ = 4. The result in 
(\ref{H15}) is indicative the objects
\be    \eqalign{
 &(\gamma^{\mu} ){}_{a \, b} \, \mathscr{A}{}_{i \, j}  
~~,~~  (\gamma^{\mu} \wedge \gamma^{\nu}){}_{a \, b} \, \mathscr{A}{}_{i \, j}~~,
\cr
&C_{a \, b} \, \delta {}_{i \, j} ~~~~~~~,~~ (\gamma^{5}){}_{a \, b} \, \delta{}_{i \, j}  
~~~~~,~~  (\gamma^{5} \gamma^{\mu}){}_{a \, b} \, \delta{}_{i \, j}~~~~~, \cr
&C_{a \, b} \, \mathscr{S}{}_{i \, j}^{(T)} ~~~~,~~ (\gamma^{5}){}_{a \, b} \, 
\mathscr{S}{}_{i \, j}^{(T)} ~~,~~  (\gamma^{5} \gamma^{\mu}){}_{a \, b} \, 
\mathscr{S}{}_{i \, j}^{(T)}  ~~,
} \label{H16}
\ee
are all anti-symmetrical under the exchange of the pair of indices $a \, i$ with $b 
\, j$.  They are antisymmetric and their number is given by the RHS in (\ref{H15}) 
indicates these are in the defining representation of the group SO(4$\cal N$).

The operator equation
\be
[ \, {\rm D}{}_{a \, i}  ~,~ {\rm D}{}_{b \, j} \, ]~=~  ( \, \mathscr{H}{}_{{a \, i} \, {b \, j}} \, )
~~~,
\label{H17}
\ee
implies the matrices in (\ref{H16}) must form a representation of the generators 
of SO(4$\cal N$).  The RHS of (\ref{H17}) can be decomposed in the basis in (\ref{
H16}) and re-expressed as
\be  \eqalign{ {~~~~}
[ \, {\rm D}{}_{a \, i}  ~,~ {\rm D}{}_{b \, j} \, ] ~&=~ i\,
(\gamma^{\mu} ){}_{a \, b} \, ( \, \mathscr{A}{}_{i \, j} \cdot  \mathscr{H}{}_{\mu} \,)
~+~ i\,  (\gamma^{\mu} \wedge \gamma^{\nu}){}_{a \, b} \, (\, \mathscr{A}{}_{i \, j} 
\cdot \mathscr{H}{}_{\mu \, \nu} \, )  \cr
&~~~~+~ i \, C_{a \, b} \, (\, \mathscr{S}{}_{i \, j} \cdot { {  \mathscr{H}}} \, ) ~+~ 
(\gamma^{5}){}_{a \, b} \, ( \,  \mathscr{S}{}_{i \, j} \cdot  {\Tilde {  \mathscr{H}}} \, )
  \cr
&~~~~+~ 
(\gamma^{5} \gamma^{\mu}){}_{a \, b} \, (\,  \mathscr{S}{}_{i \, j} \cdot  {\Tilde { 
\mathscr{H}}}{}_{\mu} \, ) ~~~,
} \label{H18}
\ee
This equation defines the operators ($ \mathscr{A}{}_{i \, j} \cdot  \mathscr{H}{}_{
\mu}$), ($ \mathscr{A}{}_{i \, j} \cdot  \mathscr{H}{}_{\mu \nu}$), ($ \mathscr{S}{}_{
i \, j} \cdot  \mathscr{H}$), $( \mathscr{S}{}_{i \, j} \cdot  {\Tilde { \mathscr{H}}} )$, and 
$( {\Tilde { \mathscr{H}}}{}_{\mu} )$ which are appropriate for $\cal N$-extended 
superspace.  Finally, for all off-shell linear representations subjected to 0-brane 
reduction in the Coulomb gauge, and node-lowering re-definitions of fields 
leading to valise supermultiplets, the final form of this equation is
\be  \eqalign{ {~~~~}
[ \, {\rm D}{}_{a \, i}  ~,~ {\rm D}{}_{b \, j} \, ] ~&=~ i\, (\gamma^{\mu} ){}_{a \, b} \, 
( \, \mathscr{A}{}_{i \, j} \cdot  \mathscr{H}{}_{\mu} \,) \, \pa_{\t} ~+~  i\,  (\gamma^{
\mu} \wedge \gamma^{\nu}){}_{a \, b} \, (\, \mathscr{A}{}_{i \, j} \cdot \mathscr{H}{
}_{\mu \, \nu} \, )  \, \pa_{\t} \cr
&~~~~+~ C_{a \, b} \, (\, \mathscr{S}{}_{i \, j} \cdot { {  \mathscr{H}}} \, )  \, \pa_{\t} ~+~ 
(\gamma^{5}){}_{a \, b} \, ( \,  \mathscr{S}{}_{i \, j} \cdot  {\Tilde {  \mathscr{H}}} \, )  
\, \pa_{\t}  \cr
&~~~~+~
(\gamma^{5} \gamma^{\mu}){}_{a \, b} \, (\,  \mathscr{S}{}_{i \, j} \cdot  {\Tilde { 
\mathscr{H}}}{}_{\mu} \, )  \, \pa_{\t} ~~~.
} \label{H19}
\ee
where the coefficients preceding the partial derivatives define the holoraumy tensors
acting on any valise representation of 4D, $\cal N$-extended supersymmetry.  In 
general these are representation depend and in general distinct for the bosonic 
versus fermionic fields in a supermultiplet as seen in our $\cal N$ = 1 examples.  
Finally, it is important to note the quantities $\mathscr{H}$,  $\mathscr{H}{}_{\mu}$,  
$\mathscr{H}{}_{\mu \nu}$, ${\Tilde { \mathscr{H}}}{}_{\mu}$, and $\Tilde { \mathscr{
H}}$ in (\ref{H19}) are not operators.  When (\ref{H19}) is evaluated on the bosons 
of a valise supermultiplet we conventionally write
\be  \eqalign{ {~~~~}
[ \, {\rm D}{}_{a \, i}  ~,~ {\rm D}{}_{b \, j} \, ] ~&=~ - \, i\, 2\, (\, \mathscr{B}{}_{{a i} 
\, {b j}} \,) \, \pa_{\t}  ~~~,
} \label{H20}
\ee
where $ (\, \mathscr{B}{}_{{a i} \, {b j}} \,) $ denotes a set of representation dependent 
constants that may be decomposed in the basis in (\ref{H16}), when (\ref{H19}) is 
evaluated on the fermions of a valise supermultiplet we write
\be  \eqalign{ {~~~~}
[ \, {\rm D}{}_{a \, i}  ~,~ {\rm D}{}_{b \, j} \, ] ~&=~ - \, i\, 2\, (\, \mathscr{F}{}_{{a i} \, 
{b j}} \,) \, \pa_{\t}  ~~~,
} \label{H21}
\ee
where $ (\, \mathscr{F}{}_{{a i} \, {b j}} \,) $ denotes a set of representation dependent 
constants that may once more be decomposed in the basis in (\ref{H16}). 

In contrast to the objects shown in (\ref{H16}), the objects below
\be    \eqalign{
&(\gamma^{\mu} ){}_{a \, b} \, \delta{}_{i \, j}  
~~,~~  (\gamma^{\mu} \wedge \gamma^{\nu}){}_{a \, b} \, \delta{}_{i \, j}~~,
\cr
&(\gamma^{\mu} ){}_{a \, b} \, \mathscr{S}{}_{i \, j}^{(T)}  
~~,~~  (\gamma^{\mu} \wedge \gamma^{\nu}){}_{a \, b} \, \mathscr{S}{}_{i \, j}^{(T)} ~~,
\cr
&C_{a \, b} \, \mathscr{A}{}_{i \, j} ~~,~~ (\gamma^{5}){}_{a \, b} \, \mathscr{A}{}_{i \, j}  
~~,~~  (\gamma^{5} \gamma^{\mu}){}_{a \, b} \, \mathscr{A}{}_{i \, j}~~,
} \label{H22}
\ee
are all symmetric under the exchange of the pairs of indices $a \, i$ $\leftrightarrow$ 
$b \, j$.  It should be noted all of these are actually determined by the relations 
between the quantities in (\ref{H16}) and the SO(4$\cal N$) generators.  In other 
words, a set of generators for SO(4$\cal N$) can be used to define the quantities
in (\ref{H22}) via definitions that arise from (\ref{H16}).

In terms of the quantities in (\ref{H22}), we can write
\be  \eqalign{ {~}
\{ \, {\rm D}{}_{a \, i}  ~,~ {\rm D}{}_{b \, j} \, \} ~&=~ i\, (\gamma^{\mu}){}_{a \, b} \, \delta
{}_{i \, j} \,  \mathscr{K}{}_{\mu} \, \pa_{\t} ~+~  i\,  (\gamma^{\mu} \wedge \gamma^{\nu}
){}_{a \, b} \,  \delta{}_{i \, j} \, \mathscr{K}{}_{\mu \, \nu} \, \pa_{\t} \cr
&~~~~+~  i\, (\gamma^{\mu} ){}_{a \, b} \, ( \, 
\mathscr{S}{}_{i \, j}^{(T)} \cdot  \mathscr{K}{}_{\mu}^{(T)}  \,) \, \pa_{\t} ~+~  i\,  
(\gamma^{\mu} \wedge \gamma^{\nu}){}_{a \, b} \, (\, \mathscr{S}{}_{i \, j}^{(T)}   
\cdot \mathscr{K}{}_{\mu \, \nu}^{(T)}  \, )  \, \pa_{\t} \cr
&~~~~+~ C_{a \, b} \, (\, \mathscr{A}{}_{i \, j} \cdot { {  \mathscr{K}}} \, )  \, \pa_{\t} ~+~ 
(\gamma^{5}){}_{a \, b} \, ( \,  \mathscr{A}{}_{i \, j} \cdot  {\Tilde {  \mathscr{K}}} \, )  
\, \pa_{\t}  \cr
&~~~~+~
(\gamma^{5} \gamma^{\mu}){}_{a \, b} \, (\,  \mathscr{A}{}_{i \, j} \cdot  {\Tilde { \mathscr{
K}}}{}_{\mu} \, )  \, \pa_{\t} ~~~,
} \label{H23}
\ee
for some matrices of coefficients $\mathscr{K}{}_{\mu}$, $\mathscr{K}{}_{\mu \, \nu}$, 
${ {  \mathscr{K}}}{}^{(T)} $, $\mathscr{K}{}_{\mu}^{(T)} $, ${{\mathscr{K}}}$, $ {\Tilde { 
\mathscr{K}}}{}_{\mu}$, and $  {\Tilde {  \mathscr{K}}}$.  From this expression it is seen 
the condition that the 0-brane representation should satisfy and off-shell SUSY algebra 
is equivalent to the requirement that only $\mathscr{K}{}_{0}$ should be non-vanishing.

The end result of the discussion in this chapter is the notion of covariant
0-brane holoraumy tensors, presented in the previous chapters for the case 
of off-shell 4D, $\cal N$ = 1 supermultiplets, can easily be extended to four 
dimensional theories where $\cal N$ $>$ 1.

\section{Conclusion, Observations and Summary} 
$~~~$ 
By this work, we have been led to reason ``SUSY 
Holography'' exists between 4D, $\cal N$-extended 0-brane SUSY valise 
supermultiplets on one side of a correspondence and 1D, $4 {\cal N}$-extended 
adinkra-based valise supermultiplets on the other side is due to the identity 
$\left[ GL(4, {\rm R}) \otimes GL({\cal N}, {\rm R})\right]{}_A $ $=$ $SO(4 
{\cal N})$.  The 0-brane models realize the left hand side of this correspondence
making manifest 4D Lorentz and O($\cal N$) symmetries, and the adinkra 
models realize the right hand side making manifest connections to 
``Garden Algebras,'' graph theory and error-correcting codes.  

All our work on the adinkra substructure of 4D, $\cal N$ = 1 theories inform
us that all such irreducible supermultiplets appear to be in the spinor 
representations of SO(4) which can be subject to linear restrictions that involve
error-correcting codes.  The spinor representations of SO(4) contain an 
SU(2) $\otimes$ SU(2) subgroup.  To distinguish between these two copies 
of SU(2), we have in the past used the notation SU${}_{\alpha}$(2) $\otimes$ 
SU${}_{\beta}$(2) \cite{G-1}.

The SU${}_{\alpha}$(2) factor corresponds apparently to the angular
momentum associated with representations in SO(1,3) theories.  Thus,
in our work we have found the SU${}_{\alpha}$(2) is closely related to 
the spin angular momentum representations contained in the super
multiplet and the corresponding adinkra network.  On the other hand, the 
SU${}_{\beta}$(2) representation factor is apparently related to an extension 
of R-symmetry.  This ``extended SU(2) R-symmetry'' is only apparent in one 
dimensional valise representations and is usually broken in higher 
dimensional representations.  This behavior is very different from what 
is seen in the spinning particle with extended SUSY \cite{HPPT}.

In a classical work  \cite{HPPT} it was shown the Hilbert space of the $N
$-extended spinning particle leads to a spectrum of states whose spin is 
related to $N$ (the number of supercharges) by the relation $spin$ = 
(1/2)$N$.  On the other hand, within the adinkra network approach with
$N$ = 4 (i.e.\ only four supercharges), we have found the ``size'' of the 
SU${}_{\alpha}$(2) representations contained in the corresponding 
adinkra network determines the maximum spin in the supermultiplet.
In the network-based adinkra approach, changing the number of link
colors in the adinkra, amounts to changing the number of space-time
supercharges on the 0-brane equivalent.

This raises the very intriguing question of whether it is possible to
construct a spinning particle-like action where an O(4) symmetry
is the one seen in the adinkra network-based approach appears?

Our next observation is the methodology we have discussed in this paper
can be extended to form similar analyses of on-shell supersymmetical theories.
Owing to the fact that such theories do {\em {not}} form representations of the
``Garden Algebra'' as noted in \cite{G-1}, any such analysis is expected to more
complicated than that presented with this work for the off-shell theories.

We have long been aware the definition of the L-matrices and R-matrices
required by 1D, $N$-extended SUSY implies the quantites
\be \eqalign{
\gamma{}_{\rm I} ~=~ 
\left[\begin{array}{cc}
~0 & ~ {\rm L}_\rI \\
~ {\rm R}_\rJ & ~0 \\
\end{array} \right] 
} \label{JHproj3}
\ee
define a set of SO(4) $\gamma$ matrices.  Furthermore for these SO(4) 
$\gamma$ matrices, a SO(4) `chirality' $\gamma{}^{\rm{CH}}$-matrix can 
be defined by
\be \eqalign{
\gamma{}^{\rm{CH}} ~=~ 
\left[\begin{array}{cc}
~{\bm {\rm I}}{}_4 & ~ 0 \\
~ 0 & ~- \, {\bm {\rm I}}{}_4 \\
\end{array} \right] 
} \label{JHproj4}
\ee
so
\be  {
{\rm L}_{\rI} ~=~ \fracm 12 \, 
\left[ \,   
{\bm {\rm I}}{}_8 \,+\, \gamma{}^{CH}  \, 
\right] \, \gamma{}_{\rm I}  ~~~,~~~ {\rm R}_{\rI} ~=~ \fracm 12 \, 
\left[ \,   
{\bm {\rm I}}{}_8 \,-\, \gamma{}^{CH}  \, 
\right] \, \gamma{}_{\rm I}  ~~~.
} \label{JHproj5}
\ee

The equation in (\ref{JHproj1}) implies the $\Tilde V$-matrices occur as 
projections arising from the covariant fermionic holoraumy tensor $(\mathscr{
F}{}_{{a} \, {b}}) $.  Using the equations of (\ref{VH3}), (\ref{VH4}), (\ref{covVAL1
}), and (\ref{BH2}), the ``Adinkra/$\gamma$-matrix Equation'' can be written as
\be \eqalign{
&\mathscr{P} \left[ \,  \left( {\rm R}{}^{\Delta} \right){}_{\Hat {\Lambda}}{\,}_{a} 
 \, \left( {\rm L}{}_{\Delta}\right){}_{b}{}^{\Hat {\Delta}} ~-~
  \left( {\rm R}{}^{\Delta} \right){}_{\Hat {\Lambda}}{\,}_{b} 
 \, \left( {\rm L}{}_{\Delta}\right){}_{a}{}^{\Hat {\Delta}} 
\, \right]   ~=~  \cr 
&{~~~~} 
\left[ \,
 (\,{\rm R}_\rI\,)_\hi{}^j\>(\, {\rm L}_\rJ\,)_j{}^\hk ~-~ (\,{\rm R}_\rJ\,)_\hi{}^j\>(\,{\rm 
 L}_\rI\,)_j{}^\hk  \, \right] ~~~,
} \label{JHproj2z}
\ee
where the projection operator simply has the effect of changing the
labels on the indices according to $\mathscr{P}: ({\Hat {\Lambda}}, \, {\Hat 
{\Delta}})$ $\to$ $(\hi , \, \hk)$ and $\mathscr{P}: (a , \, b) $ $\to$ $({\rm I} , {\rm J})$.

The left hand side of (\ref{JHproj2z}) involves some Fierz identities of the 
$\gamma$-matrices for SO(1,3) while on the right hand side there are the 
chiral parts of the SO(4) generators calculated in the eight dimensional 
spinor representation.   Identities of this type appear to be the fundamental 
reason SUSY Holography has worked in all our studies completed thus far.  We
are not aware of any prior statements in the literature on the existence of
identities of this character.

The final observation we make concerns the results reported in equations
(\ref{M1}) - (\ref{VTthetan}).  The matrix calculated on the 0-brane reduction
side is only a 5 $\times$ 5 matrix as there only exist the five representations
(CS), (VS), (TS), (AVS), and (ATS)\footnote{The CS actually has two `variant
representations \cite{Varep}, while the (VS) and (AVS) each possess \\ $~~~~~~$ 
a single such representation.}.  On the adinkra side, however, if one does
not invoke any equivalence relations, there are 1,536 separate representations
as found in \cite{permutadnk}.  

A total 1,180,416 inner products  would need to be calculated on the adinkra 
side in order to test the proposition that the only entries in (\ref{M6}) are either 
0, 1, or $-1/3$.  This is the most severe test of the concept of ``SUSY Holography'' 
proposed to date.  However, if there are entries among the 1,180,416 inner 
products that do {\it {not}} correspond to either  0, 1, or $-1/3$, the metric on the 
space can be used as a filter to reject the corresponding network-based adinkras 
as candidates that correspond to being ``shadows'' of the higher dimensional 
supermultiplets.

Let us end by noting that the concept of supersymmetrical 0-brane holoraumy 
tensors does not depend on the existence of adinkra networks.  The results in 
chapters four and five are wholly independent of the networks as all calculations 
in these chapter can be completed without ever knowing any properties of the 
networks.   This means that 0-brane holoraumy tensor and the metric defined in 
(\ref{M2z}) provide {\it {new}} tools for exploring the representation theory of 
supersymmetrical theories.

In future works, we will report on our continuing efforts, remaining conscious
of a warning from Aristotle.

 \vspace{.05in}
 \begin{center}
\parbox{4in}{{\it ``The so-called Pythagoreans, who were the first to 
take up mathematics, not only advanced this subject, but saturated 
with it, they fancied that the principles of mathematics were the 
principles of all things.''}\,\,-\,\, Aristotle $~~~~~~~~~$}
 \parbox{4in}{
 $~~$}  
 \end{center}
 
  \noindent
{\bf Acknowledgements}\\[.1in] \indent
We would like to acknowledge Professors Kevin Iga, Tristan H\" ubsch, 
and Stefan Mendez-Diaz for conversations.  This work was partially supported 
by the National Science Foundation grants PHY-0652983 and PHY-0354401. 
This research was also supported in part the University of Maryland Center for 
String \& Particle Theory.  Additional acknowledgment is given by M.\ Calkins, 
and D.\ E.\ A.\ Gates, to the Center  for String and Particle Theory, as well as 
recognition for their participation in 2013 \& 2014 SSTPRS (Student Summer 
Theoretical Physics Research Session) programs.  Some adinkras were drawn 
with the aid of the {\em  Adinkramat\/ }~\copyright\,2008 by G.~Landweber. and 
others by T.\ H\" ubsch.    

\newpage

\setcounter{equation}{0}
\noindent
{\bf \Large {Appendix A: Covariant 0-Brane Bosonic Holoraumy Tensors}} 
$~~~$ 
 
In this appendix, we briefly give the results for the calculations of the
bosonic holoraumy tensors for four of the supermultiplets seen below,
$$ \eqalign{
( \,  \mathscr{B}{}_{ab}{}^{(CS)}\, ){}_{i}^{~j} ~&= ~ C_{ab} (\gamma^{12})_i^{~j} 
+ i (\gamma^5)_{ab} (\gamma^{23})_{i}^{~j} - i (\gamma^5 \gamma^0)_{ab} 
(\gamma^{31})_i^{~j}  ~~~, \cr   
( \,  \mathscr{B}{}_{ab}{}^{(VS)}\, ){}_{i}^{~j} ~&=~- i (\gamma^5 \gamma^1)_{ab} 
(\gamma^{23})_i^{~j} + i (\gamma^5 \gamma^2)_{ab} (\gamma^{12})_i^{~j} - i 
(\gamma^5 \gamma^3)_{ab} (\gamma^{31})_i^{~j} ~~~,  \cr   
( \,  \mathscr{B}{}_{ab}{}^{(TS)}\, ){}_{i}^{~j} ~&=~-i (\gamma^5\gamma^1)_{ab} 
(\gamma^{12})_i^{~j} - i (\gamma^5\gamma^2)_{ab} (\gamma^{23})_{i}^{~j} - 
i (\gamma^5 \gamma^3)_{ab} (\gamma^{31})_i^{~j} ~~~, \cr   
( \,  \mathscr{B}{}_{ab}{}^{(TCS)}\, ){}_{i}^{~j} ~&=~ - i (\gamma^5 \gamma^1
)_{ab} (\gamma^{31})_i^{~j} - i (\gamma^5 \gamma^2)_{ab} (\gamma^{12}
)_i^{~j} + i (\gamma^5 \gamma^3)_{ab} (\gamma^{23})_i^{~j}  ~~~.
}  
\eqno(A.1) $$ 
As it is not yet clear what role of the bosonic holoraumy tensors
play, we have not concentrated upon them in the current work.  In the
future we plan to take up this question. 
$$~~$$

\setcounter{equation}{0}
\noindent
{\bf \Large {Appendix B: }} 
$~~~$

The standard Chiral Supermultiplet formulation is given by
$$
\eqalign{   {~~~~~~~~~~~}
{\rm D}_a A  ~&=~ \psi_a  ~~~~~~~~~~\,~~~~~~~~,~~~
{\rm D}_a B  ~=~ i\, ( \gamma^5 )_a{}^b \psi_b  ~~~~~~~~\,~~,~~~~~~~~ 
 \cr
{\rm D}_a \psi_b  ~&=~ i\, ( \gamma^{\mu} )_{ab}  \left( \,\partial_{
\mu} A  \, \right) - ( \gamma^5 \gamma^{\mu})_{ab}  \left( \, 
\partial_{\mu} B  \, \right)  \cr
& {~~~~~} - i C_{ab}  F   ~+~ 
( \gamma^5 )_{ab}  \, G  
~~~~~~~~~~~~~~\,~~~~~~~~~~~~~~~~\,~~~, \cr
{\rm D}_a F  ~&=~ ( \gamma^{\mu})_a{}^b \,  \pa_{\mu} \psi_b  ~~~~\,~~, ~~~
{\rm D}_a G  ~=~ i\, ( \gamma^5 \gamma^{\mu} )_a{}^b \, \pa_{\mu} \psi_b   
~~\,~~~, ~~~~~~~~~~~~~
}  \eqno(B.1)
$$
which leads to a Chiral Supermultiplet Valise formulation with 1D, $N$ = 
4 SUSY given by,
$$
\eqalign{
{\rm D}_a A  ~&=~ \psi_a  ~~~~~~~~~~\,~~~~,~~~
{\rm D}_a B   ~=~ i \, ( \gamma^5 )_a{}^b \psi_b  ~~~~~~~~~~,~~~  \cr
{\rm D}_a F  ~&=~ ( \gamma^0)_a{}^b \, \psi_b  ~~~~\,~~, ~~~
{\rm D}_a G   ~=~ i\, ( \gamma^5 \gamma^0 )_a{}^b \, \psi_b   
~~~~\,~~, \cr
{\rm D}_a \psi_b  ~&=~ i\, ( \gamma^0 )_{ab}  \left( \,\partial_{
\tau} A  \, \right) - ( \gamma^5 \gamma^0)_{ab}  \left( \, 
\partial_{\tau} B  \, \right)  \cr
& {~~~~~} - i C_{ab}  \left( \,  \partial_{\tau} F  \, \right) + 
( \gamma^5 )_{ab}  \left( \, \partial_{\tau} G   \, \right) 
~~~~~~~~~~~~~~\,~~~.
}
 \eqno(B.2)
$$
Next in a similar manner we begin with the 4D, $N$ = 1 Vector Supermultiplet 
formulation
$$  \eqalign{
{\rm D}{}_a A_{\mu} & ~=~ (\gamma_{\mu})_a{}^b \lambda_b  
 ~~~~~~~~~~~~~~~~~~~~~~~~~~~~~\,~~~, \cr
{\rm D}{}_a \lambda_b & ~=~ - i \, \fracm 12 \, (  \gamma^{\mu}
\gamma^{\nu} )_{ab} \,  F{}_{\mu \, \nu}  \, ~+~ (\gamma^5
)_{ab} \,   {\rm d} ~~~,  \cr  
{\rm D}_a {\rm d} & ~=~  i (\gamma^5 \gamma^{\mu})_a{}^b \, \pa_{\mu} 
\lambda_b ~~~~~~~~~~~~~~\,~~~~~~~\,~~~,  
}  \eqno(B.3)
$$
which leads to a Vector Supermultiplet Valise formulation with 1D, $N$ = 4 
SUSY given by,
$$
\eqalign{
{\rm D}{}_a A_m & ~=~ (\gamma_m)_a{}^b \lambda_b  ~~~, ~~~ {\rm D}_a 
{\rm d}  \, =\,  i (\gamma^5 \gamma^0)_a{}^b \, \lambda_b 
 ~~~~~~~~~, \cr
{\rm D}{}_a \lambda_b & ~=~ - i \, (  \gamma^0 \gamma^m )_{ab} \,  \left( \,  
\pa_{\tau} A_m  \, \right)~+~ (\gamma^5)_{ab} \,  \left( \, \pa_{\tau} {\rm d} \, 
\right) ~~~.  
}  \eqno(B.4)
$$
For the 4D, $N$ = 1 Tensor Supermultiplet formulation
$$
 \eqalign{
{\rm D}_a \varphi ~&=~ \chi_a  ~~~, \cr
{\rm D}_a B{}_{\mu \, \nu } ~&=~ -\, \fracm 14 ( [\, \gamma_{\mu}
\, , \,  \gamma_{\nu} \,]){}_a{}^b \, \chi_b  ~~~, \cr
{\rm D}_a \chi_b ~&=~ i\, (\gamma^\mu){}_{a \,b}\,  \partial_\mu \varphi 
~-~  (\gamma^5\gamma^\mu){}_{a \,b} \, \e{}_{\mu}{}^{\r \, \s \, \t}
\partial_\r B {}_{\s \, \t}~~.
}  \eqno(B.5)
$$
leads to a Tensor Supermultiplet Valise formulation with 1D, $N$ = 4 SUSY
given by,
$$
\eqalign{
{\rm D}{}_a \varphi ~&=~ \chi_a   ~~~, ~~~
{\rm D}{}_a B_{m \, n}  ~=~ - \tfrac{1}{4} ([\gamma_m,\, 
\gamma_n])_a{}^b \chi_b  ~\,~~, \cr
{\rm D}{}_a \chi_b ~&=~ i (\gamma^0)_{ab} \,  \partial_{\tau}
\varphi - i  \tfrac{1}{2} (\gamma^0 \, [\gamma^m,\, \gamma^n]
)_{ab}  \, \partial_\tau B_{m \, n}  
 ~~~~.
}  \eqno(B.6)
$$
Finally, there exist a second supermultiplet which contains a spin-1 gauge
field.  The component fields shown in (B.1) and (B.2) are appropriate since
the gauge field is a normal vector.  However, it is also possible to introduce 
a axial Vector Supermultiplet (AVS) with component fields $U_{\mu}$, ${\Tilde 
{\lambda}}{}_a$, and ${\Tilde {\rm d}} $.  The main difference with the VS is 
here the component field $U_{\mu}$ corresponds to an axial vector.  
Accordingly, the D-equations are changed to the forms
$$  \eqalign{
{\rm D}{}_a U_{\mu} & ~=~ i \, (\gamma^5 \gamma_{\mu})_a{}^b {\Tilde {
\lambda}}{}_b   ~~~~~~~~~~~~~~~~~~~~~~~~~~~~~\,~~~, \cr
{\rm D}{}_a {\Tilde {\lambda}}{}_b & ~=~   \fracm 12 \, (\gamma^5  \gamma^{
\mu} \gamma^{\nu} )_{ab} \,  F{}_{\mu \, \nu}(U)  \, ~+~ i C{}_{ab} \,   {\Tilde 
{\rm d}} ~~~~~~~,  \cr  
{\rm D}_a {\Tilde {\rm d}} & ~=~  -\,  ( \gamma^{\mu})_a{}^b \, \pa_{\mu} 
{\Tilde {\lambda}}{}_b 
~~~~~~~~~~~~~~\,~~~~~~~~\,~~~\,~~~,  
}  \eqno(B.7)
$$
where $F{}_{\mu \, \nu}(U)$ simply indicates the `curl' of the $U$-gauge field.
So an axial Vector Supermultiplet Valise formulation with 1D, $N$ = 4 SUSY
is given by,
$$
\eqalign{
{\rm D}{}_a U_m & ~=~ i\, (\gamma^5 \gamma_m)_a{}^b {\Tilde {\lambda}}_b  
~~~, ~~~ {\rm D}_a {\Tilde {\rm d}}  \, =\, -\,  ( \gamma^0)_a{}^b \,  {\Tilde {
\lambda}}{}_b   ~~~~~~~~~, \cr
{\rm D}{}_a {\Tilde {\lambda}}{}_b & ~=~   (  \gamma^5  \gamma^0 
\gamma^m )_{ab} \,  \left( \,  \pa_{\tau} U_m  \, \right)
~+~ i \, C{}_{ab} \, ( \, \pa_{\tau} {\Tilde {\rm d}} \, ) 
~~~~~\,~~.  
}  \eqno(B.8)
$$

On comparing (B.3) with (B.7), it is seen the latter corresponds applying a set 
of replacements in the former according to $V$ $\to$ $U$, $\lambda_a$ $\to$  
$-\, i\, (\gamma^5){}_a {}^b {\Tilde {\lambda}}{}_b$, and d $\to$ ${\Tilde {\rm d}}$.  
In turn, this observation implies the possibility of constructing an axial Tensor 
Supermultiplet (ATS) via replacements in the (B.6) according to: $B$ $\to$ ${
\Tilde B}$, $\chi_a$ $\to$  $-\, i\, (\gamma^5){}_a {}^b {\Tilde {\chi }}{}_b$, and 
$\varphi$ $\to$ ${\Tilde {\varphi }}$.

Let us emphasize the bosonic and fermionic functions appearing in (B.2), (B.4), 
(B.6), and (B.7) depend solely on a single real continuous parameter $\tau$.  
Though these field variable have indices, these should be regarded as isospin 
indices as there are no spatial coordinate in these four equations.  In a similar 
manner, the matrices multiplying such indices are also to be thought of as linear 
operators acting in isospin space.  It should be noted it is consistent to assign 
to all the bosonic fields in (B.2), (B.4), (B.6), and (B.7) the same engineering 
dimensions.  Likewise this can also be done for all the fermionic fields in the 
same four equations, but of course the engineering dimension for the bosonic 
fields are distinct from that of the fermionic fields.
$$~~$$

\setcounter{equation}{0}
\noindent
{\bf \Large {Appendix C: Detailed Derivation}} 
$~~~$ 

In this appendix, we provide a detailed derivation of the steps led from
the equations in (\ref{Khy3}) to those in (\ref{Khy8b}).  As indicated in
(\ref{Khy3}) we begin with the equations
$$ \eqalign{
{\rm D}_a \, A{}_{\mu} ~&=~  (\gamma_\mu){}_a {}^b \,  \l_b  ~~~, \cr
{\rm D}_a \l_b ~&=~   - \,i \, \fracm 14 ( [\, \gamma^{\mu}\, , \,  \gamma^{\nu} 
\,]){}_a{}_b \, (\,  \partial_\mu  \, A{}_{\nu}    ~-~  \partial_\nu \, A{}_{\mu}  \, )
~+~  (\gamma^5){}_{a \,b} \,    {\rm d} ~~,  \cr
{\rm D}_a \, {\rm d} ~&=~  i \, (\gamma^5\gamma^\mu){}_a {}^b \, 
\,  \partial_\mu \l_b  ~~~, \cr
}    \eqno(C.1)  $$
as our starting point.  The first of these sets of equations in (C.1) simply gives 
us
$$ \eqalign{
{\rm D}_a \, A{}_{0} ~&=~  (\gamma_0){}_a {}^b \,  \l_b  ~~~, ~~~~
{\rm D}_a \, A{}_{1} ~=~  (\gamma_1){}_a {}^b \,  \l_b  ~~~, \cr
{\rm D}_a \, A{}_{2} ~&=~  (\gamma_2){}_a {}^b \,  \l_b  ~~~, ~~~
{\rm D}_a \, A{}_{3} ~=~  (\gamma_3){}_a {}^b \,  \l_b  ~~~, \cr
}   \eqno(C.2) $$
and the second set yields
$$ \eqalign{
{\rm D}_a \l_b ~&=~   - \,i \, \fracm 12 ( [\, \gamma^{0}\, , \, \gamma^{1} 
\,]){}_a{}_b \, (\,  \partial_0  \, A{}_{1}    ~-~  \partial_1 \, A{}_{0}  \, )
  - \,i \, \fracm 12 ( [\, \gamma^{1}\, , \, \gamma^{2} 
\,]){}_a{}_b \, (\,  \partial_1  \, A{}_{2}     ~-~  \partial_2 \, A{}_{1}  \, )
 ~~  \cr
~&~ ~~~~   - \,i \, \fracm 12 ( [\, \gamma^{0}\, , \, \gamma^{2} 
\,]){}_a{}_b \, (\,  \partial_0  \, A{}_{2}    ~-~  \partial_2 \, A{}_{0}  \, )
 - \,i \, \fracm 12 ( [\, \gamma^{2}\, , \, \gamma^{3} 
\,]){}_a{}_b \, (\,  \partial_2  \, A{}_{3}     ~-~  \partial_3 \, A{}_{2}  \, )
 ~~  \cr
~&~ ~~~~   - \,i \, \fracm 12 ( [\, \gamma^{0}\, , \, \gamma^{3} 
\,]){}_a{}_b \, (\,  \partial_0  \, A{}_{3}    ~-~  \partial_3 \, A{}_{0}  \, )
 - \,i \, \fracm 12 ( [\, \gamma^{3}\, , \, \gamma^{1} 
\,]){}_a{}_b \, (\,  \partial_3  \, A{}_{1}     ~-~  \partial_1 \, A{}_{3}  \, )
 ~~  \cr
~&~ ~~~~ +~  (\gamma^5){}_{a \,b} \,    {\rm d} ~~.  \cr
} \eqno(C.3)
$$

Next in (C.1) we delete entirely the first equation and go to the Coulomb 
gauge in (C.2).  This leaves us with
$$ \eqalign{ {~~~~~~~~~}
{\rm D}_a \, A{}_{1} ~=~  (\gamma_1){}_a {}^b \,  \l_b  ~~~, ~~~
{\rm D}_a \, A{}_{2} ~=~  (\gamma_2){}_a {}^b \,  \l_b  ~~~, ~~~
{\rm D}_a \, A{}_{3} ~=~  (\gamma_3){}_a {}^b \,  \l_b  ~~~,
}  \eqno(C.4)
$$
and in (C.3) we use both the gauge condition and the restrictions on the
coordinate dependence to find
$$ \eqalign{
{\rm D}_a \l_b ~&=~  
  + \,i \, \fracm 12 ( [\, \gamma^{2}\, , \, \gamma^{0} 
\,]){}_a{}_b \, (\,  \partial_0  \, A{}_{2}   \, )   + \,i \, \fracm 12 ( [\, \gamma^{2}\, 
, \, \gamma^{1} \,]){}_a{}_b \, (\,  \partial_1  \, A{}_{2}   \, )
 ~~  \cr
~&~ ~~~~   + \,i \, \fracm 12 ( [\, \gamma^{3}\, , \, \gamma^{0} 
\,]){}_a{}_b \, (\,  \partial_0  \, A{}_{3}   \, )
 + \,i \, \fracm 12 ( [\, \gamma^{3}\, , \, \gamma^{1} 
\,]){}_a{}_b \, (\,   \partial_1 \, A{}_{3}  \, )
 ~~  \cr
~&~ ~~~~ +~  (\gamma^5){}_{a \,b} \,    {\rm d} 
 - \,i \, \fracm 12 ( [\, \gamma^{0}\, , \, \gamma^{1} 
\,]){}_a{}_b \, (\,  \partial_0  \, A{}_{1}  \, )
~~.  \cr
}   \eqno(C.5)
$$
With a bit more algebra
$$ \eqalign{
{\rm D}_a \l_b ~&=~  
+ \,i \, ( \gamma^{2} \gamma^{0} ){}_a{}_b \, (\,  \partial_0  \, A{}_{2}   \, )   
+ \,i \,(  \gamma^{2} \gamma^{1}){}_a{}_b \, (\,  \partial_1  \, A{}_{2}   \, )
 ~~  \cr
~&~ ~~~~   + \,i \,  (  \gamma^{3} \gamma^{0}){}_a{}_b \, (\,  \partial_0  \, 
A{}_{3}   \, ) + \,i \, ( \gamma^{3} \gamma^{1}){}_a{}_b \, (\,   \partial_1 \, 
A{}_{3}  \, ) ~~  \cr
~&~ ~~~~ +~  (\gamma^5){}_{a \,b} \,    {\rm d}  - \,i \,  ( \gamma^{
 0} \gamma^{1} ){}_a{}_b \, {\widehat {\rm F}} ~~,  \cr
}   \eqno(C.6)
$$
where we have introduced the notation
$$
{\widehat {\rm F}} ~=~ (\,  \partial_0  \, A{}_{1}   \, )   ~~~.
 \eqno(C.7)
$$
Finally the last equations of (C.1) become
$$  {
{\rm D}_a \, {\rm d} ~=~  i \, (\gamma^5\gamma^0){}_a {}^b \, 
\,  \partial_0 \l_b  ~+~  i \, (\gamma^5\gamma^1){}_a {}^b \, 
\,  \partial_1 \l_b   ~~~.  }    \eqno(C.8)
$$
The latter two equations of (C.4), along with (C.6), and (C.7) are
covariant with respect to the two dimensional Lorentz symmetry of the
space with coordinates $x^0$ and $x^1$.  For (C.6) and (C.7) we
must take into consideration the Coulomb gauge condition.

We can make one final change of notation 
$$
A_2 ~\equiv~ {\widehat A}    ~~,~~ A_3 ~\equiv~ {\widehat B} ~~,~~ 
\l_a ~\equiv~ {\widehat \psi}_a ~~,~~ {\rm d} ~\equiv~ {\widehat G}
~~~,  \eqno(C.9)
$$
to obtain
$$
\eqalign{
{\rm D}_a \, {\widehat A} ~&=~  (\gamma_2){}_a {}^b \, {\widehat \psi}_b  
~~,~~ {\rm D}_a \, {\widehat B} ~=~  (\gamma_3){}_a {}^b \, {\widehat 
\psi}_b  ~~~~~~~~~~\,~~~~~\,~~, ~~~  \cr
 {\rm D}_a {\widehat \psi}_b ~&=~  i \, ( \gamma^{2}  \gamma^{0} ){}_a{}_b 
\, (\,  \partial_0  \, {\widehat A}   \, )   + \,i \,  (  \gamma^{2} \gamma^{1} ){}_a
{}_b \, (\,   \partial_1 \, {\widehat A}  \, )
 ~~  \cr
~&~ ~~~~+~   i \, (  \gamma^{3} \gamma^{0} ){}_a{}_b \, (\,  \partial_0
\, {\widehat B}    \, )  + \,i \,  ( \gamma^{3} \gamma^{1} ){}_a{}_b \, (\,  
\partial_1  \, {\widehat B}  \, )  \cr
~&~ ~~~~   - \,i \,  ( \gamma^{0} \gamma^{1} ){}_a{}_b \, {\widehat F} ~+~  
(\gamma^5){}_{a \,b} \, {\widehat G}  ~~,   \cr
{\rm D}_a \, {\widehat F} ~&=~  (\gamma_1){}_a {}^b \, (\, \partial_0 
{\widehat \psi}_b \, ) ~-~  (\gamma_0){}_a {}^b \, ( \, \partial_1 {\widehat 
\psi}_b \, ) ~~,~~  \cr
{\rm D}_a \, {\widehat G} ~&=~  i \, (\gamma^5\gamma^0){}_a {}^b \, 
\, ( \partial_0{\widehat \psi}_b  )  ~+~  i \, (\gamma^5\gamma^1){}_a {}^b \, 
\, ( \partial_1{\widehat \psi}_b  ) 
~~~. \cr
} \eqno(C.10)
$$
This is a formulation of the  $\cal N$ = 2 twisted chiral supermultiplet (TCS) 
in two dimensions.  The process above is precisely the one that led to the 
discovery of the twisted chiral supermultiplet.
$$~~$$

\noindent
{\bf \Large {Appendix D: Comparison of Defining Conditions} }
$~~~$

The equations on the first line of (\ref{Khy10}) can be used to derive a first 
order spinorial differential equation on $A$ and $B$.   We note
$$ {
{\rm D}_a \, {B} ~=~  i \, (\gamma^5){}_a {}^b \, { \psi}_b  ~~~\Rightarrow ~~~ 
{\psi}_b   ~=~ - \, i (\gamma^5){}_b {}^c \,  {\rm D}_c \, {B} ~~~,
}    \eqno(D.1)
$$
and this final result can be substituted into the first equation in (\ref{Khy11})
to derive
$$ {
{\rm D}_a \, {A} ~+~  i \, (\gamma^5){}_a {}^b \,  {\rm D}_b \, {B} ~=~ 0 ~~~.
}    \eqno(D.2)
$$
By using similar arguments based on  (\ref{Khy8b}) we find a first order spinorial 
differential equation imposed upon $\widehat A$, and $\widehat B$
$$ {
{\rm D}_a \, {\widehat B} ~=~   (\gamma_3){}_a {}^b \, {\widehat \psi}_b  ~~~\Rightarrow
~~~ {\widehat \psi}_b   ~=~  (\gamma_3){}_b {}^c \,  {\rm D}_c \, {\widehat B} ~~~,
}    \eqno(D.3)
$$
which leads to
$$ {
{\rm D}_a \, {\widehat A} ~-~   (\gamma_2 \, \gamma_3  ){}_a {}^b \,  {\rm D}_b 
\, {\widehat B} ~=~ 0 ~~~.}    \eqno(D.4)
$$
The condition in (D.2) corresponds to the real formulation of what is known as
the ``chirality condition'' and in a similar manner (D.4) corresponds to the real 
formulation of what is known as the ``twisted chirality condition.''

The equation in (D.1) implies
$$
 \eqalign{
 & \to {~}  
 {\rm D}_b {\big (}  \, A ~+~ i B  \, {\big)} ~=~  {\big [}  \, ({\rm { I}}
 ){}_b{}^c  ~-~    (\gamma^5){}_b{}^c \, {\big]}  \,  \psi_c  
  ~\,~~~~~~~~\,~~\,, ~~
  \cr
 & \to {~}  
\fracm 12 \,  {\big [}  \, ({\rm { I}})  ~+~    (\gamma^5) \, {\big]}{}_a{}^b \,  
 {\rm D}_b {\big (}  \, A ~+~ i B  \, {\big)} ~=~   0  
   ~\,~~~~~~~~~\,~~\,, ~~   \cr
 & \to {~}   
\fracm 12 \, 
  \left[\begin{array}{c}
{\rm D}_{1} ~+~ i \, {\rm D}{}_4 \\
{\rm D}_{2} ~-~ i \, {\rm D}{}_3 \\
i \, (\, {\rm D}_{2}  ~-~ i \, {\rm D}{}_3  \,)  \\
-\, i \, (\, {\rm D}_{1}  ~+~ i \, {\rm D}{}_4  \,)  \\
\end{array}\right] {\big (}  \, {A} ~+~ i {B} \, {\big)} ~=~  0   ~~~,
} \eqno(D.5)
$$
so there are only two linearly independent defining conditions
$$  {
\left[\begin{array}{c}
{\rm D}_{1} ~+~ i \, {\rm D}{}_4 \\
\end{array}\right] {\big (}  \, {A} ~+~ i {B} \, {\big)} ~=~  0 ~~,~~ 
\left[\begin{array}{c}
{\rm D}_{2} ~-~ i \, {\rm D}{}_3 \\
\end{array}\right] {\big (}  \, {A} ~+~ i {B} \, {\big)} ~=~  0  
~~~~,}  \eqno(D.6)
$$
and of course the complex conjugates of these equations.

The equation in (D.4) implies
$$
\eqalign{ {~~~}
{\rm D}_a \, {\widehat A} ~&=~  {\widehat \psi}_a  ~~~~~~~~~,~~~~
{\rm D}_a \, {\widehat B} ~=~  - \, (\gamma_2 \gamma_3){}_a {}^b \, 
{\widehat \psi}_b   ~\,~~~~~~~~\,~~\,, ~~   \cr
 & \to {~}   
 {\rm D}_b {\big (}  \, {\widehat A} ~+~ i {\widehat B} \, {\big)} ~=~  {\big [}  \, 
 ({\rm { I}}){}_b{}^c  ~-~  i\, (\gamma^2 \gamma^3){}_b{}^c \, {\big]} \, \psi_b
   ~\,~\,~~~\,, ~~   \cr
 & \to {~}   
\fracm 12 \,   {\big [}  \, ({\rm { I}})  ~+~  i\,  (\gamma^2 \gamma^3) \, {\big]} {
}_a{}^b {\rm D}_b {\big (}  \, {\widehat A} ~+~ i {\widehat B} \, {\big)} ~=~  0  
~~~~~~~~ ,~~    \cr
 & \to {~}   
\fracm 12 \,   \left[\begin{array}{c}
{\rm D}_{1} ~+~ i \, {\rm D}{}_4 \\
{\rm D}_{2} ~+~ i \, {\rm D}{}_3 \\
-\, i \, (\, {\rm D}_{2}  ~+~ i \, {\rm D}{}_3  \,)  \\
-\, i \, (\, {\rm D}_{1}  ~+~ i \, {\rm D}{}_4  \,)  \\
\end{array}\right]{\big (}  \, {\widehat A} ~+~ i {\widehat B} \, {\big)} ~=~  0 
~~~.
}  \eqno(D.7)
$$
Once more there are only two linearly independent defining conditions,
$$  {
\left[\begin{array}{c}
{\rm D}_{1} ~+~ i \, {\rm D}{}_4 \\
\end{array}\right]{\big (}  \, {\widehat A} ~+~ i {\widehat B} \, {\big)} 
~=~  0  ~~,~~
\left[\begin{array}{c}
{\rm D}_{2} ~+~ i \, {\rm D}{}_3 \\
\end{array}\right]{\big (}  \, {\widehat A} ~+~ i {\widehat B} \, {\big)} 
~=~  0  ~~,  }  \eqno(D.8)
$$
and of course the complex conjugates of these equations also.  
The only difference between (D.6) and (D.8) is the replacement
D${}_3$ $\to$ $-$ D${}_3$.  If we write the two equations in 
(D.6) in the forms of
$$  {
{\Bar {\rm D}}{}_+ {\big (}  \, {A} ~+~ i {B} \, {\big)} ~=~  0 ~~,~~ 
{\Bar {\rm D}}{}_- {\big (}  \, {A} ~+~ i {B} \, {\big)} ~=~  0  
~~~~,}  \eqno(D.9)
$$
utilizing the 2D complex light supercovariant derivatives ${\Bar {\rm D}}{}_+$, 
and ${\Bar {\rm D}}{}_-$ then we find (D.8) takes the form
$$  {
{\Bar {\rm D}}{}_+ {\big (}  \, {\widehat A} ~+~ i {\widehat B} \, {\big)} 
~=~  0  ~~,~~
{ {\rm D}}{}_- {\big (}  \, {\widehat A} ~+~ i {\widehat B} \, {\big)} 
~=~  0  ~~,  }  \eqno(D.10)
$$
which is in accord with the definitions of the chiral and twisted chiral 2D, $\cal N$
= 2 supermultiplets given in 1984.

$$~~$$
{\bf \Large {Appendix E: }} 
$~~~$

In the work of \cite{G-1}, we introduced shorthand notations for 4 $\times$ 4 matrices
${\a}^1$, ${\a}^2$, ${\a}^3$, ${\b}^1$, ${\b}^2$, ${\b}^3$ where
$$ \eqalign{
&~~~~{\a}^1 =~ \s^2 \otimes \s^1 ~~~~~\,~~~~~,~~~~~ {\a}^2 = 
{\bf I}{}_2  \otimes \s^2  ~~   ~~~~~~~~~\,~~,~~~~{\a}^3 = \s^2 \otimes \s^3 ~~~~~~~~~~~, \cr
 &~~~~~~~=~ i \, (3)_b \, (14)(23) ~~~,  ~~~~~~~~=~ i \, (5)_b \, (12)(34)
  ~~\,~~,  ~~~~~~~=~ i \, (9)_b \, (13)(24)  ~~,   \cr
 &~~~~{\b}^1 =~ \s^1 \otimes \s^2 ~~~~~\,~~~~~,~~~~~ {\b}^2 = 
 \s^2 \otimes {\bf I}{}_2   ~~   ~~~~~~~~~~~~,~~\,~{\b}^3 = \s^3 \otimes \s^2~~~~~~~~~~~, \cr
 &~~~~~~~=~ i \, (5)_b \, (14)(23) ~~~,  ~~~~~~~~=~ i \, (3)_b \, (13)(24)
  ~~~\,~,  ~~~~~~~=~ i \, (9)_b \, (12)(34)  ~~.   \cr 
}  \eqno(E.1)
$$
where the outer product conventions are in \cite{G-1} and the Boolean
factor/permutation conventions are in \cite{redux}.  These matrices
satisfy the identities
$$  \eqalign{ {~~~~~~~}
\a^{\Hat I} \, \a^{\Hat K} ~&=~ \delta{}^{{\Hat I} \, {\Hat K}} \,  {\bm {\rm I}}{}_4
~+~ i \, \epsilon{}^{{\Hat I}  \, {\Hat K} \, {\Hat L}} \, \a^{\Hat K} ~~~,~~~ \b^{\Hat 
I} \, \b^{\Hat K} ~=~ \delta{}^{{\Hat I} \, {\Hat K}} \,  {\bm {\rm I}}{}_4 ~+~ i \, 
\epsilon{}^{{\Hat I}  \, {\Hat K} \, {\Hat L}} \, \b^{\Hat K}  ~~,   \cr
& {~~~~~~~~~} {\rm {Tr}} \big( \,  \a^{\Hat I}  \, \big) ~=~ {\rm {Tr}} \big( \, \b^{\Hat 
I}  \, \big) ~=~ 4 ~~,~~ \big[  \,   \a^{\Hat I} ~,~   \b^{\Hat K} \, \big]  ~=~ 0 ~~~.
}   \eqno(E.2)
$$
These will be useful in the following discussion of this chapter.

Now we can use the adinkra in (Adnk-A) to find the associated L-matrices (and 
R-matrices which are simply the transposed versions of the L-matrices).  Using 
the notation in (E.1), and the rainbow assignment of (black, green, red, blue) 
colors to corresponding link numbers (1, 2, 3, 4) respectively,  for the L-matrices 
yields
$$  \eqalign{
{\rm L}_1^{(Adnk-A)} ~&=~ {\bf I}{}_4  ~~,~~ {\rm L}_2^{(Adnk-A)} ~=~  i\,  
\b^{3} ~~,~~ \cr
{\rm L}_3^{(Adnk-A)} ~&=~ -\, i \,  \b^{2}  ~~,~~ {\rm L}_4^{(Adnk-A)} ~=~ -\, i \, 
\b^{1} ~~,}
\eqno(E.3)
$$
associated with (Adnk-A).  These imply the corresponding ${\Tilde V}$ matrices 
associated with (Adnk-A) take the forms
$$  \eqalign{
{\Tilde V}_{1 \, 2}^{({\rm {Adnk-A}})} ~&=~  i\,  \b^{3}  ~~~~~~,~~ {\Tilde V}_{2 \, 3
}^{({\rm {Adnk-A}})} ~=~ \, i \, \b^{1}  ~~~~~,~ ~  {\Tilde V}_{3 \, 4}^{({\rm {Adnk-A
}})} ~=~   i\,  \b^{3}  ~~, \cr
{\Tilde V}_{1 \, 3}^{({\rm {Adnk-A}})} ~&=~  -\,  i\,  \b^{2}  ~~,~ ~    {\Tilde V}_{2 \, 
4}^{({\rm {Adnk-A}})} ~=~  -\,  i\,  \b^{2}   ~,  \cr
{\Tilde V}_{1 \, 4}^{({\rm {Adnk-A}})} ~&=~  -\,  i\,  \b^{1}  ~~,~~
}   \eqno(E.4)
$$
and these equations further imply
$$ \eqalign{  {~~~~~~~~}
{\Tilde V}_{1 \, 2}^{({\rm {Adnk-A}})} ~&=~ {\Tilde V}_{3 \, 4}^{({\rm {Adnk-A}})} ~~,~~
{\Tilde V}_{2 \, 3}^{({\rm {Adnk-A}})} ~=~ -\, {\Tilde V}_{1 \, 4}^{({\rm {Adnk-A}})} ~~,~~
\cr
{\Tilde V}_{1 \, 3}^{({\rm {Adnk-A}})} ~&=~ {\Tilde V}_{2 \, 4}^{({\rm {Adnk-A}})}  ~~~.
}   \eqno(E.5)
$$
By use of the adinkra in (Adnk-B), it is seen one only needs to make the 
replacement of L${}_3$ $\to$ $- \, $L${}_3$ 
to find
$$  \eqalign{
{\rm L}_1^{(Adnk-B)} ~&=~ {\bf I}{}_4  ~~,~~ {\rm L}_2^{(Adnk-B)} ~=~  i\,  \b^{3} ~~,
~~  \cr
{\rm L}_3^{(Adnk-B)} ~&=~ \, i \,  \b^{2}  ~~,~~ {\rm L}_4^{(Adnk-B)} ~=~ -\, i \, \b^{1} 
~~,~~ 
~~.}
\eqno(E.6)
$$
and this leads to the corresponding $\Tilde V$ matrices given by
$$
 \eqalign{ {~~~~~~~~}
{\Tilde V}_{1 \, 2}^{({\rm {Adnk-B}})} ~&=~  i\,  \b^{3}  ~~~~~~~,~ ~{\Tilde 
V}_{2 \, 3}^{({\rm {Adnk-B}})} ~=~ - \, i \, \b^{1}  ~~~,~ ~ 
 {\Tilde V}_{3 \, 4}^{({\rm {Adnk-B}})} ~=~ -\,  i\,  \b^{3}  ~~, \cr
{\Tilde V}_{1 \, 3}^{({\rm {Adnk-B}})} ~&=~  \,  i\,  \b^{2}  ~~~~~\,~,~~    
 {\Tilde V}_{2 \, 4}^{({\rm {Adnk-B}})} ~=~  -\,  i\,  \b^{2}   ~~~,  \cr
{\Tilde V}_{1 \, 4}^{({\rm {Adnk-B}})} ~&=~  -\,  i\,  \b^{1}  ~~~,~~
}   \eqno(E.7)
$$
and these equations further imply
$$
\eqalign{  {~~~~~~~~}
{\Tilde V}_{1 \, 2}^{({\rm {Adnk-B}})} ~&=~ - \, {\Tilde V}_{3 \, 4}^{({\rm {Adnk-B}})} ~~,~~
{\Tilde V}_{2 \, 3}^{({\rm {Adnk-B}})} ~=~ {\Tilde V}_{1 \, 4}^{({\rm {Adnk-B}})} ~~,~~ \cr
{\Tilde V}_{1 \, 3}^{({\rm {Adnk-B}})} ~&=~  -\, {\Tilde V}_{2 \, 4}^{({\rm {Adnk-B}})}  ~~~.
}  \eqno(E.8)
$$
We can further see
$$
\eqalign{  {~~~~~~~~}
{\Tilde V}_{1 \, 2}^{({\rm {Adnk-A}})} ~&=~ - \, {\Tilde V}_{1 \, 2}^{({\rm {Adnk-B}})} ~~,~~
{\Tilde V}_{1 \, 3}^{({\rm {Adnk-A}})} ~=~ - \, {\Tilde V}_{1 \, 3}^{({\rm {Adnk-B}})} ~~,~~
\cr
{\Tilde V}_{1 \, 4}^{({\rm {Adnk-A}})} ~&=~   {\Tilde V}_{1 \, 4}^{({\rm {Adnk-B}})}  ~~~.
}  \eqno(E.9)
$$

From the results shown in (E.8) and (E.9) it is clear
$$
{{\Tilde V}_{I \, J}^{({\rm {Adnk-A}})} \, {\Tilde V}_{K \, L}^{({\rm {Adnk-B}})} ~-~  {\Tilde 
V}_{K \, L}^{({\rm {Adnk-B}})}   \, {\Tilde V}_{I \, J}^{({\rm {Adnk-A}})} ~\ne~ 0  ~~~.
} \eqno(E.10)
$$

\end{document}